\documentclass[fleqn,usenatbib]{mnras}
\usepackage[T1]{fontenc}
\DeclareRobustCommand{\VAN}[3]{#2}
\let\VANthebibliography\thebibliography
\def\thebibliography{\DeclareRobustCommand{\VAN}[3]{##3}\VANthebibliography}

\usepackage{graphicx}	
\usepackage{amsmath}	
\usepackage{amssymb}	
\usepackage{gensymb}
\usepackage{xcolor}     
\usepackage[singlelinecheck=off,labelfont=bf,font=small]{caption}
\usepackage{subcaption}
\usepackage{contour}
\contourlength{2pt}
\usepackage{longtable}
\usepackage[percent]{overpic}
\usepackage{newtxtext,newtxmath}

\DeclareUnicodeCharacter{2212}{-}

\title[Imaging polarimetry survey of SNe Ia]{An imaging polarimetry survey of Type Ia supernovae: are peculiar extinction and polarization properties produced by circumstellar or interstellar matter?}

\author[M. Chu et al.]{Matthew R. Chu$^{1}$,
Aleksandar Cikota$^{2,3}$,
Dietrich Baade$^{4}$, 
Ferdinando Patat$^{4}$,
\newauthor Alexei V. Filippenko$^{5,6}$,
J. Craig Wheeler$^{7}$,
Justyn Maund$^{8}$,
Mattia Bulla$^{9}$,
Yi Yang$^{5}$,
\newauthor Peter H\"oflich$^{10}$,
Lifan Wang$^{11}$
\\
$^{1}$Department of Physics, University of California, Berkeley, CA 94720-7300, USA\\
$^{2}$European Organisation for Astronomical Research in the Southern Hemisphere (ESO), Alonso de Cordova 3107, Vitacura, Casilla 19001, Santiago de Chile, Chile \\
$^{3}$E.O. Lawrence Berkeley National Laboratory, 1 Cyclotron Rd., Berkeley, CA, 94720, USA\\
$^{4}$European Organisation for Astronomical Research in the Southern Hemisphere (ESO), Karl-Schwarzschild-Str. 2, 85748 Garching b. M\"{u}nchen, Germany\\
$^{5}$Department of Astronomy, University of California, Berkeley, CA 94720-3411, USA\\ 
$^{6}$Miller Institute for Basic Research in Science, University of California, Berkeley, CA 94720, USA\\
$^{7}$Department of Astronomy, University of Texas, Austin, TX 78712-1205, USA\\
$^{8}$Department of Physics and Astronomy, University of Sheffield, Hicks Building, Hounsfield Road, Sheffield S3 7RH, UK\\
$^{9}$The Oskar Klein Centre, Department of Astronomy, Stockholm University, AlbaNova, SE-10691 Stockholm, Sweden\\
$^{10}$Department of Physics, Florida State University, Tallahassee, FL 32306−4350, USA \\
$^{11}$Department of Physics, Texas A\&M University, College Station, TX 77843, USA
}

\date{Accepted XXX. Received YYY; in original form ZZZ}

\pubyear{2021}

\begin{document}
\label{firstpage}
\pagerange{\pageref{firstpage}--\pageref{lastpage}}
\maketitle

\begin{abstract}
Some highly reddened Type Ia supernovae (SNe~Ia) display low total-to-selective extinction ratios ($R_V \lesssim 2$) in comparison to that of typical Milky Way dust ($R_V \approx 3.3$), and polarization curves that rise steeply to blue wavelengths, with peak polarization values at short wavelengths ($\lambda_{\rm max} < 0.4$ \textmu m) in comparison to the typical Galactic values ($\lambda_{\rm max} \approx 0.55$ \textmu m). Understanding the source of these properties could provide insight into the progenitor systems of SNe~Ia. We aim to determine whether they are the result of the host galaxy's interstellar dust or circumstellar dust. This is accomplished by analysing the continuum polarization of 66 SNe Ia in dust-rich spiral galaxies and 13 SNe Ia in dust-poor elliptical galaxies as a function of normalised galactocentric distance. We find that there is a general trend of SNe~Ia in spiral galaxies displaying increased polarization values when located closer to the host galaxies' centre, while SNe~Ia in elliptical host galaxies display low polarization. Furthermore, all highly polarized SNe~Ia in spiral host galaxies display polarization curves rising toward blue wavelengths, while no evidence of such polarization properties is shown in elliptical host galaxies. This indicates that the source of the peculiar polarization curves is likely the result of interstellar material as opposed to circumstellar material. The peculiar polarization and extinction properties observed toward some SNe Ia may be explained by the radiative torque disruption mechanism induced by the SN or the interstellar radiation field.
\end{abstract}




\section{Introduction}
\label{sect:intro}

Type Ia supernovae (SNe~Ia) are used as standardisable candles on cosmological scales; as such, measuring their redshift and apparent brightness can quantify the expansion of the observable Universe \citep{1998AJ....116.1009R, 1999ApJ...517..565P}. To understand the evolution (if any) of their peak luminosity with redshift, it is imperative to accurately identify their progenitor systems. These SNe are explosions of carbon-oxygen white dwarfs (WDs) near the Chandrasekhar limit, but identifying their exact progenitor systems from current observational results is not yet conclusive \citep{2014ARA&A..52..107M}. However, the nature of the progenitor systems may be encrypted in dust along lines of sight to SNe~Ia, since single-degenerate and double-degenerate models may imply different circumstellar environments \citep{2018PhR...736....1L}.

Prior works reveal that some highly-reddened SNe~Ia, with $E(B-V) \gtrsim 1$  mag, display two unusual characteristics: (i) low total-to-selective extinction values of $R_V \lesssim 2$ \citep[e.g.,][]{2008A&A...487...19N, 2009ApJ...699L.139W, 2010AJ....139..120F, 2014ApJ...789...32B, 2015MNRAS.453.3300A} in comparison to typical Milky Way (MW) dust with $R_V = 3.32 \pm 0.18$ \citep{2016ApJ...821...78S} or dust in the Large Magellanic Cloud with a wide range of $R_V$ values between 3 and 6 \citep{2017AJ....154..102U}, and (ii) polarization curves steeply growing toward blue wavelengths and peaking at wavelengths below 0.4\,$\mu$m \citep{patat2015, zelaya2017}, compared to typical Galactic Serkowski-like polarization curves that peak at $\sim 0.55$ \textmu m \citep{serk1975}. In general, low $R_V$ values are correlated with smaller dust grains \citep{2003ARA&A..41..241D}. Along the sight lines toward SNe~Ia, a high abundance of small dust grains may be a result of rotational disruption of larger dust grains by radiative torques in strong radiation fields, a phenomenon proposed by \citet{2019NatAs...3..766H} and \citet{2021ApJ...907...37H}. Furthermore, the wavelength of peak polarization, $\lambda_\mathrm{max}$, is found to depend on the dust grains' size distribution \citep{serk1975,2009ApJ...696....1D,2013ApJ...779..152H,2014ApJ...790....6H,2017ApJ...836...13H}. In the case of an enhanced abundance of small grains, the peak moves toward blue wavelengths. There is also a correlation of the interstellar extinction law with the wavelength of maximum polarization. For typical dust in the Milky Way, \citet{1978A&A....66...57W} deduced $R_V = (5.6 \pm 0.3) \lambda_{\rm max}$, where $\lambda_{\rm max}$ is in \textmu m. On the other hand, the steeply rising polarization curves can also be explained by Rayleigh scattering from nearby dust, which produces polarization curves proportional to $\lambda^{-4}$ \citep{2013ApJ...775...84A,patat2015}. Thus, the steeply rising polarization curves toward blue wavelengths can be explained by linear dichroism of aligned dust grains located in the interstellar medium (ISM) with a high abundance of small dust grains located, or alternatively by scattering from nearby circumstellar dust within the progenitor system.

A reason to believe that the peculiar polarization curves are a result of scattering from circumstellar material (CSM) is the similarity between polarization curves of SNe~Ia and of protoplanetary nebulae \citep[PPNe;][]{cikota2017}. Depending on the nature of the companion of the exploding WD, several varieties might produce CSM. Asymptotic giant branch (AGB) stars, after evolving into PPNe, display polarization profiles similar to those observed toward SNe~Ia \citep{patat2015}. The curves in PPNe are shown to be produced by CSM scattering \citep{opp2005}. Therefore, the common continuum polarization properties between PPNe and SNe~Ia might be an indication that some SNe~Ia explode within a PPN after the WD (formed from the initially more-massive star) merges with the core of the companion AGB star during the common-envelope phase (core-degenerate scenario, \citealt{2011MNRAS.417.1466K}; see also \citealt{2017NatAs...1E.117J, 2019NewAR..8701535S, 2019MNRAS.490.2430S, 2021MNRAS.502..176C, 2020ApJ...900..140H, 2021arXiv210209326W, 2021arXiv210612140A}).

Research conducted by \citet{bulla2018,bulla2018_2} suggests the contrary, that the dust responsible for the observed extinction in some SNe~Ia is the result of interstellar material. Assuming a spherical dust shell and using Monte-Carlo simulations, they demonstrated that the evolution of the SN light curves, particularly the colour excess $E(B-V)$, can place strong constraints on the distance between dust and the SN. \citet{bulla2018_2} find a time-variable colour excess for 15 out of their sample of 48 SNe~Ia, and constrain dust distances between 0.013\,pc and 45\,pc. The constant colour excess inferred for the rest of the sample was consistent with dust located at distances $\gtrsim 0.5$\,pc. They also find that SNe with relatively nearby dust ($< 1$\,pc) are located close to the nuclei of their host galaxies, where (in the dusty high-density region) interaction between dust and the SN radiation is more likely. Furthermore, for SNe showing time-variable reddening, \citet{bulla2018_2} notice a possible preference for low $R_V$ values. This may suggest that cloud-cloud collisions driven by the radiation pressure of SNe produce a high abundance of small dust grains, as proposed by \citet{2017ApJ...836...13H} (see also \citealt{giang2020}, who predict time-varying polarization, extinction, and colour of SNe~Ia owing to rotational disruption of dust grains). 

\citet{2013ApJ...779...38P} found a correlation between the strength of the diffuse interstellar band (DIB) at 5780\,\AA\ along the sight lines to SNe~Ia and extinction $A_V$. Because DIBs are characteristic of ISM in our Galaxy, and not of CSM, they suggest that the dust responsible for the extinction toward some SNe~Ia is predominantly located in the ISM of the host galaxy. \citet{2013ApJ...779...38P} also found that one quarter of their sample of 32 SNe~Ia display anomalously large Na~I column densities with blueshifted Na~I~D lines, suggesting that there is outflowing circumstellar gas around some SNe. Interestingly, not all SNe that display strong Na I D lines suffer from dust reddening, and some SNe with independent evidence of CSM do not display strong Na I D lines \citep{2013ApJ...779...38P}.

This work aims to determine whether the peculiar dust properties observed toward some SNe~Ia are produced by dust in the circumstellar or interstellar medium. Our main goal is to answer whether many SN\,Ia progenitors reside in environments with peculiar dust properties, are they surrounded by peculiar dust as a tracer of the evolution of the mass donor, or do the explosions radiatively modify nearby interstellar dust.
We investigate the continuum polarization of a sample of SNe~Ia and look for relations between the degree of polarization of the SN and the galactocentric distance, for different morphological galaxy types.

\begin{figure}
\includegraphics[trim=10mm 0mm 20mm 40mm, width=8.9cm, clip=true]{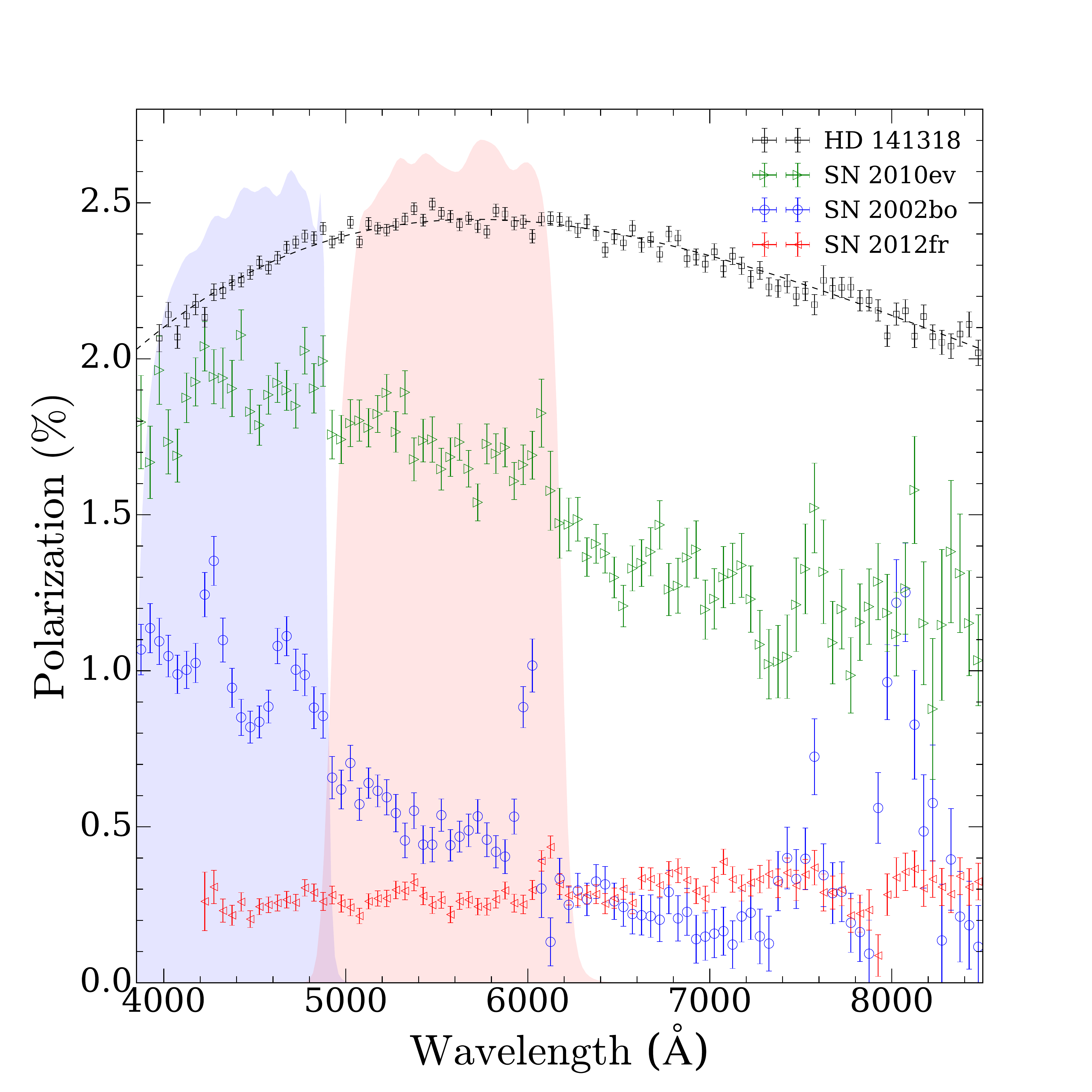}
\vspace{-5mm}
\caption{Examples of polarization curves observed toward SNe Ia, compared to a typical Serkowski-like polarization curve observed toward HD 141318 \citep{2018A&A...615A..42C}. SN\,2010ev, SN\,2002bo, and SN\,2012fr were observed with FORS1 or FORS2 at a phase of $-1.1$, $-1.4$, and $-6.9$ days relative to peak brightness, respectively \citep{2019MNRAS.490..578C}.
Also shown are the FORS2 $B$- and $V$-band filter profiles.}
\label{fig:figBVpassbands}
\end{figure}

Polarimetry allows us to examine the dust along the line of sight ``in one shot,'' independent of the epoch of the SN~Ia, because the intrinsic continuum polarization of SNe~Ia is generally negligible ($\lesssim 0.2$\%; see \citealt{2008ARA&A..46..433W}). In addition, the line polarization is $<1$\%, so the polarized lines do not significantly contaminate continuum measurements (Fig.~\ref{fig:figBVpassbands}; see also examples of polarization curves in \citealt{zelaya2017}, their Figs. 2 and 3). In contrast, to determine the reddening using photometry, multiple epochs are required to fit the light curve with intrinsic SN templates. Furthermore, from the slope of the polarization curve (derived from measurements in two different passbands), we can determine whether the polarization curve is consistent with the normal Serkowski-like curve and likely a result of linear dichroism in nonspherical grains (located in the ISM), or rising toward blue wavelengths and possibly induced by scattering from nearby clouds (dust in the CSM).

High polarization along sightlines toward SNe~Ia located far away from the centres of their host galaxies, where we expect low amounts of interstellar dust, or in elliptical galaxies, which are known to be dust poor \citep{2012ApJ...748..123S, cikota2016}, would provide strong evidence for polarization arising from dust within CSM and support, for example, the core-degenerate scenario \citep{2003ApJ...594L..93L, 2011MNRAS.417.1466K, 2012MNRAS.419.1695I, 2013MNRAS.428..579I}. On the other hand, if we do not observe high polarization values for SNe~Ia in dust-poor elliptical galaxies, and/or if there is a dependence of the steeply rising polarization curves on galactocentric distance in dust-rich spiral galaxies, this would support the hypothesis that the peculiar polarization curves are produced by host-galaxy ISM.

In Sect.~\ref{sect:targets} we describe the observations and the SN sample. Sect.~\ref{sect:reduction} explains the data reduction, the interstellar polarization removal, the determination of the normalised galactocentric distances of the SNe in their host galaxies, and the morphological classification of the host galaxies. The results are presented and discussed in Sect.~\ref{sect:results} and summarised in Sect.~\ref{sect:summary}.

\section{Targets and observations}
\label{sect:targets}

We acquired imaging polarimetry of 68 SNe~Ia with the FOcal Reducer and low dispersion Spectrograph (FORS2, \citealt{1998Msngr..94....1A}) mounted at the European Southern Observatory’s Very Large Telescope (ESO's VLT). The SNe were observed during the interval 2018--2021 (ESO observing periods P101, P102, and P104) to create a statistical sample of SN~Ia polarization measurements (Prog. IDs 0101.D-0190(A), 0102.D-0163(A), and 0104.D-0175(A), PI Cikota). 
Additionally, we included the complete sample of 35 archival SNe~Ia from \citet{2019MNRAS.490..578C} that were observed with FORS1/2 in spectropolarimetry mode between 2001 and 2015.

\subsection{Target selection}

Every year hundreds of SNe~Ia are being discovered by transient surveys such as the Zwicky Transient Facility \citep[ZTF;][]{2019PASP..131a8002B}, Pan-STARRS \citep{2016arXiv161205560C}, and {\it GAIA} \citep{2016A&A...595A...1G}, and classified by follow-up surveys such as the Public ESO Spectroscopic Survey for Transient Objects \citep[PESSTO;][]{2015A&A...579A..40S}. For instance, there were 420 SNe~Ia discovered in 2016 (according to the Transient Name Server TNS\footnote{https://www.wis-tns.org}). Of these, 210 were brighter than 18\,mag at peak, and 106 were brighter than 17\,mag. 
For this experiment we have been monitoring the TNS and the Latest Supernovae Page\footnote{https://www.rochesterastronomy.org/snimages/} for recently classified SNe, visible from the southern hemisphere, and selected the brightest ($m_V \lesssim 18$\,mag) SNe~Ia that exploded in the most nearby host galaxies of any morphological type.

\subsection{Observing strategy}
\label{sect:obs_strategy}

SNe~Ia brighter than $m_V \approx 18$\,mag were observed in imaging polarimetry (IPOL) mode through the $B$ filter, and the data were instantaneously reduced. If an observation toward a SN~Ia showed significant polarization ($p \gtrsim 1.0$\% in P101 and P102; $p \gtrsim 0.5$\% in P104), we observed the SN additionally in the $V$ band, to determine the slope of the polarization curve.
The reason that we lowered the follow-up condition from $p \approx 1.0$\% in P101 and P102 to $p \approx 0.5$\% in P104 was to increase the sample of SNe for which we know the slope of the polarization curves. During P101 and P102, only 3 out of 42 observed SNe displayed polarization values higher than 1\%, and 6 additional SNe had polarization values between 0.5\% and 1\% in the $B$ band. However, based on the sample of \citet{zelaya2017}, we initially expected that $\sim 1/3$ of SNe will have polarization values higher than 1\% in the $B$ band.

Observations in $B$ ($\rm \lambda_{\rm eff}=440$\,nm) and $V$ ($\rm \lambda_{\rm eff}=557$\,nm) are sufficient to determine whether the polarization curve is ``normal'' or steeply rising toward the blue (see, e.g., Fig. 2 of \citealt{patat2015}, and the determination of the slope by \citealt{zelaya2017}, their Fig. 2). The observing log is presented in Table~\ref{tab:obslog}. For 12 SNe we conducted follow-up observations in the $V$ band. SN\,2019vv was mistakenly observed only in the $V$ band (instead of in the $B$ band).

\section{Data reduction and methods}
\label{sect:reduction}

\subsection{Imaging polarimetry}
\label{sect:reduction_impol}

The imaging polarimetry mode of FORS2 \citep{1998Msngr..94....1A} splits the image through a Wollaston prism into two beams with perpendicular polarizations: the ordinary (o) beam and the extraordinary (e) beam. The multi-object spectroscopy (MOS) slitlets strip mask prevents overlaps of the split beams. The targets were observed in the upper frame (CHIP1) centred on the optical axis of the telescope. The polarization is determined by conducting aperture photometry of the target in the ordinary and extraordinary beams using the DAOPHOT.PHOT package \citep{1987PASP...99..191S} in IRAF. The normalised Stokes parameters ($q$ and $u$) were calculated following the FORS2 User Manual \citep{fors2} using the Fourier transform of the normalised flux differences measured at four half-wave retarder plate angles ($\theta$) of $0^\circ$, $22.5^\circ$, $45^\circ$, and $67.5^\circ$ (Eq.~\ref{eq:stokes}):
\begin{equation}
\label{eq:stokes}
\begin{array}{l}
q = \frac{2}{N} \sum_{i=0}^{N-1} F (\theta_i) \cos (4\theta_i), \\ 
u = \frac{2}{N} \sum_{i=0}^{N-1} F (\theta_i) \sin (4\theta_i).
\end{array}
\end{equation}
$F (\theta_i)$ is the normalised flux difference between the ordinary ($f^o$) and extraordinary ($f^e$) beams (Eq.~\ref{eq:normfluxdiff}):
\begin{equation}
\label{eq:normfluxdiff}
F (\theta_i) = \frac{f^o (\theta_i) - f^e (\theta_i)}{f^o (\theta_i) + f^e (\theta_i)}.
\end{equation}
In Eq.~\ref{eq:stokes}, $N$ ranges over the aforementioned half-wave retarder plate angles, $\theta_i = 22.5\degree \times i$ , with $0 \leq i \leq 3$.

The normalised Stokes parameters were corrected for the residual retardance chromatism from the superachromatic half-wave plate, using the wavelength-dependent retardance offset, $\Delta \theta (\lambda)$, tabulated in \citet{fors2}:
\begin{equation}
\label{eq:stokes-corr}
\begin{array}{l}
q_0 = q \cos 2 \Delta \theta (\lambda) - u \sin 2 \Delta \theta (\lambda), \\
u_0 = q \cos 2 \Delta \theta (\lambda) + u \sin 2 \Delta \theta (\lambda).
\end{array}
\end{equation}
Finally, the polarization is calculated as
\begin{equation}
\label{eq:pol}
p = \sqrt{q_0^2 + u_0^2}
\end{equation}
and the polarization angle as
\begin{equation}
\label{eq:polang}
\theta_0 = \frac{1}{2} \arctan \frac{u_0}{q_0}.
\end{equation}

The uncertainties in the normalised flux differences, the normalised Stokes parameters, and the polarization percent and angle were calculated by propagating the flux uncertainty of the aperture-photometry measured in the ordinary and extraordinary beams, which were extracted along with the flux measurements.
The polarization bias was corrected following Eq.~3 of \citet{1997ApJ...476L..27W}.

The imaging-polarimetry results in the $B$ and $V$ bands are listed in Table~\ref{tab:PB}.
Furthermore, we calculated the average Stokes $q$ and $u$ in the $B$ (4000 -- 4800 \AA) and $V$ (5200 -- 6000 \AA) bands for the 35 archival SNe observed in spectropolarimetry mode and published by \citet{2019MNRAS.490..578C}. The polarization measurements of these SNe are included in the analysis in this work and given in Table~\ref{tab:archivalSNe}.

\subsection{Interstellar polarization removal}

We estimated the MW interstellar polarization (ISP) along the sight lines toward the SNe~Ia using field stars in the imaging-polarimetry data. Using SExtractor \citep{1996A&AS..117..393B}, we extracted all stars in the field with a ``detection threshold'' $> 50$ standard deviations of the background noise and selected those with ``ellipticity'' $< 0.3$ and ``flags'' $< 1$. This excludes (i) galaxies and unreliable objects with neighbours that are bright and close enough to significantly bias the photometry, (ii) objects originally blended with another one, (iii) saturated objects, (iv) truncated objects, or (v) objects with incomplete or corrupt aperture data. We measured the flux of the selected stars in the ordinary and extraordinary beams using IRAF's DAOPHOT.PHOT package and calculated the polarization as described in Sect.~\ref{sect:reduction_impol}. 

There is strong linear instrumental polarization across the FORS2 field, which shows a high degree of axial symmetry and increases from $< 0.03$\% on the optical axis to $\sim 1.4$\% at the edges of the field \citep{2006PASP..118..146P}. The instrumental polarization is well characterised, within 0.05\% of the absolute value \citep{2006PASP..118..146P}. To correct for the instrumental polarization we apply a cubic polynomial of the degree of polarization as a function of the distance from the optical axis determined by \citet{2020A&A...634A..70G}:
\begin{multline}
\label{eq:polynomial}    
p_{B_\mathrm{inst}}(r) = \left( 1.114 \times 10^{-5} \right) r - \left( 1.275 \times 10^{-8} \right) r^2 \\ + \left( 9.153  \times 10^{-12} \right) r^3 ,
\end{multline}
and estimate the polarization angle as
\begin{equation}
\label{eq:polangle} 
\alpha = {\rm arctan} ((y-y_c)/(x-x_c)) ,
\end{equation}
where $p_{B_\mathrm{inst}}$ is the instrumental polarization in the $B$ band, $r$ is the distance from the optical axis $(x_c,y_c)$, and $(x,y)$ are the stellar coordinates in pixels.

\begin{figure}
\includegraphics[trim=10mm 0mm 20mm 0mm, width=4.2cm, clip=true]{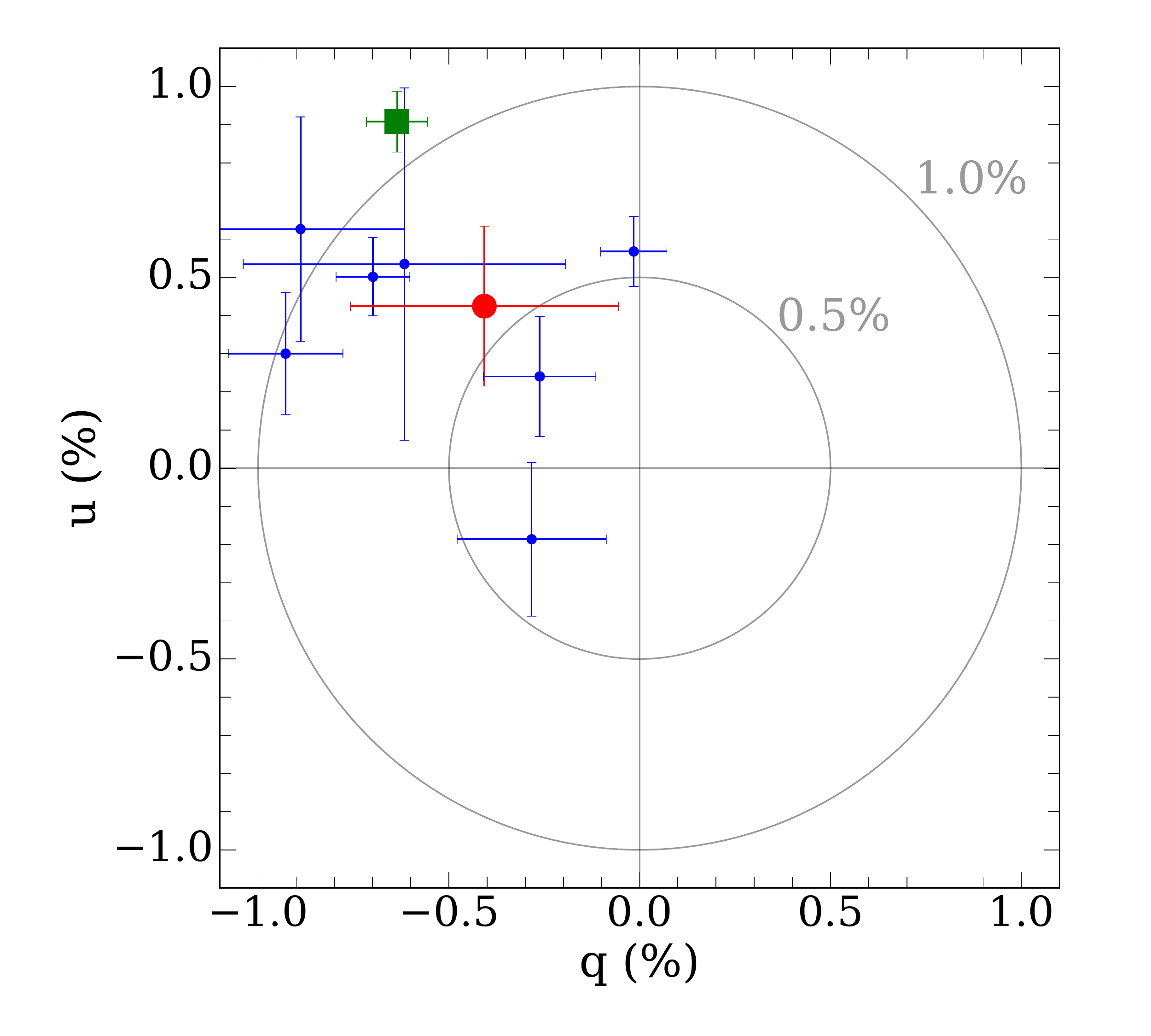}
\includegraphics[trim=10mm 0mm 20mm 0mm, width=4.2cm, clip=true]{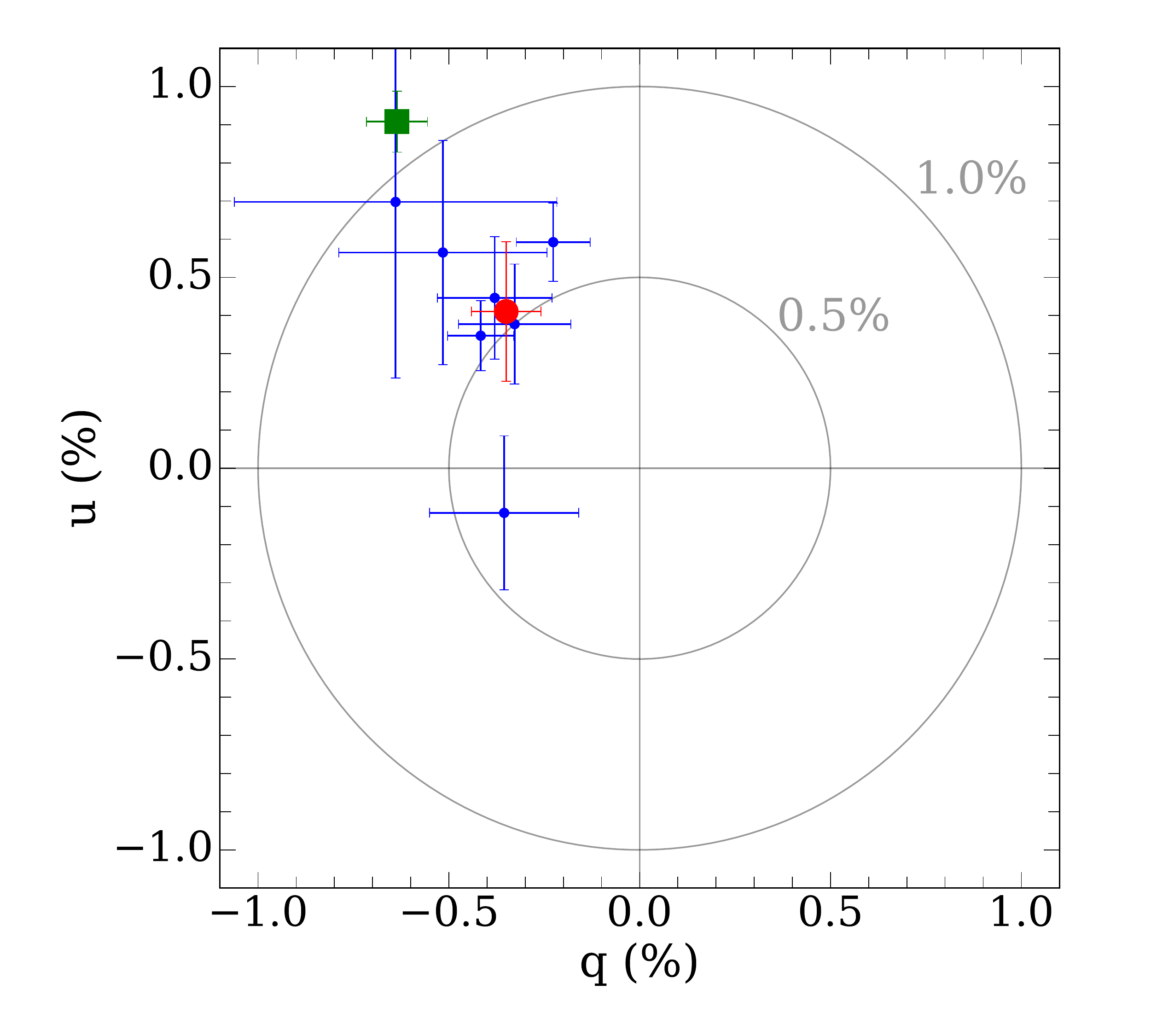}
\vspace{-7mm}
\caption{Interstellar polarization estimate using field stars in the case of SN\,2018koy. The small blue dots are the individual field-star measurements, and the large red dot shows their weighted mean. The green square displays the polarization of SN\,2018koy. The left plot shows the Stokes q and u parameter measurements uncorrected for instrumental polarization, and on the right, the same measurements are shown after correcting for the instrumental polarization. }
\label{fig:ISPcorrection}
\end{figure}

An example of field stars before and after correcting for the instrumental polarization is shown in Fig.~\ref{fig:ISPcorrection}. By determining the weighted mean of these field stars in the $q$--$u$ plane, we estimate the average interstellar polarization (produced by MW dust) for each field. The uncertainty of the average ISP is typically $\sim$0.3-0.5\% and varies from field to field, depending on the scatter and number of individually measured stars. We finally inspect the $q$--$u$ plots (e.g., shown in Fig.~\ref{fig:ISPcorrection}) and in the case of significant ISP subtract the average interstellar polarization from the SN polarization measurements:
\begin{equation}
\label{eq:stokes-corr}
\begin{array}{l}
q_\mathrm{SN} = q_0 - q_\mathrm{ISP}, \\
u_\mathrm{SN} = u_0 - u_\mathrm{ISP}.
\end{array}
\end{equation}

The estimated average value of the interstellar polarization for a field is often close to the measured polarization of the SN or located in the same quadrant in the $q$--$u$ plane, as shown in Fig.~\ref{fig:ISPcorrection} for the case of SN\,2018koy. If there is only one field star, or 2--3 field stars with inconsistent measurements in the $q$--$u$ plane, we skip the ISP correction. Furthermore, if the average ISP is close to zero and has large error bars, to avoid introducing unnecessary systematic errors, we also do not subtract the ISP. We indicate in Table~\ref{tab:PB} (``ISPcorr'' column) whether the ISP has been subtracted.

\subsection{Normalised SN position}

We determined the normalised distances of the SNe from their host galaxy centres in terms of flux percentile of the total host galaxy flux. This can be achieved by fitting elliptical isophotes to the host galaxies \citep{1987MNRAS.226..747J} and determining the flux value contained inside of the ellipse containing the SN. This value we then compared to the total flux of the galaxy and calculated the flux percentile.

To cross-match our SN sample to host galaxies, we queried our sample in the TNS for listed host galaxies. For those SNe with listed host galaxies, the host-galaxy coordinates, redshifts, and morphological classifications were acquired directly from the NASA/IPAC Extragalactic Database (NED; \citealt{1991ASSL..171...89H}), or if the information was not available in NED, cross-referenced with the HyperLEDA database \citep{Makarov2014_HyperLEDA}. For SNe without listed host galaxies in TNS or in the literature, a query in the NED database of the local region was performed and the closest appropriate listed galaxy was considered to be the host galaxy. Following this, visual checks and comparison of the SNe and host redshifts were performed to ensure accuracy. 

To fit elliptical isophotes to the host galaxies, we used archival images which are not contaminated by the SN. We obtained archival FITS \citep{1981A&AS...44..363W} images of the host galaxies in visual bands ($V$, $g$, or equivalent) from the Pan-STARRS survey \citep{2016arXiv161205560C}, the DESI Legacy Sky Survey \citep{2019AJ....157..168D}, the Sloan Digital Sky Survey \citep[SDSS;][]{2020ApJS..249....3A}, or the Digitized Sky Surveys\footnote{http://archive.eso.org/dss/dss/} (DSS). 
In these images, ellipses were fit to the galaxies at constant levels of brightness using the \texttt{photutils.isophote} library \citep{larry_bradley_2020_4044744}. This returns a list of isophote objects with attributes of semimajor axis, ellipticity, angle of rotation, the mean intensity value along the elliptical path (hereafter intensity), and the sum of all pixels inside the ellipse. The algorithm also provides a $k$-sigma clipping option for cleaning deviant sample points at each isophote. This improves the convergence stability against any nonelliptical structure such as stars, spiral arms, H~II regions, and defects.
The fit method is especially sensitive to initial conditions, and for those galaxies in which a fit could not converge or converged on visually clearly incorrect isophotes, the galaxy centre and ellipse attributes were determined manually to ensure accurate measurements of the intensity and flux values with \texttt{photutils.aperture.EllipticalAperture}.

\begin{figure}
\centering
\begin{overpic}[width=\linewidth, trim=15mm 0mm 20mm 15mm]{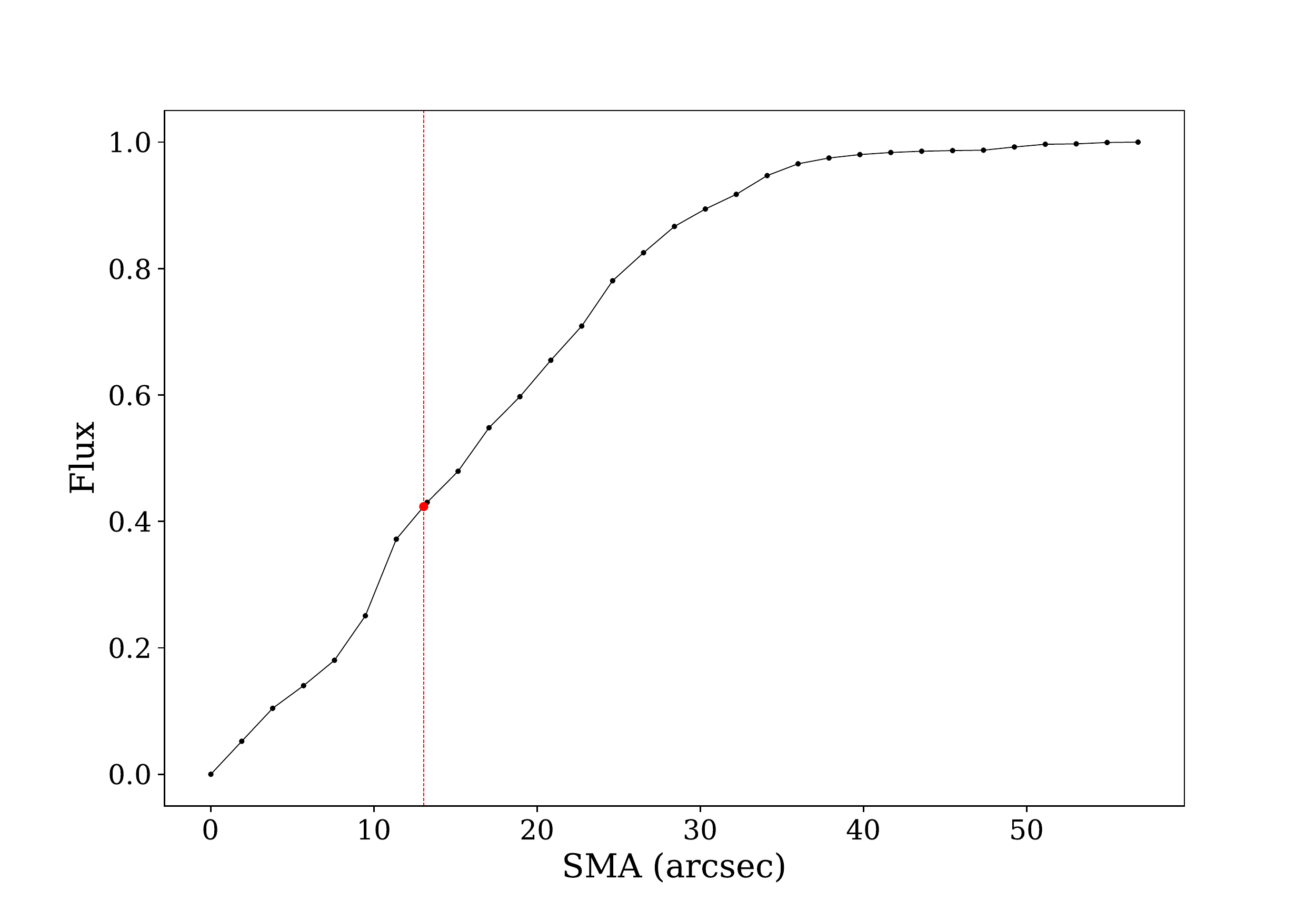}
     \put(60,11.5){\includegraphics[scale=0.36]{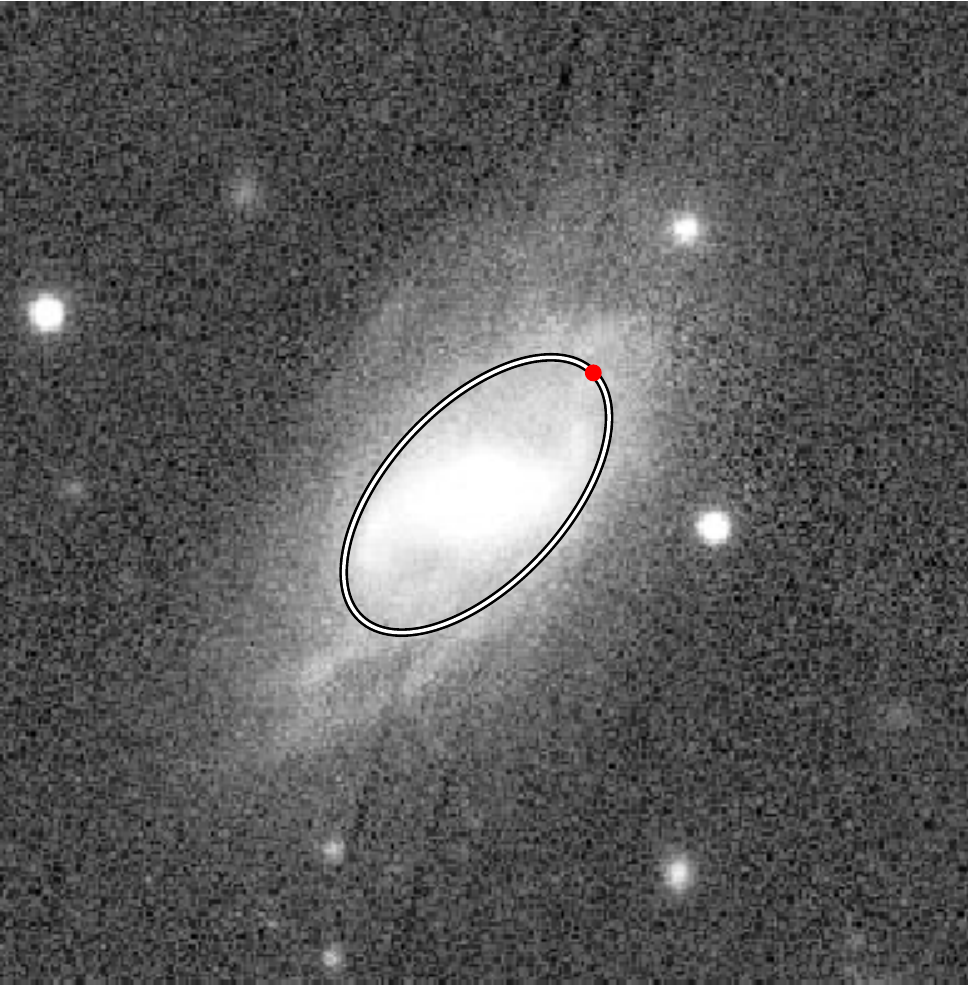}}  
  \end{overpic}
\vspace{-7mm}
\caption{Normalised flux curve of growth for MCG -01-35-011 (the host galaxy of SN\,2018evt) as a function of semimajor axis (SMA). The interpolated semimajor axis of the isophote containing SN\,2018evt is indicated with the red line. The inset shows the host galaxy MCG -01-35-011 and the isophote containing SN\,2018evt. The position of the SN is marked with the red dot.}
\label{fig:sampleflux}
\end{figure}

As the isophotes get larger, the surface brightness of the galaxies is expected to decrease to zero because of the relative scarcity of stars, and thus the total-flux curve of growth will plateau at a constant flux value. As an example, Fig.~\ref{fig:sampleflux} displays a curve of growth as a function of the semimajor axis for the host galaxy of SN\,2018evt, MCG -01-35-011. The inset in the figure shows an example of the fitted isophote containing SN\,2018evt. For the Pan-STARRS and DESI Legacy images, the background noise in the images was negligible compared to the brightness of the galaxies, so the flux curve of growth did approach an apparent plateau. However, for the DSS images, the background level was measured by calculating the average of a manually selected area, and subtracted to ensure convergence, as the background was relatively high. 

The size of the semimajor axis at the SN location can be interpolated from the pair of closest isophotes between which the SN is enclosed using the SN's coordinates. This allows us to determine the fraction of the total flux contained within the isophote containing the SN, relative to the total flux of the galaxy. This value is the flux percentile, which is the normalised metric for the SN position in the host galaxy. For example, a SN located at the 95$\rm ^{th}$ flux percentile is located at the edge of the galaxy, with 95$\%$ of the total flux contained inside of the isophote at which the SN lies. 

The advantage of parameterising the SN locations in terms of total-flux fractions is that the total-flux fractions determined by fitting isophotes do not depend on the galaxies' inclinations, in contrast to normalised galactocentric distances using half-light radii or de Vaucouleurs radii.

\subsection{Host-galaxy morphological classification}

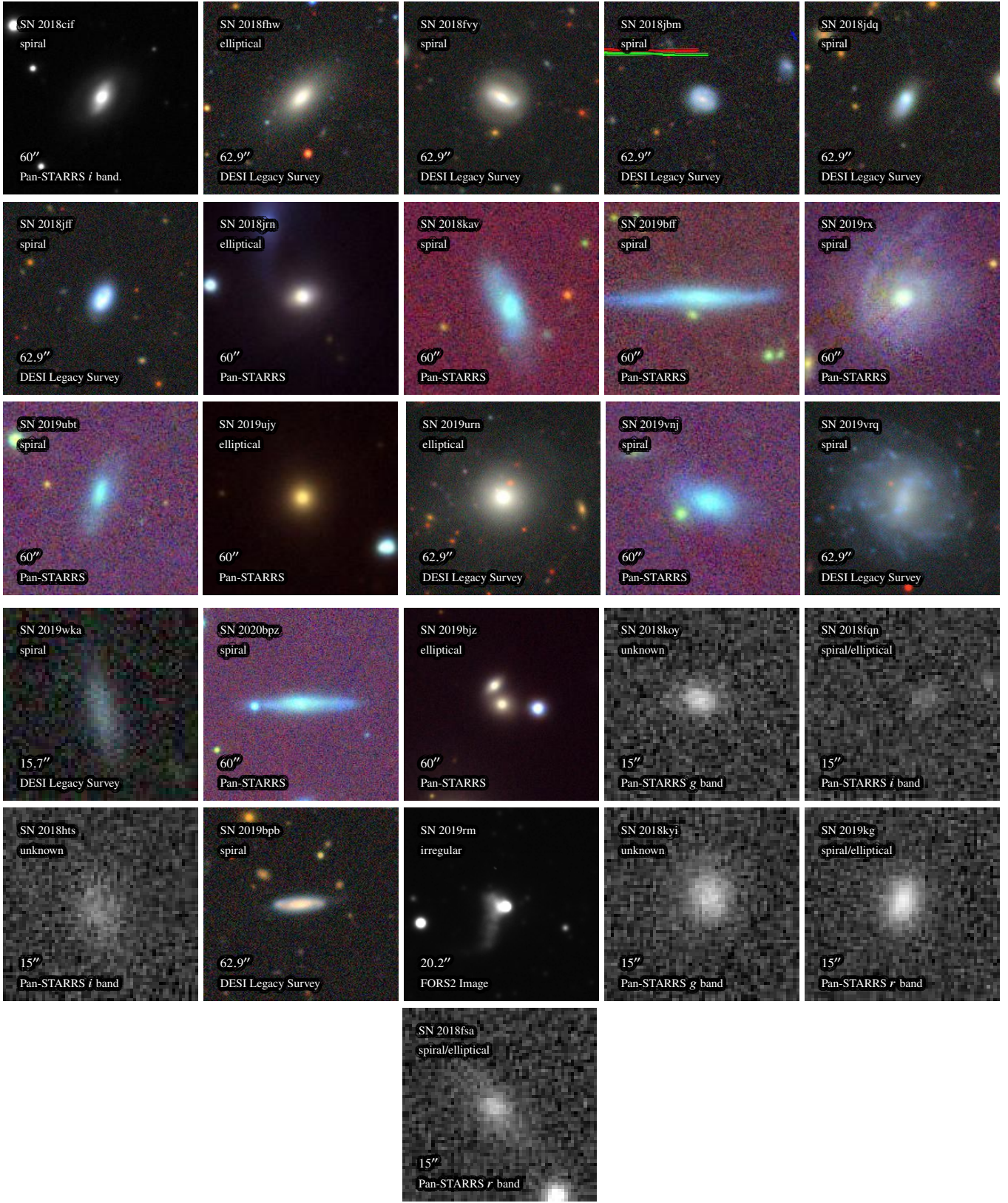
\begin{figure*}
    \centering
    \begin{subfigure}[t]{0.195\textwidth}
    \centering
    \includegraphics[width=\textwidth]{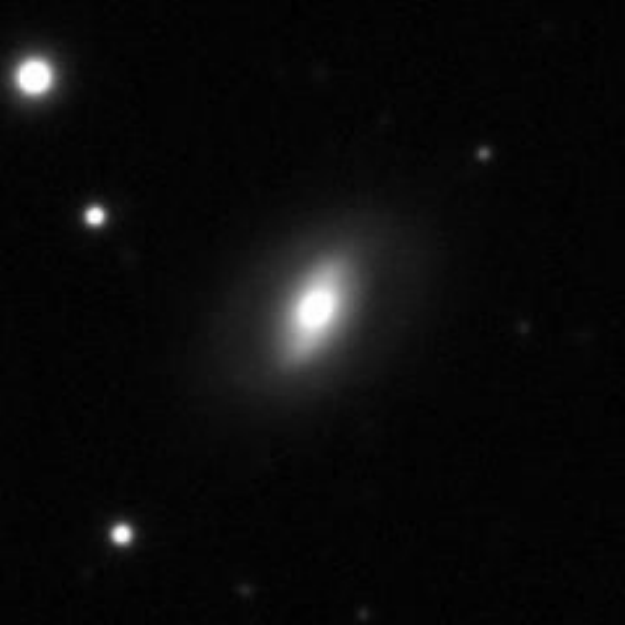}
    \scriptsize
    \put(-90,85){\color{white}\contour{black}{SN 2018cif}}
    \put(-90,75){\color{white}\contour{black}{spiral}}
    \put(-90,17){\color{white}\contour{black}{$60 \arcsec$}}
    \put(-90,7){\color{white}\contour{black}{Pan-STARRS $i$ band.}}
    \end{subfigure}
    \hfill
    \begin{subfigure}[t]{0.195\textwidth}
    \centering
    \includegraphics[width=\textwidth]{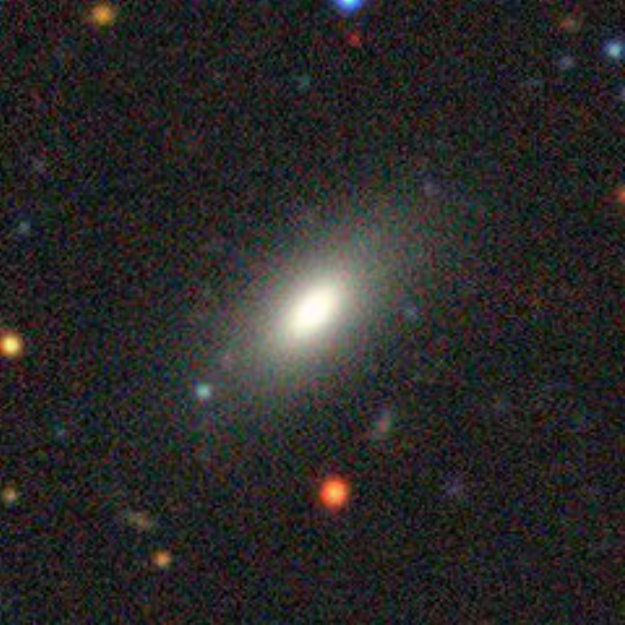}
    \scriptsize
       \put(-90,85){\color{white}\contour{black}{SN 2018fhw}}
       \put(-90,75){\color{white}\contour{black}{elliptical}}
    \put(-90,17){\color{white}\contour{black}{$62.9\arcsec$}}
    \put(-90,7){\color{white}\contour{black}{DESI Legacy Survey}}
    \end{subfigure}
    \hfill
    \begin{subfigure}[t]{0.195\textwidth}
    \centering
    \includegraphics[width=\textwidth]{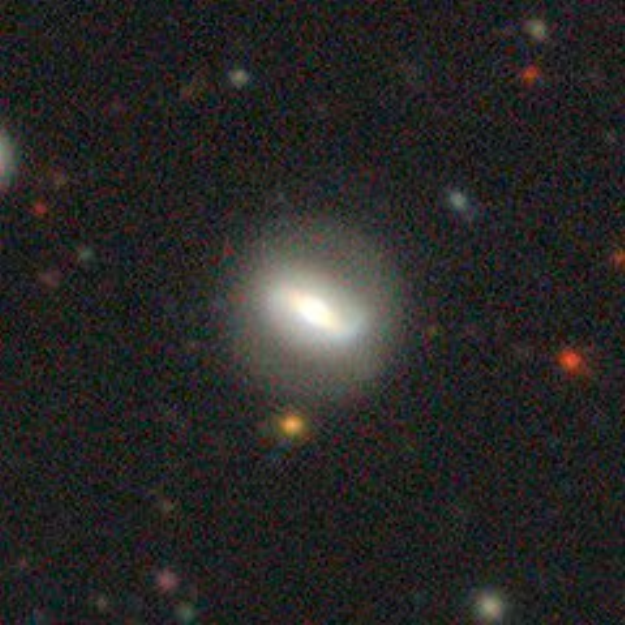}
    \scriptsize
       \put(-90,85){\color{white}\contour{black}{SN 2018fvy}}
       \put(-90,75){\color{white}\contour{black}{spiral}}
    \put(-90,17){\color{white}\contour{black}{$62.9\arcsec$}}
    \put(-90,7){\color{white}\contour{black}{DESI Legacy Survey}}
    \end{subfigure}
    \hfill
    \begin{subfigure}[t]{0.195\textwidth}
    \centering
    \includegraphics[width=\textwidth]{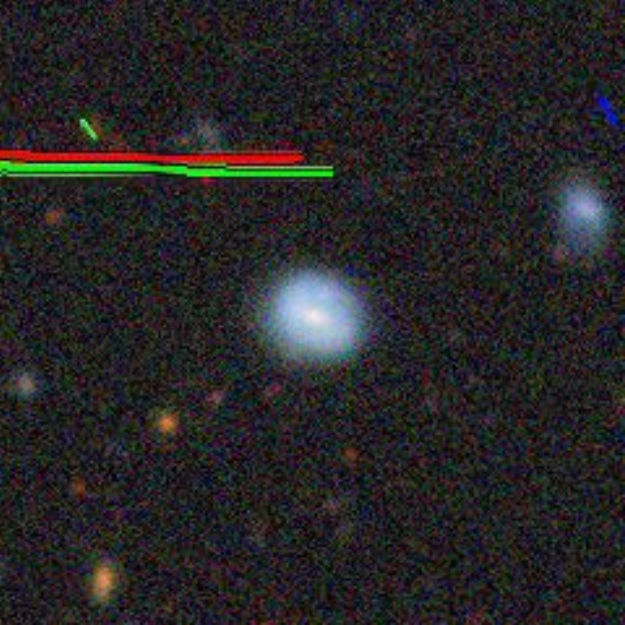}
    \scriptsize
       \put(-90,85){\color{white}\contour{black}{SN 2018jbm}}
       \put(-90,75){\color{white}\contour{black}{spiral}}
    \put(-90,17){\color{white}\contour{black}{$62.9\arcsec$}}
    \put(-90,7){\color{white}\contour{black}{DESI Legacy Survey}}
    \end{subfigure}
     \hfill
    \begin{subfigure}[t]{0.195\textwidth}
    \centering
    \includegraphics[width=\textwidth]{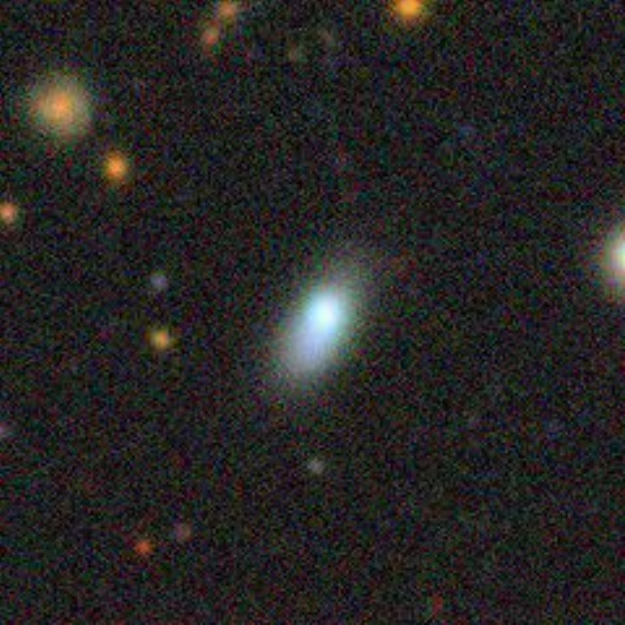}
    \scriptsize
       \put(-90,85){\color{white}\contour{black}{SN 2018jdq}}
       \put(-90,75){\color{white}\contour{black}{spiral}}
    \put(-90,17){\color{white}\contour{black}{$62.9\arcsec$}}
    \put(-90,7){\color{white}\contour{black}{DESI Legacy Survey}}
    \end{subfigure}
    \\
       \vspace{0.1cm}
    \begin{subfigure}[t]{0.195\textwidth}
    \centering
    \includegraphics[width=\textwidth]{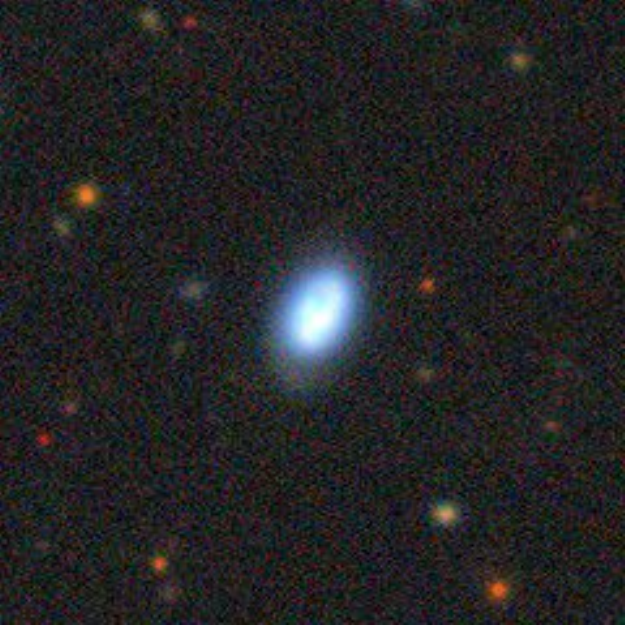}
    \scriptsize
       \put(-90,85){\color{white}\contour{black}{SN 2018jff}}
       \put(-90,75){\color{white}\contour{black}{spiral}}
    \put(-90,17){\color{white}\contour{black}{$62.9\arcsec$}}
    \put(-90,7){\color{white}\contour{black}{DESI Legacy Survey}}
    \end{subfigure}
    \hfill
    \begin{subfigure}[t]{0.195\textwidth}
    \centering
    \includegraphics[width=\textwidth]{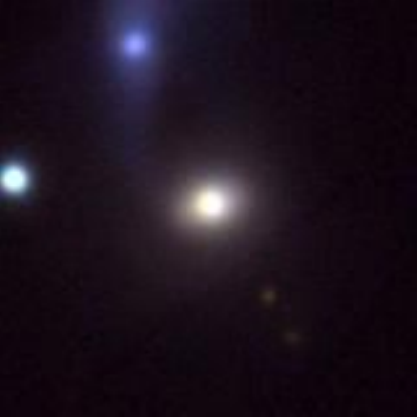}
    \scriptsize
       \put(-90,85){\color{white}\contour{black}{SN 2018jrn}}
       \put(-90,75){\color{white}\contour{black}{elliptical}}
    \put(-90,17){\color{white}\contour{black}{$60 \arcsec$}}
    \put(-90,7){\color{white}\contour{black}{Pan-STARRS}}
    \end{subfigure}
    \hfill
    \begin{subfigure}[t]{0.195\textwidth}
    \centering
    \includegraphics[width=\textwidth]{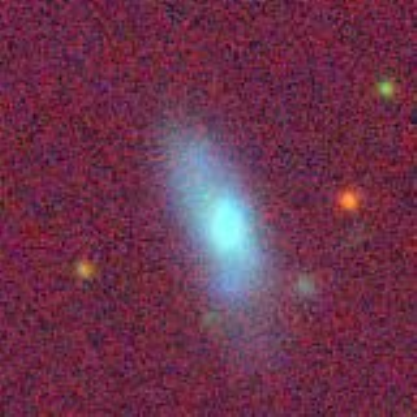}
    \scriptsize
       \put(-90,85){\color{white}\contour{black}{SN 2018kav}}
       \put(-90,75){\color{white}\contour{black}{spiral}}
    \put(-90,17){\color{white}\contour{black}{$60 \arcsec$}}
    \put(-90,7){\color{white}\contour{black}{Pan-STARRS}}
    \end{subfigure}
    \hfill
    \begin{subfigure}[t]{0.195\textwidth}
    \centering
    \includegraphics[width=\textwidth]{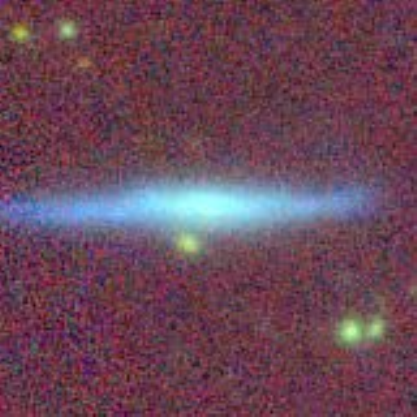}
    \scriptsize
       \put(-90,85){\color{white}\contour{black}{SN 2019bff}}
       \put(-90,75){\color{white}\contour{black}{spiral}}
    \put(-90,17){\color{white}\contour{black}{$60 \arcsec$}}
    \put(-90,7){\color{white}\contour{black}{Pan-STARRS}}
    \end{subfigure}
    \hfill
    \begin{subfigure}[t]{0.195\textwidth}
    \centering
    \includegraphics[width=\textwidth]{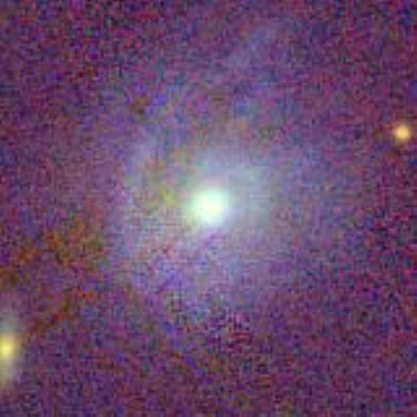}
    \scriptsize
       \put(-90,85){\color{white}\contour{black}{SN 2019rx}}
       \put(-90,75){\color{white}\contour{black}{spiral}}
    \put(-90,17){\color{white}\contour{black}{$60 \arcsec$}}
    \put(-90,7){\color{white}\contour{black}{Pan-STARRS}}
    \end{subfigure}
    \\
       \vspace{0.1cm}
    \begin{subfigure}[t]{0.195\textwidth}
    \centering
    \includegraphics[width=\textwidth]{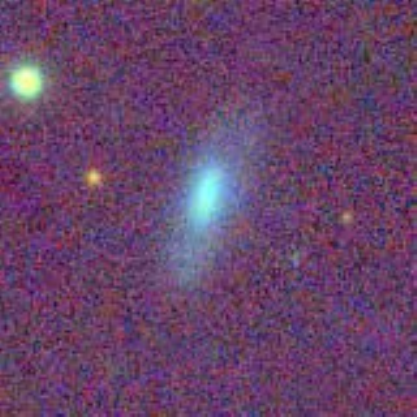}
    \scriptsize
       \put(-90,85){\color{white}\contour{black}{SN 2019ubt}}
       \put(-90,75){\color{white}\contour{black}{spiral}}
    \put(-90,17){\color{white}\contour{black}{$60 \arcsec$}}
    \put(-90,7){\color{white}\contour{black}{Pan-STARRS}}
    \end{subfigure}
    \hfill
    \begin{subfigure}[t]{0.195\textwidth}
    \centering
    \includegraphics[width=\textwidth]{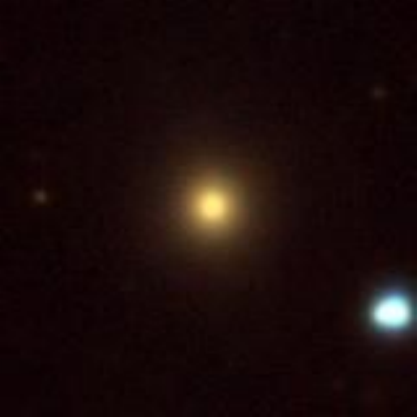}
    \scriptsize
       \put(-90,85){\color{white}\contour{black}{SN 2019ujy}}
       \put(-90,75){\color{white}\contour{black}{elliptical}}
    \put(-90,17){\color{white}\contour{black}{$60 \arcsec$}}
    \put(-90,7){\color{white}\contour{black}{Pan-STARRS}}
    \end{subfigure}
    \hfill
    \vspace{0.1cm}
    \begin{subfigure}[t]{0.195\textwidth}
    \centering
    \includegraphics[width=\textwidth]{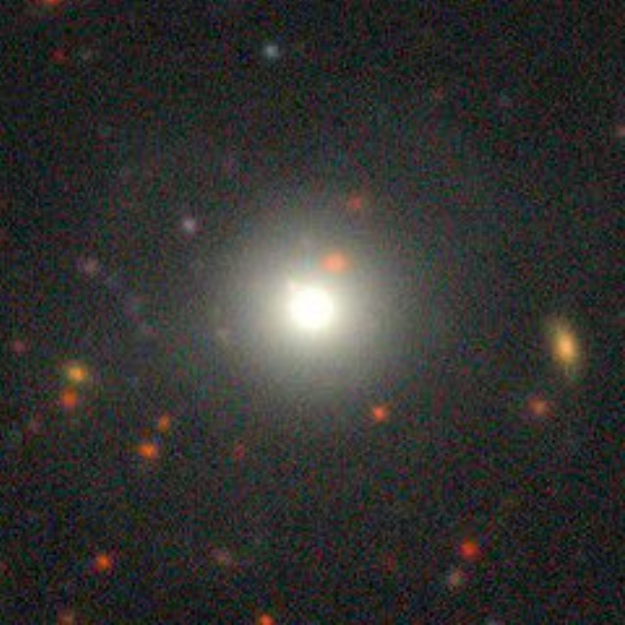}
    \scriptsize
       \put(-90,85){\color{white}\contour{black}{SN 2019urn}}
       \put(-90,75){\color{white}\contour{black}{elliptical}}
    \put(-90,17){\color{white}\contour{black}{$62.9 \arcsec$}}
    \put(-90,7){\color{white}\contour{black}{DESI Legacy Survey}}
    \end{subfigure}
    \hfill
    \begin{subfigure}[t]{0.195\textwidth}
    \centering
    \includegraphics[width=\textwidth]{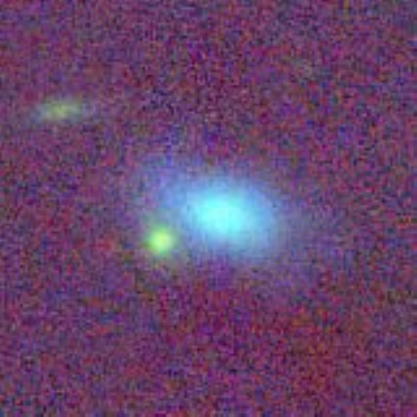}
    \scriptsize
       \put(-90,85){\color{white}\contour{black}{SN 2019vnj}}
       \put(-90,75){\color{white}\contour{black}{spiral}}
    \put(-90,17){\color{white}\contour{black}{$60 \arcsec$}}
    \put(-90,7){\color{white}\contour{black}{Pan-STARRS}}
    \end{subfigure}
    \hfill
    \begin{subfigure}[t]{0.195\textwidth}
    \centering
    \includegraphics[width=\textwidth]{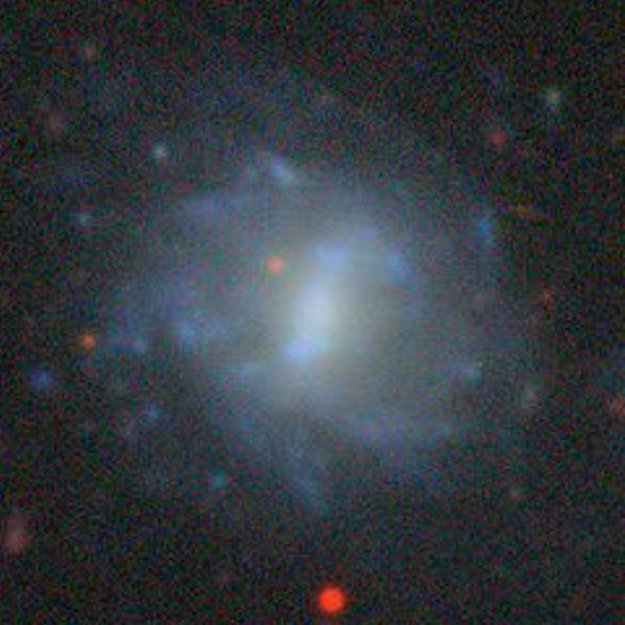}
    \scriptsize
       \put(-90,85){\color{white}\contour{black}{SN 2019vrq}}
       \put(-90,75){\color{white}\contour{black}{spiral}}
    \put(-90,17){\color{white}\contour{black}{$62.9\arcsec$}}
    \put(-90,7){\color{white}\contour{black}{DESI Legacy Survey}}
    \end{subfigure}
    \\
       \vspace{0.1cm}
    \begin{subfigure}[t]{0.195\textwidth}
    \centering
    \includegraphics[width=\textwidth]{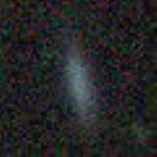}
    \scriptsize
       \put(-90,85){\color{white}\contour{black}{SN 2019wka}}
       \put(-90,75){\color{white}\contour{black}{spiral}}
    \put(-90,17){\color{white}\contour{black}{$15.7 \arcsec$}}
    \put(-90,7){\color{white}\contour{black}{DESI Legacy Survey}}
    \end{subfigure}
    \hfill
    \begin{subfigure}[t]{0.195\textwidth}
    \centering
    \includegraphics[width=\textwidth]{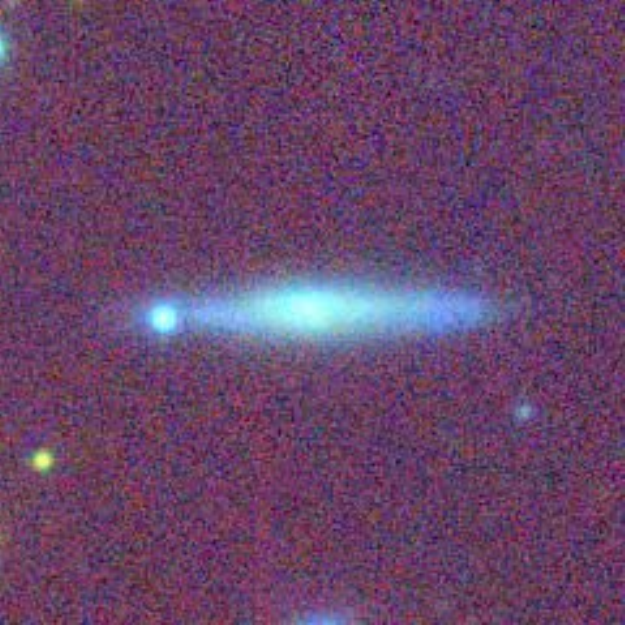}
    \scriptsize
       \put(-90,85){\color{white}\contour{black}{SN 2020bpz}}
       \put(-90,75){\color{white}\contour{black}{spiral}}
    \put(-90,17){\color{white}\contour{black}{$60 \arcsec$}}
    \put(-90,7){\color{white}\contour{black}{Pan-STARRS}}
    \end{subfigure}
    \hfill
    \begin{subfigure}[t]{0.195\textwidth}
    \centering
    \includegraphics[width=\textwidth]{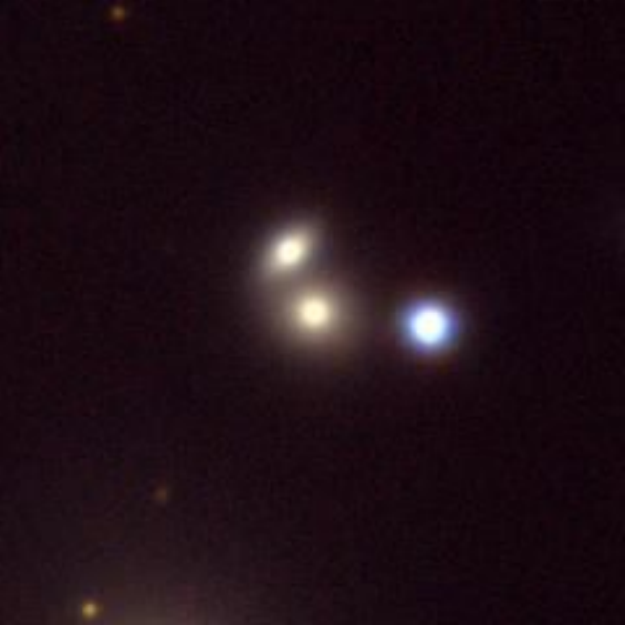}
    \scriptsize
       \put(-90,85){\color{white}\contour{black}{SN 2019bjz}}
       \put(-90,75){\color{white}\contour{black}{elliptical}}
    \put(-90,17){\color{white}\contour{black}{$60 \arcsec$}}
    \put(-90,7){\color{white}\contour{black}{Pan-STARRS}}
    \end{subfigure}
    \hfill
    \begin{subfigure}[t]{0.195\textwidth}
    \centering
    \includegraphics[width=\textwidth]{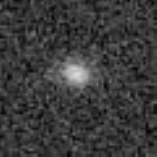}
    \scriptsize
       \put(-90,85){\color{white}\contour{black}{SN 2018koy}}
       \put(-90,75){\color{white}\contour{black}{unknown}}
    \put(-90,17){\color{white}\contour{black}{$15 \arcsec$}}
    \put(-90,7){\color{white}\contour{black}{Pan-STARRS $g$ band}}
    \end{subfigure}
    \hfill
    \begin{subfigure}[t]{0.195\textwidth}
    \centering
    \includegraphics[width=\textwidth]{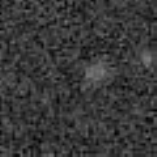}
    \scriptsize
       \put(-90,85){\color{white}\contour{black}{SN 2018fqn}}
       \put(-90,75){\color{white}\contour{black}{spiral/elliptical}}
    \put(-90,17){\color{white}\contour{black}{$15 \arcsec$}}
    \put(-90,7){\color{white}\contour{black}{Pan-STARRS $i$ band}}
    \end{subfigure}
    \\
       \vspace{0.1cm}
    \begin{subfigure}[t]{0.195\textwidth}
    \centering
    \includegraphics[width=\textwidth]{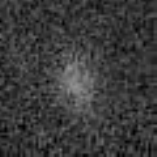}
    \scriptsize
       \put(-90,85){\color{white}\contour{black}{SN 2018hts}}
       \put(-90,75){\color{white}\contour{black}{unknown}}
    \put(-90,17){\color{white}\contour{black}{$15 \arcsec$}}
    \put(-90,7){\color{white}\contour{black}{Pan-STARRS $i$ band}}
    \end{subfigure}
    \hfill
    \begin{subfigure}[t]{0.195\textwidth}
    \centering
    \includegraphics[width=\textwidth]{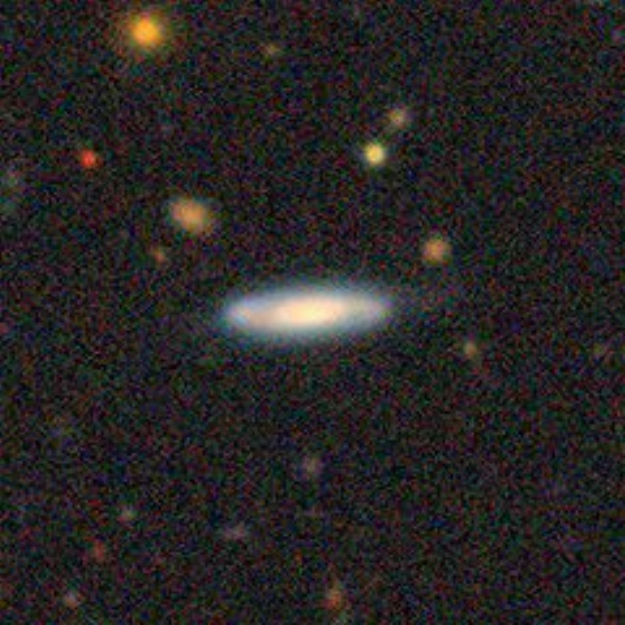}
    \scriptsize
       \put(-90,85){\color{white}\contour{black}{SN 2019bpb}}
       \put(-90,75){\color{white}\contour{black}{spiral}}
    \put(-90,17){\color{white}\contour{black}{$62.9 \arcsec$}}
    \put(-90,7){\color{white}\contour{black}{DESI Legacy Survey}}
    \end{subfigure}
    \hfill
    \begin{subfigure}[t]{0.195\textwidth}
    \centering
    \includegraphics[width=\textwidth]{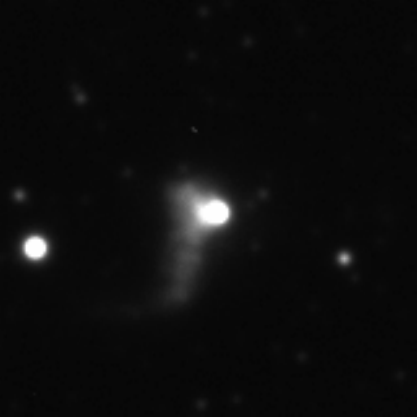}
    \scriptsize
       \put(-90,85){\color{white}\contour{black}{SN 2019rm}}
       \put(-90,75){\color{white}\contour{black}{irregular}}
    \put(-90,17){\color{white}\contour{black}{$20.2 \arcsec$}}
    \put(-90,7){\color{white}\contour{black}{FORS2 Image}}
    \end{subfigure}
    \hfill
    \begin{subfigure}[t]{0.195\textwidth}
    \centering
    \includegraphics[width=\textwidth]{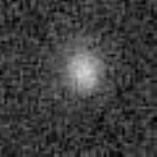}
    \scriptsize
       \put(-90,85){\color{white}\contour{black}{SN 2018kyi}}
       \put(-90,75){\color{white}\contour{black}{unknown}}
    \put(-90,17){\color{white}\contour{black}{$15 \arcsec$}}
    \put(-90,7){\color{white}\contour{black}{Pan-STARRS $g$ band}}
    \end{subfigure}
    \hfill
    \begin{subfigure}[t]{0.195\textwidth}
    \centering
    \includegraphics[width=\textwidth]{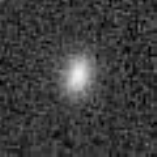}
    \scriptsize
       \put(-90,85){\color{white}\contour{black}{SN 2019kg}}
       \put(-90,75){\color{white}\contour{black}{spiral/elliptical}}
    \put(-90,17){\color{white}\contour{black}{$15 \arcsec$}}
    \put(-90,7){\color{white}\contour{black}{Pan-STARRS $r$ band}}
    \end{subfigure}
    \\
       \vspace{0.1cm}
    \begin{subfigure}[t]{0.195\textwidth}
    \centering
    \includegraphics[width=\textwidth]{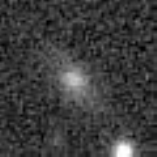}
    \scriptsize
       \put(-90,85){\color{white}\contour{black}{SN 2018fsa}}
       \put(-90,75){\color{white}\contour{black}{spiral/elliptical}}
    \put(-90,17){\color{white}\contour{black}{$15 \arcsec$}}
    \put(-90,7){\color{white}\contour{black}{Pan-STARRS $r$ band}}
    \end{subfigure}
    \hfill
    
       \caption{Host galaxies classified in this work. Indicated are the names of the SNe that occurred in the galaxies, our visual classifications, the image sources, and the widths of the images in arcseconds. Note that the image of the host galaxy of SN\,2019rm is a FORS2 acquisition image (this work), so the SN is also visible, while the SNe are not visible in all other archival images from the Pan-STARRS Sky Survey and the DESI Legacy Sky Survey. The thumbnails are centred on the coordinates of the host galaxies.}
 \label{fig:classfications}
\end{figure*}

For 26 host galaxies without a given morphological classification in the NED or HyperLEDA databases, we searched for additional references or tried to classify them visually \citep{Buta2013} using archival images from either Pan-STARRS or the DESI Legacy Survey, and from available spectra.
We classified 20 initially unclassified host galaxies, shown in Fig.~\ref{fig:classfications}. In the case of the host galaxy of SN\,2019rm, there were no archival images from Pan-STARRS, the DESI Legacy Survey, or SDSS available. We show, therefore, an acquisition image obtained with FORS2, which contains the SN. 

The host galaxy of SN\,2007if is likely a dwarf galaxy \citep{2010ApJ...713.1073S}. It is very faint and shows H$\alpha$ and [O~II] $\lambda$3727 emission lines at a heliocentric redshift of $0.07416 \pm 0.00082$ in a spectrum obtained on 2009 August 24.5 with the Low Resolution Imaging Spectrometer (LRIS) at the Keck-I 10\,m telescope on Maunakea \citep{2010ApJ...713.1073S}. 

In addition to the host of SN\,2019bjz looking like an elliptical galaxy in an image acquired by the Pan-STARRS survey, an archival SDSS spectrum also shows that the host displays a typical spectrum of an elliptical galaxy with strong absorption lines.

The host of SN\,2019rm looks irregularly shaped; however, because of the low resolution of the available images, we could not successfully classify it.  
The hosts of SN\,2018fsa, SN\,2019bpb, and SN\,2019kg display a bright core and a disk; however, the resolution of the available archival images is not sufficient to distinguish between spiral or elliptical galaxies.
Furthermore, the hosts of SN\,2018fqn, SN\,2018hts, SN\,2018koy, and SN\,2018kyi are too faint and unresolved to be reliably classified.\\

The host galaxies of the SNe, their morphological classifications, the determined flux percentiles at the SN position, and image sources used to fit the isophotes are listed in Table~\ref{tab:hostinfo}. The galaxies for which the isophote fitting did not converge (i.e., if the isophotes were defined manually) have been flagged ``No'' in the ``Fitted?'' column.

\section{Results and discussion}
\label{sect:results}


The goal of this work is to study the nature of dust, which is producing peculiar polarization and extinction properties along the sight lines to some SNe~Ia. We investigated whether the steeply rising polarization curves toward blue wavelengths with relatively high degree of polarization, which are observed toward some SNe~Ia, are produced by dust in the host-galaxy ISM or possible CSM associated with the progenitor system. If the dust in the ISM is responsible for producing the peculiar polarization curves, we would not expect to observe such polarization curves in elliptical galaxies, which are generally known to be dust poor. On the contrary, in spiral galaxies, which are dust rich, we may expect to observe an increased number of SNe with high polarization and steeply rising polarization curves closer to the galaxy centre compared to SNe in the outer parts of the spiral galaxies, because the dust abundance in spiral galaxies increases toward the galaxy centre (see, e.g., \citealt{2009ApJ...701.1965M, cikota2016, 2017A&A...605A..18C}). 

We grouped our sample of SNe Ia according to the morphological classification of the host galaxies: (i) 66 SNe that occurred in spiral galaxies, (ii) 13 in elliptical galaxies, (iii) 15 in S0 galaxies, and (iv) 7 that occurred in a galaxy with unknown morphological classification. Additionally, there is one SN that exploded in a dwarf galaxy (SN\,2007if) and one that exploded in an irregular galaxy (SN\,2019rm). Thus, we observed a factor of 5 more SNe in spiral host galaxies than in elliptical galaxies. This is an expected frequency distribution over the morphological galaxy types (for comparison, see Table~4 in \citealt{2005PASP..117..773V}).

We also analysed the polarization measurements of the SNe and distinguished between SNe that have polarization curves growing toward blue wavelengths (i.e., $p_B > p_V$), SNe with flat polarization curves (i.e. $p_B \approx p_V$), and SNe with Serkowski-like polarization curves (i.e., $p_B < p_V$). The polarization measurements for SNe in our sample that occurred before 2018 were taken from \citet{2019MNRAS.490..578C}. For these 35 archival SNe, we have spectropolarimetric measurements over the range $\sim$3800--9000\,\AA\, and the shape of the polarization curves can be visually inspected. SNe observed during 2018--2020 (this work) were observed in imaging-polarimetry mode in $B$, and eventually in $V$ if the polarization in $B$ was higher than 0.5 or 1\% (see Sect.~\ref{sect:obs_strategy}). 
Therefore, the slope of the polarization curves is not known for the SNe with low polarization in $B$.

\begin{figure*}
\includegraphics[trim=0mm 0mm 0mm 0mm, width=16.0cm, clip=true]{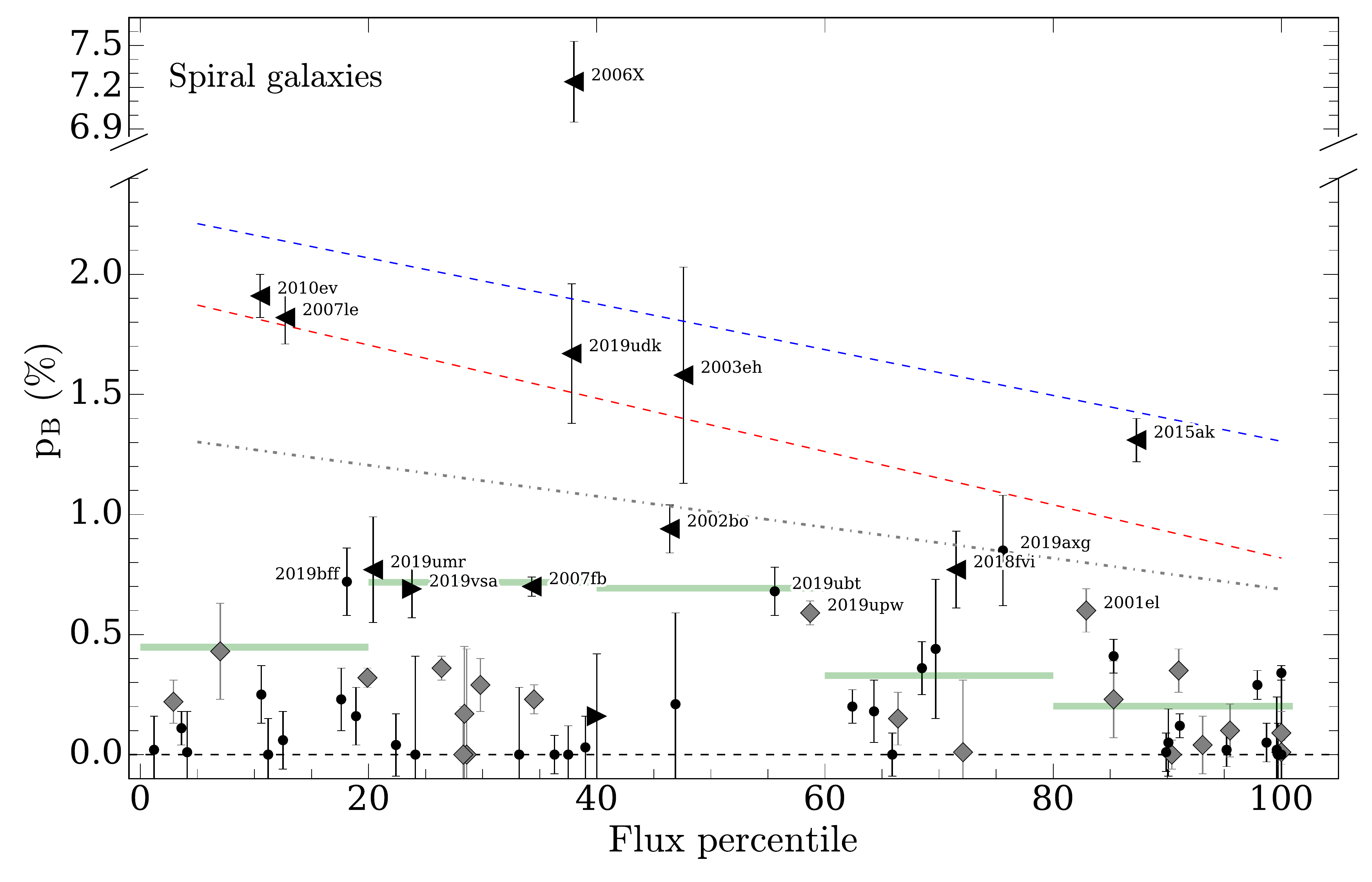}
\vspace{-2mm}
\caption{Polarization in the $B$ band as a function of the host-galaxy flux percentile at the SN position. Shown here is the sample of SNe Ia that occurred in spiral galaxies. The flux percentile represents the normalised galactocentric distance of the SNe in their host galaxies. For example, a SN at a position of the 90$^{\rm th}$ flux percentile is located near the edge of the galaxy, with 90\% of the total flux contained inside of the isophote at which the SN lies. The grey diamonds represent SNe with flat polarization curves, triangles pointing left are SNe with rising polarization curves toward blue wavelengths, and triangles pointing right mark SNe with polarization curves rising toward red wavelengths. The black dots represent SNe with only one measurement in the $B$ band (so the slope of the polarization curves is not known).
We fit a linear function to all SNe with polarization $p_B \gtrsim 0.5$\% (red line), and to SNe with polarization curves rising toward blue wavelengths and with polarization $p_B \gtrsim 0.5$\% (blue line). The grey dot-dashed line shows the same fit after excluding SN\,2006X.
The green bars represent the average polarization value of all SNe within bins of 20 percentile width.}
\label{fig:fraction-P-S}
\end{figure*}

\begin{figure}
\includegraphics[trim=0mm 0mm 0mm 0mm, width=8.5cm, clip=true]{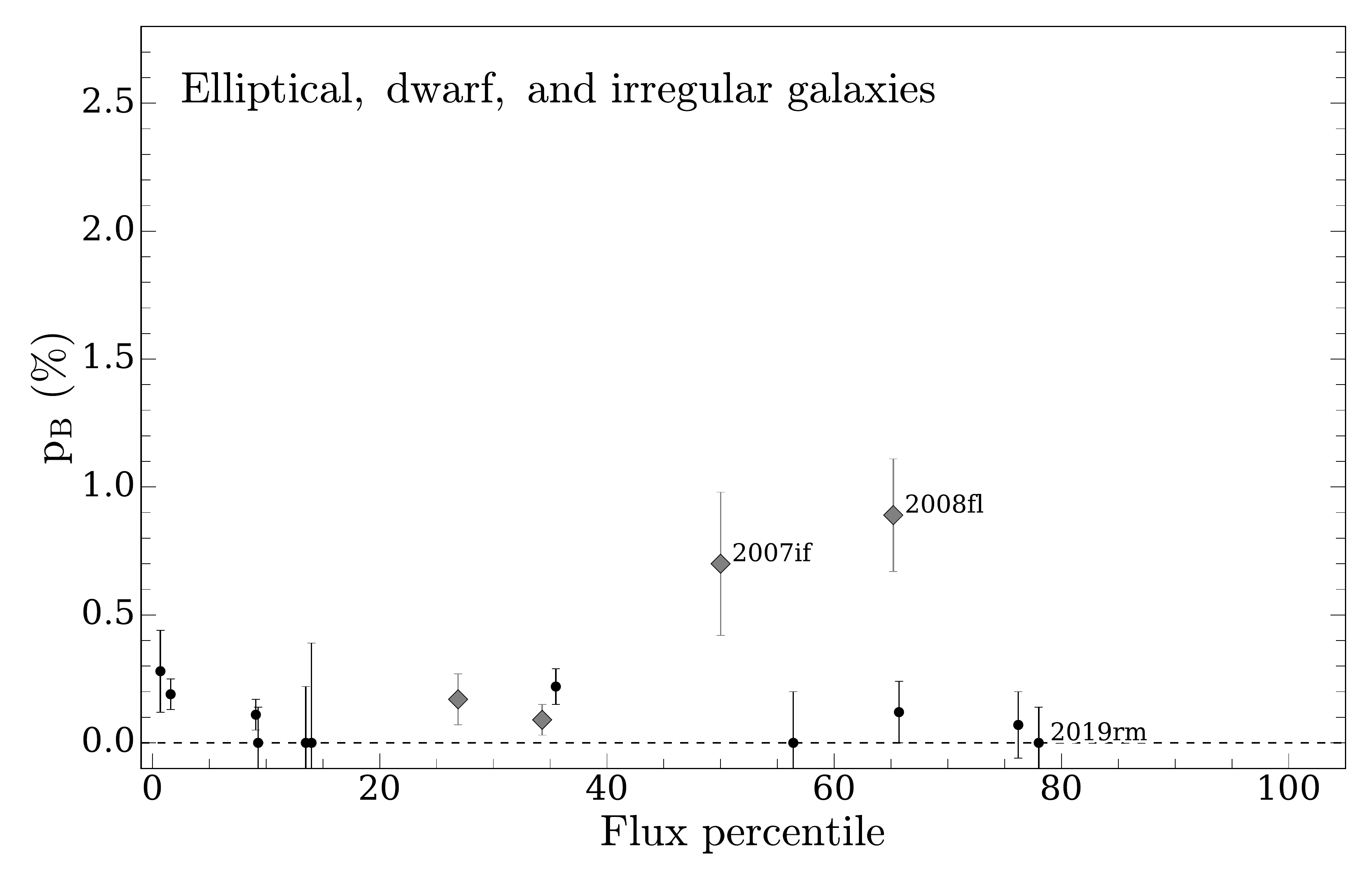}\\
\includegraphics[trim=0mm 0mm 0mm 0mm, width=8.5cm, clip=true]{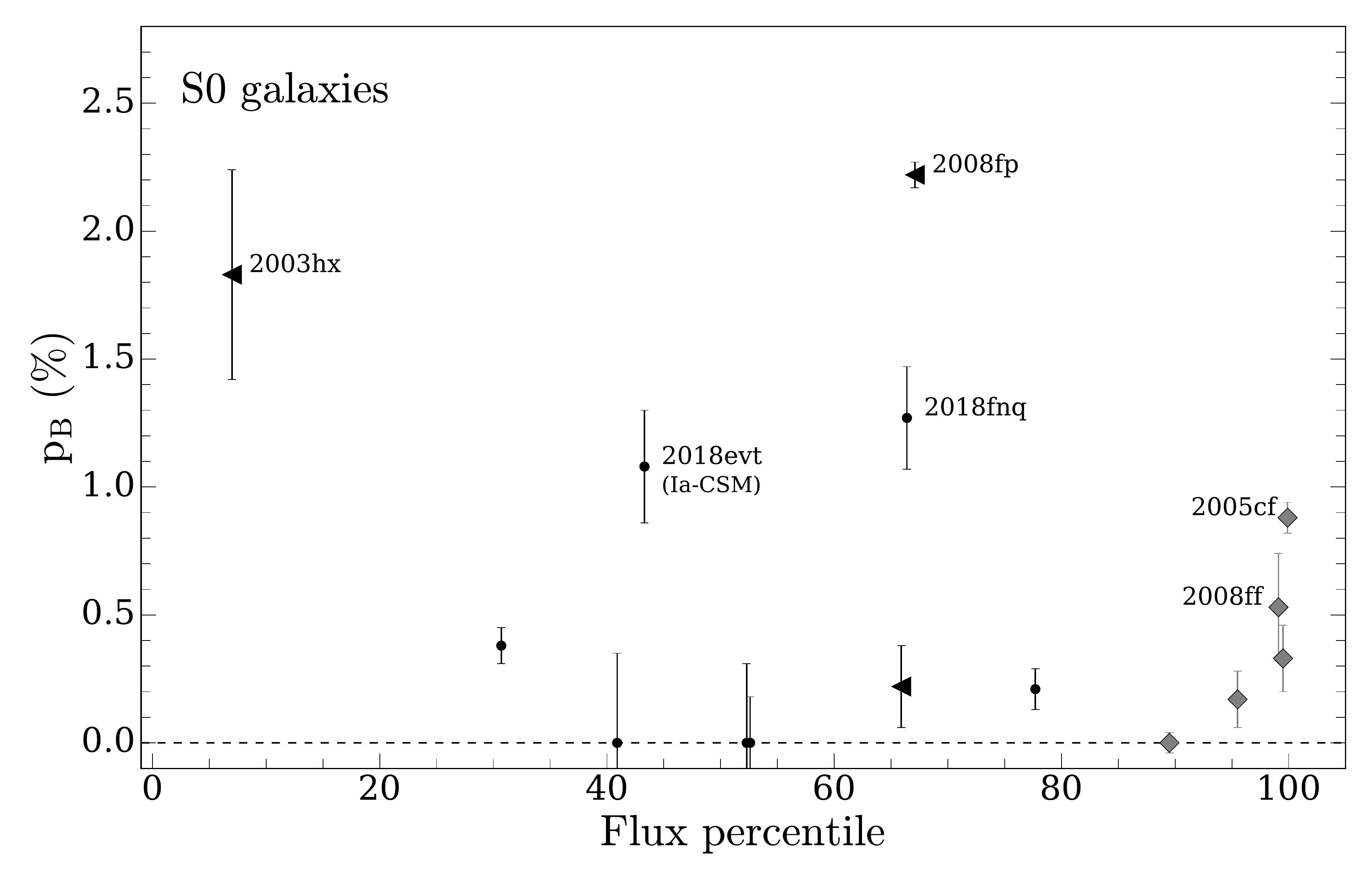}\\
\includegraphics[trim=0mm 0mm 0mm 0mm, width=8.5cm, clip=true]{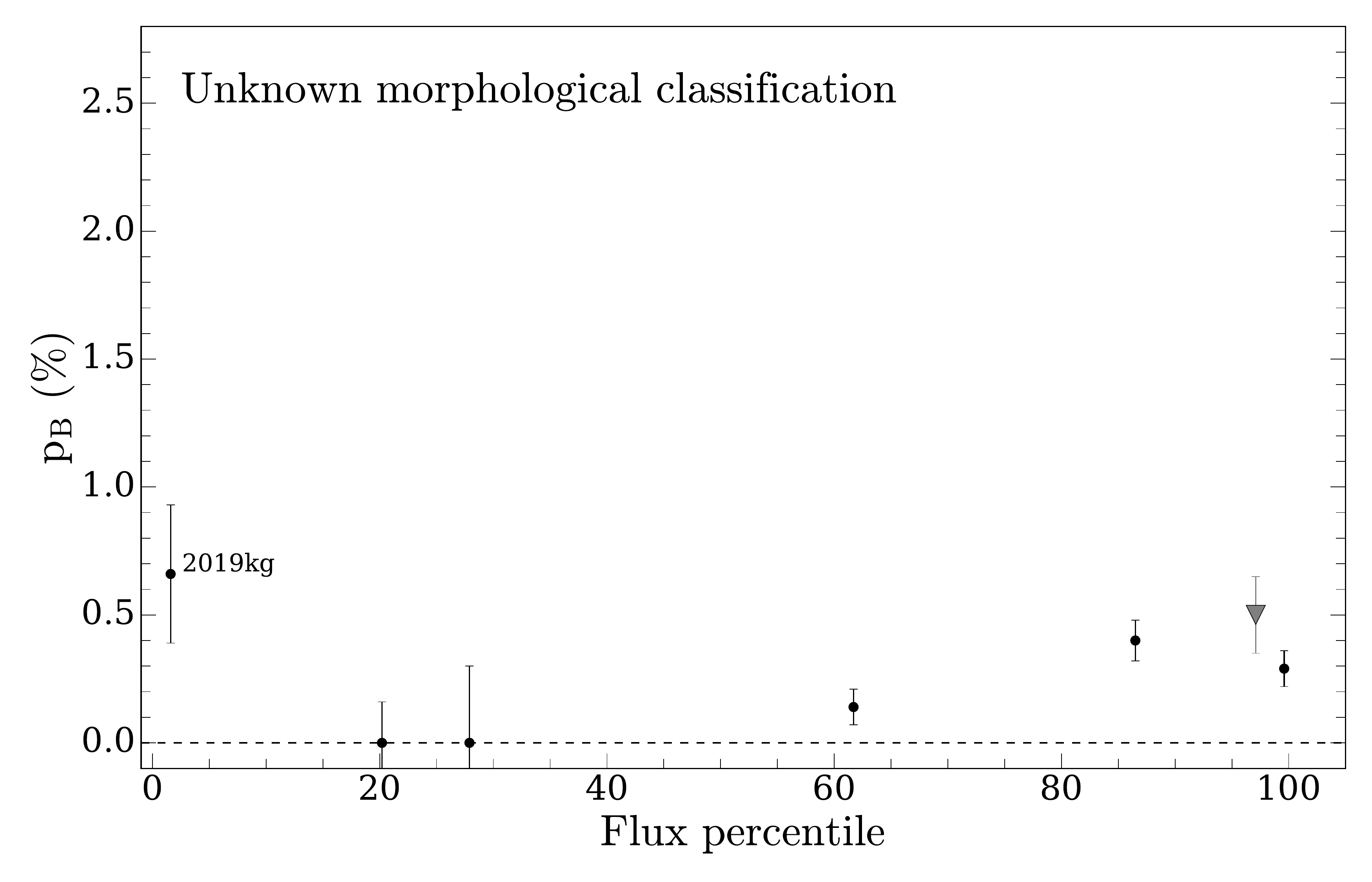}\\
\vspace{-5mm}
\caption{SN~Ia polarization in $B$ vs. host-galaxy flux percentile at the SN position for different types of host galaxies. The flux percentile represents the location of the SN in the galaxy. The grey diamonds mark SNe with flat polarization curves, and the black dots are SNe with only one measurement (in the $B$ band). \textit{Top panel:} SNe in elliptical galaxies, SN\,2019rm in a irregular galaxy, and SN\,2007if in an dwarf galaxy. \textit{Middle panel:} SNe in S0 galaxies. \textit{Bottom panel:} The remaining SNe in morphologically unclassified galaxies.}
\label{fig:fraction-P-rest}
\end{figure}

The sample is presented in Fig.~\ref{fig:fraction-P-S} (SNe in spiral galaxies) and in Fig.~\ref{fig:fraction-P-rest} (all other SNe). The figures show the degree of polarization measured in the $B$ band as a function of the fraction of flux within the isophote containing the SN relative to the total flux of the galaxy. Thus, the flux fraction represents the projected normalised location of the SN in the galaxy.  
It is known that the scale height of SNe~Ia can be relatively high \citep[e.g.,][]{1997ApJ...483L..29W}, meaning that SNe~Ia in spiral galaxies also explode far away from the galaxy plane. Therefore, depending on the scale height, the projected distance from the centre may be different from the real distance. In other words, SNe~Ia can appear to be projected very close to the galaxy centre, but in reality be very far from it. Furthermore, they can be located at the near or the far side of the galaxy, i.e. the line of sight may or may not pass through the galaxy.
The black dots represent SNe with only one measurement in $B$; thus, there is no information on the shape of the polarization curve. Most of these SNe have low polarization and are not relevant for our study, and hence have not been observed in the $V$ band as part of our survey (see Sect.~\ref{sect:obs_strategy}).

\subsection{SNe~Ia in spiral vs. elliptical galaxies}
\label{sect:SpvsEll}
Most relevant to consider for our study is the difference between the degree of polarization of SNe in spiral vs. elliptical galaxies. Figure~\ref{fig:fraction-P-S} shows SNe~Ia that exploded in spiral host galaxies. Sixteen out of 66 SNe had polarization higher than 0.5\% in $B$, and 10 of them have polarization curves rising toward blue wavelengths. Particularly noticeable is that all SNe with very high polarization ($p_B \gtrsim 1$\%) have polarization curves rising toward blue wavelengths. 

The figure also reveals that the polarization tends to be higher closer to the galaxy centre (at low flux percentiles). 
To demonstrate the trend of polarization as galactocentric distance decreases, we fit a linear function to SNe with polarization $p_B \gtrsim 0.5$\% (red line). The same trend is also visible if we only include SNe with polarization curves rising toward blue wavelengths (blue line), and if we exclude SN\,2006X (grey dot-dashed line).
Furthermore, the green bars represent the average polarization value of all SNe within 20 percentile wide bins. SNe located at the edges of their host galaxies (at distances above the 80$\rm ^{th}$ percentile) display low polarization, and SNe located close to the centre tend to have higher polarization. Because the polarization is higher close to the galactic centre, where there is more dust, the polarization is likely produced by the ISM. 
We note that the average polarization of all SNe within the first bin (closest to the galactic center, 0--20 percentile) is smaller than in the second bin (20--40 percentile). This may be due to a selection bias. Highly extincted SNe close to the galactic center are harder to discover than SNe with low dust-extinction levels.
Inspection of the acquisition images of SN\,2015ak, which may appear as an outlier with high polarization located far from the centre of its host galaxy, shows that it is actually located in the spiral arm (which are generally dust rich) of a strongly barred galaxy.

Figure~\ref{fig:fraction-P-rest} (top panel) shows the $B$-band polarization for 15 SNe in elliptical, dwarf, and irregular galaxies. 13 SNe occurred in elliptical galaxies, and they all have low polarization values, except SN\,2008fl, and SN\,2007if which are outliers. 

SN\,2008fl was observed with FORS in spectropolarimetry mode, and displays a very flat (wavelength-independent) polarization curve with $p_B =  0.89 \pm 0.22$\% and $p_V = 1.05 \pm 0.09$\%. Furthermore, despite the absence of available imaging-polarization data preventing us from estimating the Galactic ISP using field stars, the MW reddening at the position of SN\,2008fl is $E(B-V) = 0.157 \pm 0.006$\,mag \citep{2011ApJ...737..103S}. This implies a maximum polarization in the optical of $\sim 1.4$\% \citep{serk1975}. An inspection of the flux spectra of SN\,2008fl \citep{2019MNRAS.490..578C} in the observer's frame also reveals the presence of interstellar Na I D lines in the MW. No such absorption features are visible in the host galaxy's rest frame.
Therefore, the flat polarization curve in the case of SN\,2008fl is likely produced by MW ISM.

SN\,2007if, which is also shown in Fig.~\ref{fig:fraction-P-rest} and has a relatively high polarization in the $B$ band, is an overluminous SN~Ia ($M_V \approx -20.4$\,mag) that exploded in a very faint dwarf host galaxy \citep{2010ApJ...713.1073S}. SN\,2007if has been observed at four epochs between 13 and 46 days relative to peak brightness in spectropolarimetry mode \citep{2019MNRAS.490..578C}, and displays a time-independent and flat (wavelength-independent) continuum polarization of $\sim 0.7$\%. Because there is no time dependence, the polarization is likely not intrinsic but produced by foreground dust. The Galactic reddening at the position of SN\,2007if is $E(B-V) = 0.07 \pm 0.01$\,mag \citep{2011ApJ...737..103S}, which implies a maximum polarization in the optical of $\sim 0.6$\% \citep{serk1975}. The flux spectra of SN\,2007if show significant interstellar lines from the MW, but no interstellar absorption lines from the host galaxy. \citet{2010ApJ...713.1073S} derived EW(Na I D) = 0.51$^{+0.04}_{-0.05}$ \AA\ for the Milky Way dust absorption, which corresponds to $E(B-V) = 0.072$ mag and is consistent with \citet{2011ApJ...737..103S}. Therefore, the polarization is likely produced only by MW dust.

Notably, none of the 13 SNe in elliptical galaxies have a polarization curve rising toward blue wavelengths. Although the sample of 13 SNe in elliptical galaxies is smaller compared to the sample of 66 SNe in spiral galaxies, $\sim 15$\% of SNe in spiral galaxies have polarization curves rising toward blue wavelengths, which is $\sim$ 2 SNe out of 13. Assuming polarization curves rising toward the blue wavelengths is an intrinsic property of some SN~Ia progenitor systems (e.g., the single-degenerate or core-degenerate systems), we would expect such polarization curves in 1--2 SNe that exploded in elliptical host galaxies. Despite the differences in the stellar populations between spiral and elliptical galaxies, all stellar populations between $\sim 0.1$ and 10\,Gyr are expected to produce PPNe \citep{2016A&A...588A..25M,2006MNRAS.368..877B}. Therefore, the core-degenerate scenario is not a hallmark only of late-type galaxies.

\subsection{Explanation of the peculiar dust properties}
\label{sect:explanationRATD}

Our results indicate that in dust-rich spiral galaxies, high degree of polarization and polarization curves steeply rising toward blue wavelengths correlate with SNe in regions with denser ISM dust, as they are relatively closer to the centres of their host galaxies \citep[see, e.g.,][]{2009ApJ...701.1965M, cikota2016, 2017A&A...605A..18C}. We also observe that in elliptical galaxies, which are generally dust poor \citep{2012ApJ...748..123S, cikota2016}, none of the SNe display polarization curves rising toward short wavelengths, and the polarization levels are low. Therefore, the steeply rising polarization curves observed toward some SNe~Ia in spiral galaxies are likely related to host-galaxy ISM, and not to CSM originating from the progenitor system. 
However, note that the sample of 13 SNe in elliptical galaxies is small compared to the sample of 66 SNe in spiral galaxies (see Sect.~\ref{sect:SpvsEll}). To unambiguously confirm the origin of the dust, the SN sample in elliptical galaxies should be increased in the future. A discovery of a SN~Ia with a steeply rising polarization curve toward blue wavelengths in an elliptical galaxy would favour CSM for the origin of the peculiar dust.

However, constraining the origin of dust does not explain the peculiar dust properties. To produce the steeply rising polarization curves, a high abundance of small dust grains is required. The dust disruption process by the RAdiative Torque Disruption (RATD) mechanism may provide a possible explanation.

\subsubsection{RATD mechanism induced by the SN radiation field}

A possible explanation is that the strong radiation field of the SNe~Ia, which explode in ISM-rich environments in their host galaxies, spin-up the nearby ($\lesssim$ 4 pc) dust grains by radiative torques until they break apart owing to centrifugal forces, as suggested by \citet{2019NatAs...3..766H}. This RATD mechanism fragments large dust grains on relatively short timescales, within days, depending on the dust distance from the SNe, and generates a high abundance of small dust grains, which can explain the low $R_V$ values and the peculiar polarization curves with polarization maxima at short (blue) wavelengths \citep{2017ApJ...836...13H, 2019NatAs...3..766H, giang2020}.

The probability that a SN event is going to happen close to an ISM dust cloud is higher at locations close to the centre of dust-rich galaxies (e.g., spiral galaxies), where the densities of dust are higher (as also suggested by \citealt{bulla2018_2}). This could explain why we observe that the SNe~Ia with steeply rising polarization curves and higher polarization are located closer to the centre of their spiral host galaxies.

The RATD mechanism can be tested by polarimetry of highly polarized SNe at very early phases. Soon after the SN explosion, the dust around the SNe is gradually disrupted by the RATD mechanism as the SN brightens, and the polarization and the dust reddening are expected to be time-evolving before the destruction reaches maximum, as predicted by \citet{giang2020}. No such time evolution has ever been found so far (see, e.g., Fig.~4 of \citealt{zelaya2017}), which can already set strong limits on the location of dust. It may imply that the dust distance is very near the SNe ($\ll 0.5$\,pc) so that the dust disruption happens very soon after the explosion, before the SN is discovered or observed, or that the assumed parameters of the RATD and/or dust model need to be better constrained. However, theoretical modeling of dust destruction is beyond the scope of this paper. More spectropolarimetry of SNe starting within $\sim$ 1 day after the explosion, especially of highly extinguished ones, will be most useful in the studies of polarization features related to dust destruction due to SN explosions.

Furthermore, the RATD is expected to work in all intense radiation fields, and therefore we expect to observe the effects of an increased abundance of small dust grains in the environments of other objects, such as core-collapse SNe, massive stars, active galactic nuclei (AGNs), gamma-ray bursts (GRBs), etc. \citep{2019NatAs...3..766H}.
\citet{2018MNRAS.479.1542Z} showed that typical extinction curves in GRB afterglows are consistent with the extinction curves observed in the Small Magellanic Cloud (SMC), and have a total-to-selective extinction ratio $R_V = 2.61 \pm 0.08$. Some core-collapse SNe, AGNs, and massive stars also display polarization curves rising toward blue wavelengths (see, e.g., \citealt{2019MNRAS.485..102S}, \citealt{2001ApJ...563..512H}, and \citealt{1992ApJ...386..562W}, respectively). However, whether the extinction properties in these objects can be attributed to dust destruction from the RATD mechanism needs to be further investigated in a separate study.

\subsubsection{RATD mechanism induced by the interstellar radiation field}

Another possibility is that dust in the interstellar medium gets disrupted by the RATD mechanism owing to the interstellar radiation field \citep{2021ApJ...907...37H}. \citet{2021ApJ...907...37H} suggest that extinction curves with $R_V$ values in the range 1.5--2.5 can be produced by the interstellar radiation field (ISRF) if grains have composite structures of tensile strength $S_{\rm max} \lesssim 10^6$\,erg\,cm$^{−3}$. The mean radiation intensity in main-sequence galaxies cannot produce high abundances of small dust grains that may explain the extremely low ($R_V \lesssim 1.5$) values observed toward some SNe~Ia. However, such low $R_V$ values may be explained if the mean intensity of these galaxies is enhanced by starbursts and the grain temperature can reach $T_d \approx 60$\,K ($U \approx 10^3$, where $U=1$ is the typical scale factor for the ISRF in the Solar neighbourhood). In that case, the extinction and polarization curves do not vary with time, in contrast to dust disruption induced by the SN radiation field \citep{giang2020}. Although the typical mean starlight intensity scale factor $U$ is low in galaxies in the local Universe ($U \lesssim 10$; \citealt{2007ApJ...663..866D}), $U$ can range up to $10^7$. 

Some galaxies have significant fractions of dust luminosity radiated from regions with $U > 10^2$ (see Tables 4 and 5 of \citealt{2007ApJ...663..866D}). The Spitzer Infrared Nearby Galaxy Survey (SINGS; \citealt{2003PASP..115..928K}) is an infrared imaging and spectroscopic survey of 75 nearby galaxies with a wide range of morphological types. Some of these SINGS galaxies have also been observed with the SCUBA camera \citep{1999MNRAS.303..659H} on the James Clerk Maxwell Telescope (JCMT), which made it possible to determine the ISRF \citep{2007ApJ...663..866D}. The median galaxy in the SINGS-SCUBA sample of 17 galaxies has 10\% of the dust luminosity originating in regions with $U > 10^2$ and $\sim 7.8$\% in regions with $U > 10^3$ (see Fig. 8 of \citealt{2007ApJ...663..866D}).

\subsubsection{Possible differences between SN Ia populations}

An alternative explanation that cannot be excluded is that the underlying progenitor populations of SNe~Ia may be host-galaxy dependent. The lack of detected high continuum polarization in early-type galaxies may be a consequence of some systematic difference in the age or metallicity of the progenitors.
Thus, some progenitor populations may imply different CSM environments and produce the peculiar extinction and polarization properties due to CSM scattering (see Sect.~\ref{sect:intro}).

\citet{2013Sci...340..170W} found that SNe~Ia with high Si~II velocity (HV SNe~Ia) tend to explode at distances closer to the centre of the host galaxies than normal-velocity (NV) SNe~Ia and are more likely to be associated with younger stellar environments and metal-rich progenitors.
Furthermore, \citet{2019ApJ...882..120W} found that HV~SNe~Ia display a significant excess of blue flux 60--100 days past peak brightness compared to SNe with normal photospheric velocities. This excess may be attributed to light echoes by dust in the CSM. 

Therefore, \citet{2019ApJ...882..120W} suggest that the HV~SNe~Ia may arise from single-degenerate progenitors. This is also consistent with the systematically observed blueshifted outflows in the Na~I line \citep{2011Sci...333..856S,2013MNRAS.436..222M} in many SNe~Ia. In contrast, SNe~Ia that do not display HV features and possible evidence of CSM may have double-degenerate progenitor systems \citep{2011Natur.480..348L,2012Natur.489..533G,2012Natur.481..164S,2015Natur.521..332O,2019ApJ...882..120W}.

The sub-$M_{\rm Ch}$ He-detonation (double-detonation) model may also explain the characteristics of HV and NV SNe~Ia. In this scenario, the SN explosion becomes triggered by an initial detonation in the accreted He shell on the surface of the WD \citep[e.g.,][]{1982ApJ...253..798N,1982ApJ...257..780N}. 
\citet{2021ApJ...906...99L} suggest that the observed diversity in the velocity arises from projection effects: HV~SNe~Ia are observed from the He-detonation side while NV~SNe~Ia are observed from the opposite side. On the other hand, \citet{2019ApJ...873...84P} suggest that different He-shell masses of exploding sub-$M_{\rm Ch}$ WDs may produce a range of Si~II velocities. \citet{2020ApJ...895L...5P} found that HV~SNe~Ia tend to explode in massive environments with higher metallicities, while NV~SNe~Ia occur in both lower-mass and massive environments \citep[see also][]{2015MNRAS.446..354P}. This supports the possibility of multiple SN~Ia populations and that HV~SNe~Ia may originate from exploding sub-Chandrasekhar-mass WDs because higher metallicities generally produce less-massive WDs \citep{2020ApJ...895L...5P}.

Studies of SN~Ia rates in galaxies with different colour and morphological type (i.e., star-formation rates) implied that there may be two SN Ia populations: the promptly exploding and delayed SNe Ia \citep{2005ApJ...629L..85S, 2006MNRAS.370..773M,2006ApJ...648..868S}. However, \citet{2012MNRAS.426.3282M} found a continuous distribution of times between the formation of the progenitor systems and the SN explosions.

Besides different SN populations, an age dependency on SN~Ia progenitors may also exist. \citet{2020ApJ...889....8K} claim to have found a significant correlation between the SN~Ia luminosity and stellar population age. If true, SN~Ia luminosity evolution can have a high impact on cosmology, because stellar populations get younger with increasing redshift. However, the result is highly debated and unlikely to be correct \citep{2020ApJ...896L...4R,2020ApJ...903...22L,2021arXiv210706288L,2021MNRAS.503L..33Z,2021MNRAS.504L..34M}.

\subsection{Polarization angle alignment with galactic features}

\begin{figure*}
    \centering
    \includegraphics[width=0.24\textwidth]{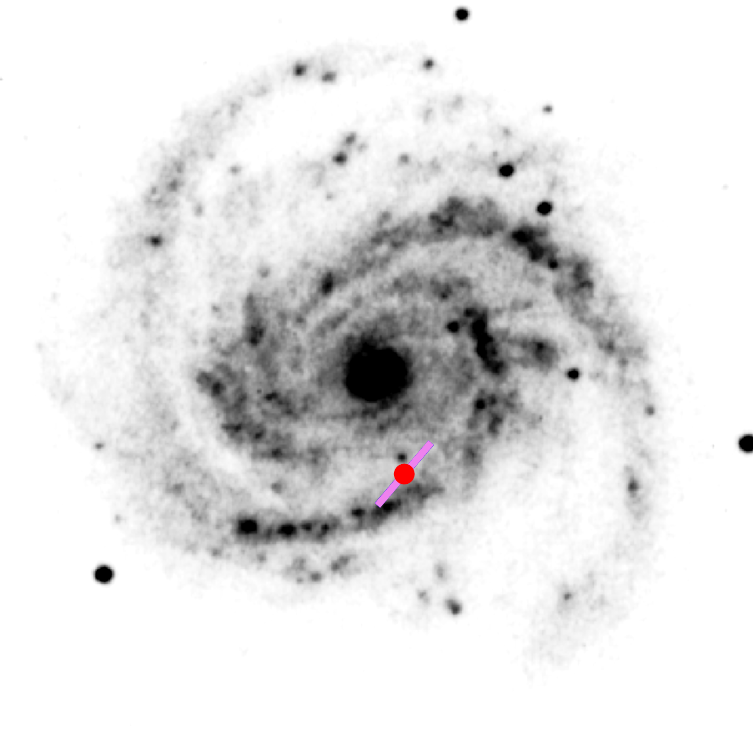}
    \put(-110,110){\color{black}\contour{white}{SN\,2006X}}
    \includegraphics[width=0.24\textwidth]{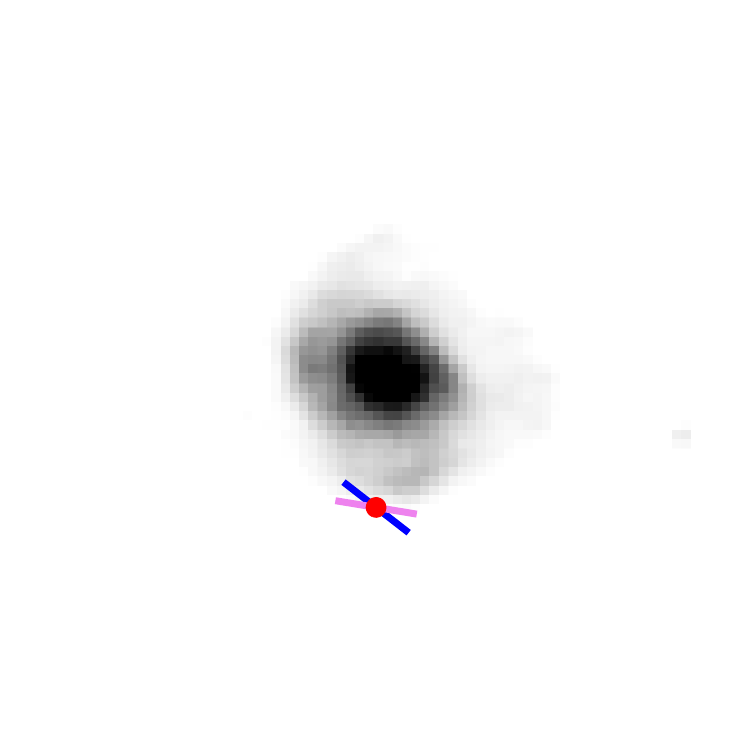}
    \put(-110,110){\color{black}\contour{white}{SN\,2018fvi}}
    \includegraphics[width=0.24\textwidth]{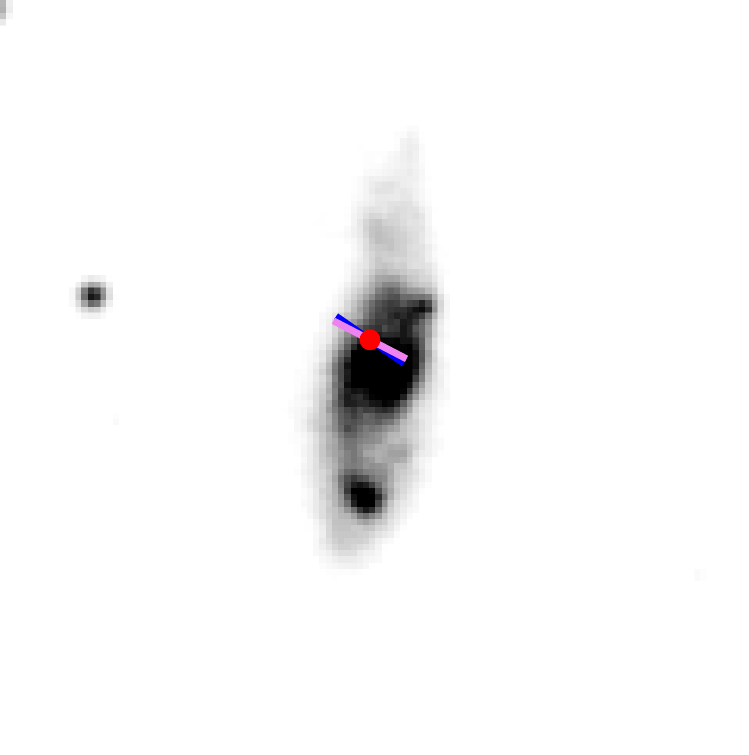}
    \put(-110,110){\color{black}\contour{white}{SN\,2019umr}}
    \includegraphics[width=0.24\textwidth]{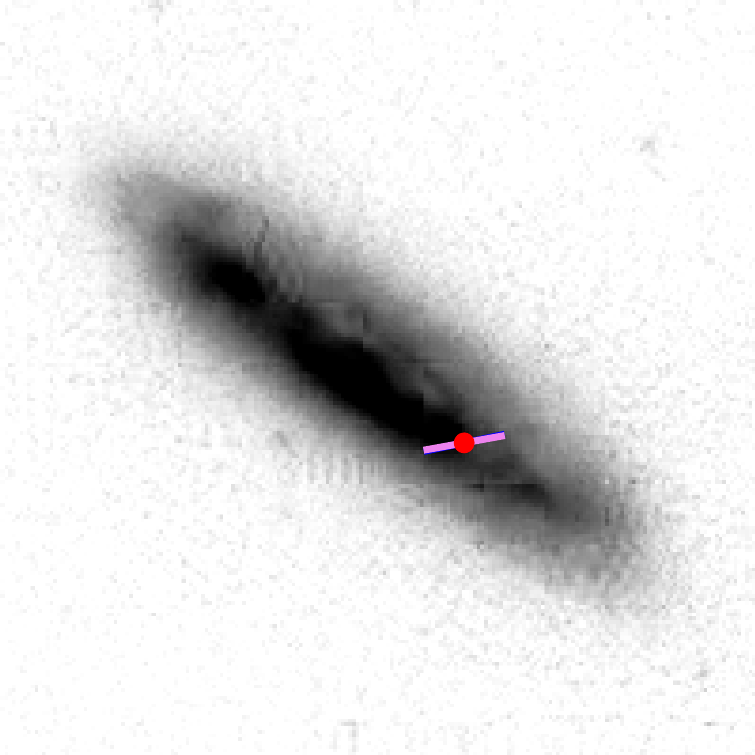}
    \put(-110,110){\color{black}\contour{white}{SN\,2019udk}}\\
   
    \caption{Polarization angles of (from left to right) SN\,2006X, SN\,2018fvi, SN\,2019umr, and SN\,2019udk. The blue and violet bars mark the polarization angles measured in the $B$ and $V$ bands, respectively. SN\,2006X is an exemplary SN that displays an almost perfect alignment of the polarization angle with the spiral arm \citep{2009A&A...508..229P, patat2015}. SN\,2018fvi shows some level of alignment with the host galaxy's features. The alignments of the polarization angles of SN\,2019umr and SN\,2019udk with possible spiral arms are debatable; it is difficult to determine the alignment in highly inclined galaxies.}
    \label{fig:pol_angle}
\end{figure*}

\citet{zelaya2017} have shown that the polarization angles of reddened SNe~Ia tend to be aligned with galactic features, notably the spiral arms of the host galaxies. Almost all SNe in their ``sodium-sample'' that display sodium absorption lines in their spectra (and thus are clearly affected by dust) display polarization angles that are aligned with major features of their host galaxies, like the spiral arms or the disk. 
The most prominent example is SN\,2006X, which displays almost perfect alignment of the polarization angle with the spiral arm (Fig.~\ref{fig:pol_angle}) in M100 \citep[see also][]{2009A&A...508..229P, patat2015}. 
It is also in general known that the dust is aligned with the magnetic fields of the host galaxies; see, e.g., \citet{1987MNRAS.224..299S}. These previous studies strongly suggest that the polarization is due to dust in the ISM. 

The alignment of dust grains is also consistent with the RATD scenario. Although the RATD mechanism disrupts the dust grains, the disrupted and spinning grains quickly realign with the magnetic field. The spinning grains that contain unpaired electrons (e.g., silicates) tend to magnetize, which is known as the Barnett effect \citep{1915PhRv....6..239B}, and then start rapidly precessing about the external magnetic field, which establishes the axis of grain alignment \citep{1976Ap&SS..43..257D, 2020ApJ...896...44L}. 

We investigated the polarization angles of three SNe~Ia observed in this work that exploded in spiral galaxies, have polarization curves rising toward blue wavelengths, and polarization higher than 0.5\% in the $B$ band: SN\,2018fvi, SN\,2019umr, and SN\,2019udk. As shown in Fig.~\ref{fig:pol_angle}, SN\,2018fvi and SN\,2019udk exhibit some level of coherence with the features of their hosts, but in the case of SN\,2019udk it is difficult to determine as the SN is located on the major axis of the galaxy. SN\,2019umr is not as well aligned as SN\,2006X with its host galaxy's features, but as noted by \citet{zelaya2017}, lower polarization values introduce higher noise and systematic error (e.g., owing to the MW's ISP), and both SN\,2018fvi and SN\,2019umr are polarized $< 1$\% compared to the polarization of nearly $2\%$ of SN\,2019udk and over 7\% of SN\,2006X. In general, it is also more difficult to determine the alignment with inclined, particularly edge-on galaxies, as features such as spiral arms are almost parallel to our line of sight in certain regions.

\subsection{The cases of SN\,2006X and SN\,2014J}

SN\,2006X was a fast-expanding and highly reddened SN~Ia that exploded in the nearby face-on spiral galaxy M100. \citet{2008ApJ...675..626W} determined the reddening $E(B-V)_{\rm host} = 1.42 \pm 0.04$\,mag with $R_V = 1.48 \pm 0.06$. SN\,2006X also displayed a steeply rising polarization curve toward blue wavelengths with a polarization maximum at $\lambda_{\rm max} \lesssim 0.4\,\mu$m and a particularly high polarization in the $B$ band, $p_B \approx 7.2$\% \citep{2009A&A...508..229P}, compared to the rest of the sample which has $p_B \lesssim 2$\%. 

Furthermore, \citet{Patat_2007} detected temporal evolution in the blueshifted Na~I~D absorption lines in the spectra of SN\,2006X, which implies that there is CSM dust around the SN, likely associated with the progenitor system. However, \citet{Patat_2007} also derived an upper limit to the CSM shell that produced the variable Na~I~D lines at the relatively low mass of $3 \times 10^{-4}\,M_\odot$. This cannot explain the bulk of the reddening toward SN\,2006X, as opposed to a deep, saturated, Na~I~D absorption line at a slightly different velocity, likely produced by a molecular cloud in the disk of the host galaxy M100 \citep[see also][]{2008A&A...485L...9C}. The evolution of the blueshifted Na~I~D lines has been explained by the authors with changes in the CSM ionization conditions induced by the variable SN radiation field.

The polarization angle is wavelength-independent (which implies that the dust grains of all sizes are well aligned), and aligned with the spiral arm. Therefore, as argued by \citet{2009A&A...508..229P} and \citet{patat2015}, it appears that the polarization along the sight line toward SN\,2006X is produced by host-galaxy ISM with a high abundance of small dust grains. 
On the other hand, it is intriguing why the dust-size distribution in the host galaxy of SN\,2006X, and other galaxies that display such peculiar polarization curves, would be so different compared to dust in the MW (\citealt{1992ApJ...386..562W}; see also \citealt{2018A&A...615A..42C}).

\citet{2019NatAs...3..766H} offered an explanation by introducing the RATD mechanism (see Sect.~\ref{sect:explanationRATD}), which we acknowledge as a possible explanation and which may also be valid for the case of SN\,2006X and other SNe with similar polarization and dust-extinction properties. The prerequisite is that the molecular cloud along the sight line is not too distant from the SN.

SN\,2014J occurred in the northern celestial hemisphere and is therefore not included in our VLT sample, but it is also a highly reddened and polarized SN Ia. The SN exploded in the nearly edge-on starburst galaxy M82, which was thought to be an irregular galaxy. However, \citet{2005ApJ...628L..33M} found spiral arms and suggested a late morphological type SBc for M82. The SN exploded relatively close to the galactic center and displays an extinction of $E(B-V)$ = 1.37 $\pm$ 0.03 mag with $R_V = 1.4 \pm 0.1$ \citep{2014ApJ...788L..21A}. The extinction can be explained by a combination of dust reddening and scattering \citep{2014MNRAS.443.2887F, 2014ApJ...792..106W, 2017ApJ...834...60Y, 2018ApJ...854...55Y}.

\citet{2014ATel.5830....1P} obtained three spectropolarimetry epochs of SN\,2014J (2014 January 28, February 3, and March 8), using the Calar Alto Faint Object Spectrograph (CAFOS) mounted at the 2.2\,m telescope in Calar Alto, Spain \citep{2011A&A...529A..57P}. The SN exhibits an extremely steeply rising polarization curve with a polarization of $4.8 \pm 0.6$\% in the $B$ band \citep{2014ApJ...795L...4K}, $6.6 \pm 0.1$\,\% at 4000\,\AA, and a wavelength-independent polarization angle (\citealt{patat2015}, see also \citealt{2016ApJ...828...24P}). The Na~I~D and K~I features observed in a high-resolution spectrum reveal that the ISM toward SN\,2014J is complex, with more than five components placed at different distances \citep{patat2015}. Therefore, \citet{patat2015} suggest that the observed extinction and polarization properties reflect the average properties of the sight line, and exclude that the bulk of reddening and polarization is produced within one single cloud having peculiar properties.
In this case, the rotational disruption by radiative torques (RATD) of dust along the sight line cannot be induced by the radiation field of the SN alone, but may be explained by RATD owing to the interstellar radiation field in this starburst galaxy (\citealt{2021ApJ...907...37H}; see also Sect.~\ref{sect:explanationRATD}).

\subsection{SNe~Ia in lenticular galaxies and morphologically unclassified galaxies}

Lenticular galaxies (S0) can have considerable amounts of dust; thus, it is not surprising that some of the SNe in our sample residing in those galaxies also exhibit high levels of polarization. Fifteen of our SNe that exploded in S0 galaxies, shown in the middle panel of Fig.~\ref{fig:fraction-P-rest}. Six of them have polarization higher than 0.5\% in the $B$ band, and two of these (SN\,2003hx and SN\,2008fp) have polarization curves rising toward blue wavelengths. SN\,2003hx and SN\,2008fp also show time-variable colour excess \citep{bulla2018_2}. \citet{bulla2018_2} inferred dust located at a distance of $\sim$0.013 and $\sim$9.4 pc from SN\,2003hx and SN\,2008fp, respectively. 

SN\,2018evt is a Type Ia-CSM SN, which exhibits strong emission of the Balmer lines most likely due to the interaction of the SN ejecta with CSM \citep[e.g.][]{2013ApJS..207....3S}. It displays a wavelength-independent continuum polarization of $\sim 0.7$\% at early epochs (Yang et al., in prep.). We have included SN\,2018evt for completeness, because it was also observed as part of this program; however, because of its nature, it should be excluded from the analysis in this work. 

SN\,2018fnq displays a polarization of $1.3 \pm 0.06$\% in the $B$ band, but only $0.21 \pm 0.07$\% in the $V$ band. The $B$ measurement may be systematically wrong; the peak values of the SN flux are only marginally below the saturation limit of the 16-bit A/D converter so that nonlinearities may have occurred.

Two objects (SN\,2005cf and SN\,2008ff) that display relatively high polarization are located near the edge of their host galaxies, where we do not expect significant amounts of ISM. SN\,2005cf displays a normal Serkowski-like polarization curve with a peak of $\sim 0.9$\% \citep{2019MNRAS.490..578C}.
The SN exploded at the edge of a tidal stream between two interacting galaxies,  MCG -01-39-003 and NGC 5917. \citet{2007MNRAS.376.1301P} reported a negligible host-galaxy reddening and nondetection of host-galaxy interstellar absorption lines. The Galactic reddening toward the SN is $E(B-V) = 0.084 \pm 0.001$\,mag \citep{2011ApJ...737..103S}, which might be just sufficient to explain the observed polarization \citep{serk1975}. \citet{2009ApJ...697..380W}, however, report clearly visible interstellar Na~I~D lines from the host galaxy and the MW in the spectrum of SN\,2005cf, and determine a reddening of $E(B-V)_{\rm host} = 0.10 \pm 0.03$\,mag, and thus a total reddening of $E(B-V) \approx 0.2$\,mag. Therefore, despite the large distance from the host-galaxy centre, CSM can likely be excluded as the origin of the relatively high polarization, which can be explained as a combination of ISM in the tidal stream and MW reddening. 

SN\,2008ff displays a wavelength-independent polarization curve \citep{2019MNRAS.490..578C} of $p = 0.5 \pm 0.2$\%. The MW reddening toward the SN is $E(B-V) = 0.044 \pm 0.001$\,mag \citep{2011ApJ...737..103S}. This amount of reddening may produce polarization up to $\sim$0.4\% \citep{serk1975}, and within the errors it may explain the observed polarization. Unfortunately, there are no detailed studies of SN\,2008ff in the literature, so the host-galaxy extinction is not determined.

SN\,2018koy, which exploded in a galaxy of unknown morphological classification, is located relatively far from the host galaxy's centre and has a significant polarization of $>$1\%. However, as shown in Fig.~\ref{fig:ISPcorrection}, the high polarization is associated with large interstellar polarization produced by MW dust.

\section{Summary and conclusions}
\label{sect:summary}

The goal of this work is to study whether the peculiar polarization curves observed toward some SNe~Ia are produced by ISM or CSM. 
In order to understand what may cause the nonstandard dust properties toward SNe~Ia, we conducted an imaging linear polarization survey of 69 SNe~Ia with the VLT and FORS2 between 2018 and 2020. Additionally, we combined the sample with archival spectropolarimetric data for 35 SNe~Ia observed with the FORS1/2 between 2001 and 2015.
We investigated whether there is a relation between the continuum degree of polarization of SNe~Ia and the galactocentric distances, for galaxies of different morphological classifications and host-galaxy dust properties. 
Our main results can be summarised as follows.

\begin{enumerate}

\item Sixteen out of 66 SNe~Ia that exploded in spiral host galaxies (Fig.~\ref{fig:fraction-P-S}) display significant polarization in the $B$ band ($p_B > 0.5$\%), and 10 of these 16 exhibit polarization curves rising toward blue wavelengths. Furthermore, all 6 SNe with $p_B > 1$\% have polarization curves rising toward blue wavelengths. The polarization of SNe closer to the galactic centre tends to be higher, while the SNe located at the edge of the galaxy (at locations $>80\rm ^{th}$ flux percentile) have low continuum polarization, on average $p_B \approx 0.2$\%. 
On the contrary, 13 SNe~Ia that exploded in elliptical host galaxies (shown in Fig.~\ref{fig:fraction-P-rest}), which are generally dust-poor, all display low polarization, $p_B \lesssim 0.3$\% (except one outlier affected by MW ISP), and none of the SNe display polarization curves rising toward short wavelengths. Therefore, the steeply rising polarization curves observed toward some SNe~Ia are likely related to dust in the host-galaxy ISM, as opposed to CSM originating from the progenitor system (see discussion in Sect.~\ref{sect:SpvsEll} and Sect.~\ref{sect:explanationRATD}). This result is consistent with the conclusion by \citet{patat2015} that the anomalous polarization curves are most probably produced by interstellar dust. Also, \citet{2014ApJ...795L...4K} concluded that the observed polarization in the case of SN\,2014J is likely predominantly caused by the interstellar dust. Furthermore, our result provides evidence that the effect of dust on SNe~Ia can be different in spiral and elliptical galaxies, which could change the SN colours and brightness in different ways, and thus introduce some systematic bias in the luminosity standardisation of these SNe.

\item All SNe~Ia (except SN\,2019vsa) with measured polarization above $\sim 0.7$\% in the $B$ band and measured slopes have blueward-rising polarization curves. This implies that for strong polarization to occur toward SNe~Ia the dust grains are almost always small.

\item Our original hope was that the observations would also reveal the reasons why dust toward SNe~Ia seems to have peculiar properties. However, this conundrum is still not unambiguously resolved. We discuss several possible explanations.

\item The peculiar, steeply rising polarization curves toward blue wavelengths may be explained through destruction of nearby dust in the ISM by the RAdiative Torque Disruption (RATD) mechanism induced by the strong radiation field of the SNe, as suggested by \citet{2019NatAs...3..766H}. The RATD mechanism generates a high abundance of small dust grains, which can explain the low $R_V$ values and the peculiar polarization curves observed toward some SNe~Ia \citep{2017ApJ...836...13H, 2019NatAs...3..766H, giang2020}. Because of the dust distribution within the host galaxies, the probability that a SN will explode close to a dust cloud is higher at locations near the centre of dust-rich galaxies such as spirals \citep{bulla2018_2}. This may explain why the steeply rising polarization curves of SNe~Ia and the relation of the degree of polarization with galactocentric distance are observed for SNe that exploded in dust-rich spiral host galaxies, in contrast to the dust-poor elliptical galaxies. 
However, there is a caveat to this model: \citet{giang2020} predicted that dust disruption by the RATD mechanism should produce temporal evolution of the continuum polarization, which has not been observed toward SNe~Ia (see, e.g., Fig.~4 of \citealt{zelaya2017}). 

\item Dust in the interstellar medium may also get disrupted by the RATD mechanism owing to the interstellar radiation field \citep{2021ApJ...907...37H}. A significant fraction of dust ($\sim 8$\%) in the SINGS galaxies is located in regions with $U > 10^3$ \citep{2007ApJ...663..866D}. Such radiation intensity may generate a high abundance of small dust grains by the RATD mechanism to explain the low $R_V$ values and polarization curves peaking at $\lambda_{\rm max} < 0.4\,\mu$m. In that case, time variability of SN~Ia extinction and polarization measurements is not expected to be observed.

\item The polarization angle alignment with galactic features (see, e.g., \citealt{zelaya2017}) is also consistent with the assumed RATD mechanism, because the disrupted and spinning dust grains can realign in the magnetic field owing to the Barnett effect \citep{1915PhRv....6..239B, 2020ApJ...896...44L}.

\item An alternative explanation, that because of some systematic differences in the age or metallicities between the progenitor populations in the early-type and late-type galaxies some progenitor populations may imply different CSM environments, cannot be excluded.

\end{enumerate}


\section*{Acknowledgements}

We thank Daniel Kasen, Jason Spyromilio, Johanna Hartke, Wolfgang Kerzendorf, and the anonymous referee for constructive discussions and comments. 
This work is based on observations collected at the European Organisation for Astronomical Research in the Southern Hemisphere under ESO programmes 0101.D-0190(A), 0102.D-0163(A), and 0104.D-0175(A); the execution in service mode of these observations by the VLT operations staff is gratefully acknowledged.
A.V.F.'s supernova group at U.C. Berkeley has received generous financial support from the Miller Institute for Basic Research in Science (where A.V.F. is a Senior Miller Fellow), the Christopher R. Redlich Fund, Gary and Cynthia Bengier, Clark and Sharon Winslow, Sanford Robertson (Y.Y. is a Bengier-Winslow-Robertson Fellow), and many additional donors. M.B. acknowledges support from the Swedish Research Council (Reg. no. 2020-03330).

This research has made use of NASA’s Astrophysics Data System Bibliographic Services, the SIMBAD database, operated at CDS, Strasbourg, France, and benefited from L.A.Cosmic \citep{2001PASP..113.1420V}, IRAF \citep{1986SPIE..627..733T}, NED, PyRAF, and PyFITS. 
NED is funded by the National Aeronautics and Space Administration (NASA) and operated by the Jet Propulsion Laboratory, California Institute of Technology. PyRAF and PyFITS are products of the Space Telescope Science Institute, which is operated by AURA, Inc., for NASA. We thank the authors for making their tools and services publicly available.  
IRAF is distributed by NOAO, which is operated by AURA, Inc., under cooperative agreement with the U.S. National Science Foundation.

The Legacy Surveys consist of three individual and complementary projects: the Dark Energy Camera Legacy Survey (DECaLS; NSF's OIR Lab Proposal ID \#2014B-0404; PIs David Schlegel and Arjun Dey), the Beijing-Arizona Sky Survey (BASS; NSF's OIR Lab Proposal ID \#2015A-0801; PIs Zhou Xu and Xiaohui Fan), and the Mayall z-band Legacy Survey (MzLS; NSF's OIR Lab Proposal ID \#2016A-0453; PI Arjun Dey). DECaLS, BASS, and MzLS together include data obtained (respectively) at the Blanco telescope, Cerro Tololo Inter-American Observatory, the NSF's National Optical-Infrared Astronomy Research Laboratory (NSF's OIR Lab); the Bok telescope, Steward Observatory, University of Arizona; and the Mayall telescope, Kitt Peak National Observatory, NSF's OIR Lab. The Legacy Surveys project is honored to be permitted to conduct astronomical research on Iolkam Du'ag (Kitt Peak), a mountain with particular significance to the Tohono O'odham Nation.
The NSF's OIR Lab is operated by the Association of Universities for Research in Astronomy (AURA) under a cooperative agreement with the National Science Foundation.

Based on photographic data obtained using The UK Schmidt Telescope. The UK Schmidt Telescope was operated by the Royal Observatory Edinburgh, with funding from the UK Science and Engineering Research Council, until 1988 June, and thereafter by the Anglo-Australian Observatory. Original plate material is copyright (c) of the Royal Observatory Edinburgh and the Anglo-Australian Observatory. The plates were processed into the present compressed digital form with their permission. The Digitized Sky Survey was produced at the Space Telescope Science Institute under US Government grant NAG W-2166.

\section*{Data Availability}

The data underlying this article are available in the article. The raw data can be accessed online from the ESO Science Archive Facility.


\bibliographystyle{mnras}
\bibliography{SNeIaImpolSurvey} 

\begin{table*}
	\centering
	\scriptsize
	\caption{\label{tab:obslog} Observing log of the imaging polarization survey. The dates denote the beginning of the observations. The right ascension ($\alpha$) and declination ($\delta$) are remeasured SN coordinates.}
	\begin{tabular}{lccccl} 
	\hline
SN name   & UT Date  & $\alpha$(J2000) & $\delta$(J2000) & Band   & Exposure (s) \\
\hline
SN\,2018cif & 2018-06-13 06:56 & 22 03 00.94 & +02 35 52.0 & b\_HIGH & 4 $\times$ 110 \\
SN\,2018cif & 2018-06-13 07:10 & 22 03 00.94 & +02 35 52.0 & v\_HIGH & 4 $\times$ 56 \\
SN\,2018evt & 2019-01-09 08:26 & 13 46 39.17 & -09 38 35.9 & b\_HIGH & 4 $\times$120 \\
SN\,2018fhw & 2018-09-07 06:27 & 04 18 06.18 & -63 36 56.5 & b\_HIGH & 4 $\times$ 276 \\
SN\,2018fhx & 2018-09-08 08:23 & 06 24 37.99 & -23 43 59.0 & b\_HIGH & 4 $\times$ 500 \\
SN\,2018fnq & 2018-09-09 00:00 & 20 12 29.98 & -44 06 35.3 & b\_HIGH & 4 $\times$ 65 \\
SN\,2018fnq & 2018-09-11 00:10 & 20 12 29.98 & -44 06 35.3 & v\_HIGH & 4$\times$28 \\
SN\,2018fqd & 2018-09-09 00:13 & 20 51 51.42 & -36 50 11.7 & b\_HIGH & 4 $\times$ 180 \\
SN\,2018fqn & 2018-09-09 00:41 & 22 39 38.18 & -15 05 09.8 & b\_HIGH & 4 $\times$ 350 \\
SN\,2018fsa & 2018-09-08 05:42 & 22 25 37.04 & -13 04 30.0 & b\_HIGH & 4 $\times$ 110 \\
SN\,2018fuk & 2018-09-09 08:58 & 05 45 07.11 & -79 23 48.2 & b\_HIGH & 4 $\times$ 200 \\
SN\,2018fvi & 2018-09-08 06:02 & 01 57 42.52 & -67 11 14.0 & b\_HIGH & 4 $\times$ 110 \\
SN\,2018fvi & 2018-09-09 04:57 & 01 57 42.52 & -67 11 14.0 & v\_HIGH & 4$\times$250 \\
SN\,2018fvy & 2018-09-09 04:04 & 01 32 16.22 & -33 06 05.3 & b\_HIGH & 4 $\times$ 280 \\
SN\,2018fzm & 2018-09-11 00:25 & 21 03 19.85 & -51 32 47.8 & b\_HIGH & 4 $\times$ 200 \\
SN\,2018gyr & 2018-10-06 00:08 & 00 49 47.12 & -61 39 13.1 & b\_HIGH & 4 $\times$ $\sim$850 \\
SN\,2018hsa & 2018-11-09 00:25 & 21 15 01.08 & -47 12 36.9 & b\_HIGH & 4 $\times$ $\sim$60 \\
SN\,2018hsy & 2018-11-06 08:15 & 09 30 29.83 & +16 20 37.3 & b\_HIGH & 4 $\times$142 \\
SN\,2018hts & 2018-11-06 01:44 & 23 24 32.94 & +17 05 37.9 & b\_HIGH &4 $\times$ 300 \\
SN\,2018htt & 2018-11-14 04:17 & 03 06 02.90 & -15 36 41.3 & b\_HIGH & 4 $\times$8 \\
SN\,2018jbm & 2018-12-03 04:07 & 01 42 17.65 & -45 24 52.1 & b\_HIGH & 4 $\times$200 \\
SN\,2018jdq & 2018-12-06 07:48 & 11 25 25.69 & -08 27 59.1 & b\_HIGH & 4 $\times$77 \\
SN\,2018jeo & 2018-12-03 07:57 & 09 04 36.91 & -19 47 09.7 & b\_HIGH & 4 $\times$90 \\
SN\,2018jff & 2018-12-05 01:11 & 23 48 06.40 & -44 58 48.7 & b\_HIGH & 4 $\times$550 \\
SN\,2018jgn & 2018-12-06 02:36 & 00 02 55.85 & -26 54 51.5 & b\_HIGH & 4 $\times$230 \\
SN\,2018jky & 2018-12-09 00:58 & 03 26 02.12 & -17 33 46.3 & b\_HIGH & 4 $\times$ 142, 2 $\times$ 90, 6 $\times$ 60, 4 $\times$ 70 \\
SN\,2018jny & 2018-12-11 06:43 & 09 24 37.76 & +25 23 42.7 & b\_HIGH & 8 $\times$ $\sim$200 \\
SN\,2018jrn & 2018-12-14 07:52 & 12 52 35.42 & -21 54 56.6 & b\_HIGH & 4 $\times$ $\sim$300 \\
SN\,2018jtj & 2019-01-08 08:13 & 12 41 03.39 & +08 04 21.8 & b\_HIGH & 4 $\times$462 \\
SN\,2018kav & 2019-01-04 01:23 & 04 50 54.70 & -17 59 11.5 & b\_HIGH & 4 $\times$230 \\
SN\,2018koy & 2019-01-04 01:48 & 05 41 13.76 & -13 13 27.1 & b\_HIGH & 4 $\times$276 \\
SN\,2018koy & 2019-01-05 01:21 & 05 41 13.76 & -13 13 27.1 & v\_HIGH & 4 $\times$90 \\
SN\,2018kyi & 2019-01-05 04:00 & 07 07 09.50 & +26 19 29.7 & b\_HIGH & 4 $\times$142 \\
SN\,2019axg & 2019-03-06 05:12 & 10 36 36.18 & -07 06 32.2 & b\_HIGH & 4 $\times$225 \\
SN\,2019bak & 2019-03-06 04:24 & 10 13 31.79 & +29 33 05.4 & b\_HIGH & 4 $\times$400 \\
SN\,2019baq & 2019-03-06 07:33 & 12 22 01.42 & +20 20 27.4 & b\_HIGH & 4 $\times$120 \\
SN\,2019bdz & 2019-03-06 06:56 & 14 48 36.92 & +06 48 52.2 & b\_HIGH & 4 $\times$105 \\
SN\,2019bff & 2019-03-08 06:59 & 14 14 31.78 & +17 13 58.7 & b\_HIGH & 4 $\times$120 \\
SN\,2019bjw & 2019-03-05 07:05 & 10 55 47.44 & +27 40 17.1 & b\_HIGH & 4 $\times$142 \\
SN\,2019bjz & 2019-03-05 08:11 & 16 02 11.80 & +15 55 05.8 & b\_HIGH & 4 $\times$77 \\
SN\,2019bkh & 2019-03-05 08:28 & 13 08 57.41 & +28 16 52.1 & b\_HIGH & 4 $\times$142 \\
SN\,2019bpb & 2019-03-12 01:50 & 05 12 09.87 & -43 55 16.6 & b\_HIGH & 4 $\times$ $\sim$ 400 \\
SN\,2019kg & 2019-02-06 04:08 & 11 42 45.75 & +21 42 53.9 & b\_HIGH & 4 $\times$142 \\
SN\,2019np & 2019-01-13 06:47 & 10 29 21.97 & +29 30 38.5 & b\_HIGH & 4 $\times$50 \\
SN\,2019rm & 2019-01-29 03:26 & 05 53 13.28 & -73 06 56.9 & b\_HIGH & 4 $\times$180 \\
SN\,2019rx & 2019-01-29 01:56 & 05 21 59.63 & -07 11 24.7 & b\_HIGH & 4 $\times$ $\sim$480 \\
SN\,2019rzm & 2019-10-23 02:24 & 01 17 21.66 & -16 04 04.7 & b\_HIGH & 4 $\times$120 \\
SN\,2019shq & 2019-10-24 08:26 & 08 44 28.67 & -31 41 25.6 & b\_HIGH & 4 $\times$142 \\
SN\,2019ubs & 2019-11-24 01:45 & 02 26 24.11 & +28 30 28.8 & b\_HIGH & 4 $\times$450 \\
SN\,2019ubs & 2019-11-27 01:24 & 02 26 24.11 & +28 30 28.8 & v\_HIGH & 4$\times$100 \\
SN\,2019ubt & 2019-11-23 02:15 & 23 29 57.48 & +29 49 46.5 & b\_HIGH & 4 $\times$750 \\
SN\,2019udk & 2019-11-25 02:13 & 00 40 51.25 & +02 50 25.4 & b\_HIGH & 4 $\times$100 \\
SN\,2019udk & 2019-11-27 01:05 & 00 40 51.25 & +02 50 25.4 & v\_HIGH & 4$\times$40 \\
SN\,2019uhz & 2019-11-23 03:20 & 01 47 13.94 & -08 27 40.4 & b\_HIGH & 4 $\times$260 \\
SN\,2019ujy & 2019-11-23 00:44 & 19 41 54.67 & -20 10 04.8 & b\_HIGH & 4 $\times$550 \\
SN\,2019ulp & 2019-11-25 05:56 & 08 28 13.09 & +00 01 29.7 & b\_HIGH & 4 $\times$150 \\
SN\,2019umr & 2019-11-25 01:26 & 22 16 04.96 & -26 40 41.9 & b\_HIGH & 4 $\times$67 \\
SN\,2019umr & 2019-11-27 00:35 & 22 16 04.96 & -26 40 41.9 & v\_HIGH & 4$\times$25 \\
SN\,2019uoo & 2019-11-24 07:15 & 08 38 45.65 & -14 40 49.1 & b\_HIGH & 4 $\times$350 \\
SN\,2019upw & 2019-11-25 01:50 & 00 26 15.49 & +21 48 36.4 & b\_HIGH & 8 $\times$ 60 \\
SN\,2019upw & 2019-11-27 00:52 & 00 26 15.49 & +21 48 36.4 & v\_HIGH & 4$\times$20 \\
SN\,2019urn & 2019-11-24 00:35 & 00 57 49.26 & -17 38 27.0 & b\_HIGH & 4 $\times$300 \\
SN\,2019vju & 2019-12-24 07:44 & 11 04 00.26 & -18 46 35.7 & b\_HIGH & 4 $\times$500 \\
SN\,2019vnj & 2019-12-21 07:46 & 11 09 01.21 & -14 38 10.1 & b\_HIGH & 4 $\times$196 \\
SN\,2019vrq & 2020-02-22 00:53 & 03 04 21.49 & -16 01 25.4 & b\_HIGH & 4 $\times$200, 4 $\times$280 \\
SN\,2019vsa & 2019-12-26 07:08 & 12 05 47.59 & -27 56 37.6 & b\_HIGH & 4 $\times$160 \\
SN\,2019vsa & 2019-12-27 07:21 & 12 05 47.59 & -27 56 37.6 & v\_HIGH & 4$\times$20 \\
SN\,2019vv & 2019-02-06 07:58 & 15 34 03.66 & +02 12 40.4 & v\_HIGH & 4 $\times$169 \\
SN\,2019wka & 2019-12-27 01:58 & 03 14 40.13 & -00 52 16.9 & b\_HIGH & 4 $\times$77 \\
SN\,2019wqz & 2019-12-27 06:40 & 11 47 18.38 & -29 11 19.7 & b\_HIGH & 4 $\times$77 \\
SN\,2020bpz & 2020-02-22 07:18 & 11 40 58.53 & -11 26 57.3 & b\_HIGH & 4 $\times$276 \\
SN\,2020ckp & 2020-02-22 02:08 & 08 36 21.31 & -25 31 57.9 & b\_HIGH & 4 $\times$142, 4 $\times$ 300 \\
SN\,2020ckp & 2020-02-23 06:18 & 08 36 21.31 & -25 31 57.9 & v\_HIGH & 4$\times$200 \\
SN\,2020clq & 2020-02-22 01:47 & 04 33 25.53 & -04 15 57.7 & b\_HIGH & 4 $\times$35 \\
SN\,2020cpt & 2020-02-24 08:12 & 14 57 30.26 & +08 23 24.4 & b\_HIGH & 4 $\times$160 \\
SN\,2020ctr & 2020-02-22 02:58 & 08 58 37.58 & +20 11 31.1 & b\_HIGH & 4 $\times$110, 4 $\times$200 \\
SN\,2020dvr & 2020-03-20 01:54 & 09 54 35.14 & -19 53 14.7 & b\_HIGH & 4 $\times$12 \\
SN\,2020dyf & 2020-03-20 07:54 & 15 30 29.68 & -12 36 06.1 & b\_HIGH & 4 $\times$100 \\
SN\,2020dyf & 2020-03-22 07:20 & 15 30 29.68 & -12 36 06.1 & v\_HIGH & 4$\times$42 \\
SN\,2020ejm & 2020-03-20 06:15 & 10 16 18.74 & -33 33 49.7 & b\_HIGH & 4 $\times$18 \\
SN\,2020ejm & 2020-03-21 03:26 & 10 16 18.74 & -33 33 49.7 & v\_HIGH & 8$\times$4 \\
		\hline
	\end{tabular}
\end{table*}

\clearpage
\onecolumn
{\scriptsize
\begin{longtable}{lcccccccccc} 
\caption{\label{tab:PB} Polarization measurements of 69 SNe~Ia in the $B$ band obtained in this work. Twelve SNe have been followed-up in the $V$ band.} \\
\hline
Name & Filter & $q$ (\%) & $u$ (\%) & $\theta(^{\circ})$ & p (\%) & $q_{\rm ISP}$ (\%) & $u_{\rm ISP}$ (\%) & p$_{\rm ISP}$ (\%) & ISPcorr & p$_{\rm corr}$ (\%) \\
\hline
\endfirsthead
\caption{continued.}\\
\hline
Name & Filter & $q$ (\%) & $u$ (\%) & $\theta(^{\circ})$ & p (\%) & $q_{\rm ISP}$ (\%) & $u_{\rm ISP}$ (\%) & p$_{\rm ISP}$ (\%) & ISPcorr & p$_{\rm corr}$ (\%) \\
\hline
\endhead
\hline
\endfoot

\multicolumn{11}{c}{SNe~Ia in spiral host galaxies}\\
\hline
SN\,2018cif & $B$ & 0.05 $\pm$ 0.07 & 0.06 $\pm$ 0.07 	&  25.1 $\pm$ 25.7	& 0.08 $\pm$ 0.07 & 0.0  $\pm$ 0.14 & -0.34  $\pm$ 0.38 & 0.34  $\pm$ 0.38 & No & 0.02 $\pm$ 0.07 \\
SN\,2018cif & $V$ &  -0.02 $\pm$ 0.09 & 0.11 $\pm$ 0.09  & 51.2 $\pm$ 23.0  & 0.12 $\pm$ 0.09 & \dots & \dots & \dots & No & 0.04 $\pm$ 0.09 \\
SN\,2018fqd & $B$ & 0.06 $\pm$ 0.22 & -0.73 $\pm$ 0.25 	&  137.3 $\pm$ 9.8	& 0.73 $\pm$ 0.25 & 0.31  $\pm$ 0.18 & -0.2  $\pm$ 0.16 & 0.37  $\pm$ 0.17 & Yes &0.44 $\pm$ 0.29 \\
SN\,2018fuk & $B$ & 0.04 $\pm$ 0.05 & 0.13 $\pm$ 0.05 	&  36.4 $\pm$ 10.5	& 0.14 $\pm$ 0.05 & 0.54  $\pm$ 0.27 & 0.02  $\pm$ 0.22 & 0.54  $\pm$ 0.27 & No & 0.12 $\pm$ 0.05 \\
SN\,2018fvi & $B$ & -0.2 $\pm$ 0.16 & 0.78 $\pm$ 0.16 	&  52.2 $\pm$ 5.7	& 0.81 $\pm$ 0.16 & 0.07  $\pm$ 0.14 & -0.25  $\pm$ 0.17 & 0.26  $\pm$ 0.17 & No & 0.77 $\pm$ 0.16 \\
SN\,2018fvi & $V$ &  -0.56 $\pm$ 0.08 & 0.19 $\pm$ 0.08  & 80.6 $\pm$ 4.1   & 0.59 $\pm$ 0.08  & \dots & \dots & \dots & No & 0.58 $\pm$ 0.08 \\
SN\,2018fvy & $B$ & -0.06 $\pm$ 0.09 & -0.07 $\pm$ 0.08 &  114.7 $\pm$ 26.2	& 0.09 $\pm$ 0.08 & \dots & \dots & \dots & No & 0.01 $\pm$ 0.08 \\
SN\,2018fzm & $B$ & 0.03 $\pm$ 0.12 & -0.27 $\pm$ 0.12 	&  138.2 $\pm$ 12.7	& 0.27 $\pm$ 0.12 & 0.17  $\pm$ 0.17 & -0.42  $\pm$ 0.3 & 0.45  $\pm$ 0.29 & Yes & 0.0 $\pm$ 0.28 \\
SN\,2018hsa & $B$ & 0.16 $\pm$ 0.06 & -0.39 $\pm$ 0.05 	&  146.2 $\pm$ 3.5	& 0.42 $\pm$ 0.05 & 0.38  $\pm$ 0.21 & -0.35  $\pm$ 0.04 & 0.52  $\pm$ 0.16 & Yes &0.02 $\pm$ 0.22 \\
SN\,2018hsy & $B$ & -0.57 $\pm$ 0.12 & -0.27 $\pm$ 0.13 &  102.7 $\pm$ 5.5	& 0.63 $\pm$ 0.12 & -0.28  $\pm$ 0.01 & -0.2  $\pm$ 0.06 & 0.34  $\pm$ 0.04 & Yes &0.25 $\pm$ 0.12 \\
SN\,2018jbm & $B$ & -0.05 $\pm$ 0.1 & 0.2 $\pm$ 0.1 	&  52.0 $\pm$ 13.9	& 0.21 $\pm$ 0.1 & 0.0  $\pm$ 0.09 & 0.3  $\pm$ 0.09 & 0.3  $\pm$ 0.09 & Yes & 0.0 $\pm$ 0.13 \\
SN\,2018jdq & $B$ & 0.39 $\pm$ 0.06 & 0.07 $\pm$ 0.06 	&  5.1 $\pm$ 4.3	& 0.4 $\pm$ 0.06 & 0.24  $\pm$ 0.12 & 0.09  $\pm$ 0.12 & 0.26  $\pm$ 0.12 & Yes &0.03 $\pm$ 0.13 \\
SN\,2018jeo & $B$ & 0.0 $\pm$ 0.05 & -0.1 $\pm$ 0.05 	&  135.0 $\pm$ 14.3	& 0.1 $\pm$ 0.05 & -0.02  $\pm$ 0.35 & -0.11  $\pm$ 0.17 & 0.11  $\pm$ 0.18 & Yes & 0.0 $\pm$ 0.33 \\
SN\,2018jff & $B$ & -0.1 $\pm$ 0.06 & 0.08 $\pm$ 0.06 	&  70.7 $\pm$ 13.4	& 0.13 $\pm$ 0.06 & -0.13  $\pm$ 0.14 & 0.17  $\pm$ 0.14 & 0.21  $\pm$ 0.14 & Yes & 0.0 $\pm$ 0.15 \\
SN\,2018jky & $B$ & 0.33 $\pm$ 0.03 & 0.08 $\pm$ 0.03 	&  6.8 $\pm$ 2.5	& 0.34 $\pm$ 0.03 & 0.09  $\pm$ 0.05 & -0.18  $\pm$ 0.05 & 0.2  $\pm$ 0.05 & No & 0.34 $\pm$ 0.03 \\
SN\,2018jny & $B$ & -0.07 $\pm$ 0.06 & -0.29 $\pm$ 0.06 &  128.2 $\pm$ 5.8	& 0.3 $\pm$ 0.06 & 0.12  $\pm$ 0.05 & 0.11  $\pm$ 0.06 & 0.16  $\pm$ 0.05 & No & 0.29 $\pm$ 0.06 \\
SN\,2018jtj & $B$ & -0.03 $\pm$ 0.07 & -0.22 $\pm$ 0.07 &  131.1 $\pm$ 9.0	& 0.22 $\pm$ 0.07 & -0.56  $\pm$ 0.08 & 0.08  $\pm$ 0.07 & 0.57  $\pm$ 0.08 & No & 0.2 $\pm$ 0.07 \\
SN\,2018kav & $B$ & -0.18 $\pm$ 0.1 & 0.2 $\pm$ 0.1 	&  66.0 $\pm$ 10.6	& 0.27 $\pm$ 0.1 & 0.04  $\pm$ 0.06 & 0.17  $\pm$ 0.1 & 0.17  $\pm$ 0.1 & Yes &0.16 $\pm$ 0.12 \\
SN\,2019axg & $B$ & -0.65 $\pm$ 0.2 & 0.68 $\pm$ 0.2 	&  66.9 $\pm$ 6.1	& 0.94 $\pm$ 0.2 & 0.07  $\pm$ 0.09 & 0.13  $\pm$ 0.16 & 0.15  $\pm$ 0.15 & Yes &0.85 $\pm$ 0.23 \\
SN\,2019bak & $B$ & 0.17 $\pm$ 0.17 & -0.04 $\pm$ 0.17 	&  173.4 $\pm$ 27.9	& 0.17 $\pm$ 0.17 & 0.42  $\pm$ 0.11 & 0.06  $\pm$ 0.54 & 0.42  $\pm$ 0.13 & No & 0.01 $\pm$ 0.17 \\
SN\,2019bdz & $B$ & 0.09 $\pm$ 0.09 & 0.3 $\pm$ 0.09 	&  36.7 $\pm$ 8.2	& 0.31 $\pm$ 0.09 & -0.05  $\pm$ 0.13 & 0.2  $\pm$ 0.06 & 0.21  $\pm$ 0.07 & Yes &0.05 $\pm$ 0.14 \\
SN\,2019bff & $B$ & -0.05 $\pm$ 0.12 & 0.91 $\pm$ 0.12 	&  46.6 $\pm$ 3.8	& 0.91 $\pm$ 0.12 & -0.17  $\pm$ 0.27 & 0.17  $\pm$ 0.06 & 0.24  $\pm$ 0.2 & Yes &0.72 $\pm$ 0.14 \\
SN\,2019bjw & $B$ & 0.14 $\pm$ 0.07 & -0.02 $\pm$ 0.07 	&  175.9 $\pm$ 14.2	& 0.14 $\pm$ 0.07 & -0.04  $\pm$ 0.02 & -0.01  $\pm$ 0.07 & 0.04  $\pm$ 0.03 & No & 0.11 $\pm$ 0.07 \\
SN\,2019bkh & $B$ & -0.11 $\pm$ 0.12 & -0.11 $\pm$ 0.12 &  112.5 $\pm$ 22.1	& 0.16 $\pm$ 0.12 & 0.05  $\pm$ 0.06 & -0.02  $\pm$ 0.03 & 0.05  $\pm$ 0.06 & No & 0.06 $\pm$ 0.12 \\
SN\,2019np  & $B$ & -0.03 $\pm$ 0.07 & 0.42 $\pm$ 0.07 	&  47.0 $\pm$ 4.8	& 0.42 $\pm$ 0.07 & \dots & \dots & \dots & No & 0.41 $\pm$ 0.07 \\
SN\,2019rx  & $B$ & -0.21 $\pm$ 0.14 & -0.14 $\pm$ 0.11 &  106.8 $\pm$ 14.9	& 0.25 $\pm$ 0.13 & -0.17  $\pm$ 0.06 & 0.17  $\pm$ 0.17 & 0.24  $\pm$ 0.13 & No & 0.18 $\pm$ 0.13 \\
SN\,2019shq & $B$ & -0.41 $\pm$ 0.36 & 0.25 $\pm$ 0.37 	&  74.3 $\pm$ 21.6	& 0.48 $\pm$ 0.36 & -0.46  $\pm$ 0.18 & -0.06  $\pm$ 0.17 & 0.46  $\pm$ 0.18 & Yes & 0.0 $\pm$ 0.41 \\
SN\,2019ubs & $B$ & 0.11 $\pm$ 0.06 & -0.72 $\pm$ 0.06 	&  139.3 $\pm$ 2.4	& 0.73 $\pm$ 0.06 & -0.19  $\pm$ 0.29 & -0.72  $\pm$ 0.07 & 0.74  $\pm$ 0.1 & Yes &0.01 $\pm$ 0.3 \\
SN\,2019ubs & $V$ &   0.01 $\pm$ 0.11 & -0.66 $\pm$ 0.11 & 135.4 $\pm$ 4.8  & 0.66 $\pm$ 0.11  & \dots & \dots & \dots & No & 0.65 $\pm$ 0.11 \\
SN\,2019ubt & $B$ & -0.19 $\pm$ 0.09 & 0.67 $\pm$ 0.1 	&  52.9 $\pm$ 4.1	& 0.7 $\pm$ 0.1 & \dots & \dots & \dots & No & 0.68 $\pm$ 0.1 \\ 
SN\,2019udk & $B$ & -1.69 $\pm$ 0.17 & -0.75 $\pm$ 0.16 &  102.0 $\pm$ 2.6	& 1.85 $\pm$ 0.17 & -0.08  $\pm$ 0.24 & -0.14  $\pm$ 0.23 & 0.16  $\pm$ 0.23 & Yes &1.67 $\pm$ 0.29 \\
SN\,2019udk & $V$ &  -1.47 $\pm$ 0.16 & -0.55 $\pm$ 0.16 & 100.3 $\pm$ 2.9  & 1.57 $\pm$ 0.16  & \dots & \dots & \dots & No & 1.56 $\pm$ 0.16 \\
SN\,2019uhz & $B$ & -0.04 $\pm$ 0.11 & -0.39 $\pm$ 0.11 &  132.1 $\pm$ 8.0	& 0.39 $\pm$ 0.11 & -0.56  $\pm$ 0.26 & 0.73  $\pm$ 0.68 & 0.92  $\pm$ 0.56 & No & 0.36 $\pm$ 0.11 \\
SN\,2019ulp & $B$ & 0.11 $\pm$ 0.13 & -0.27 $\pm$ 0.13 	&  146.1 $\pm$ 12.8	& 0.29 $\pm$ 0.13 & -0.08  $\pm$ 0.06 & 0.18  $\pm$ 0.08 & 0.2  $\pm$ 0.08 & No & 0.23 $\pm$ 0.13 \\
SN\,2019umr & $B$ & -0.29 $\pm$ 0.22 & 0.78 $\pm$ 0.22 	&  55.2 $\pm$ 7.6	& 0.83 $\pm$ 0.22 & -0.02  $\pm$ 0.15 & -0.05  $\pm$ 0.06 & 0.05  $\pm$ 0.08 & No & 0.77 $\pm$ 0.22 \\
SN\,2019umr & $V$ &  -0.4 $\pm$ 0.23  & 0.57 $\pm$ 0.23  & 62.5 $\pm$ 9.7   & 0.69 $\pm$ 0.23  & \dots & \dots & \dots & No & 0.61 $\pm$ 0.23 \\
SN\,2019upw & $B$ & -0.47 $\pm$ 0.05 & -0.36 $\pm$ 0.05 &  108.7 $\pm$ 2.4	& 0.59 $\pm$ 0.05 & \dots & \dots & \dots & No & 0.59 $\pm$ 0.05 \\
SN\,2019upw & $V$ &  -0.54 $\pm$ 0.11 & -0.25 $\pm$ 0.11 & 102.5 $\pm$ 5.1  & 0.6  $\pm$ 0.11  & \dots & \dots & \dots & No & 0.58 $\pm$ 0.11 \\
SN\,2019vju & $B$ & -0.46 $\pm$ 0.36 & -0.21 $\pm$ 0.36 &  102.3 $\pm$ 20.4	& 0.51 $\pm$ 0.36 & 0.01  $\pm$ 0.14 & -0.03  $\pm$ 0.03 & 0.03  $\pm$ 0.05 & Yes &0.21 $\pm$ 0.38 \\
SN\,2019vnj & $B$ & -0.03 $\pm$ 0.13 & 0.15 $\pm$ 0.13 	&  50.7 $\pm$ 24.3	& 0.15 $\pm$ 0.13 & \dots & \dots & \dots & No & 0.04 $\pm$ 0.13 \\
SN\,2019vrq & $B$ & 0.29 $\pm$ 0.1 & -0.05 $\pm$ 0.1 	&  175.1 $\pm$ 9.7	& 0.29 $\pm$ 0.1 & 0.41  $\pm$ 0.12 & -0.14  $\pm$ 0.04 & 0.43  $\pm$ 0.11 & Yes &0.02 $\pm$ 0.14 \\
SN\,2019vsa & $B$ & -0.32 $\pm$ 0.12 & -0.63 $\pm$ 0.12 &  121.5 $\pm$ 4.9	& 0.71 $\pm$ 0.12 & -0.17  $\pm$ 0.26 & 0.13  $\pm$ 0.17 & 0.21  $\pm$ 0.23 & No & 0.69 $\pm$ 0.12 \\
SN\,2019vsa & $V$ &  -0.43 $\pm$ 0.1  & -0.7 $\pm$ 0.1   & 119.0 $\pm$ 3.6  & 0.82 $\pm$ 0.1   & \dots & \dots & \dots & No & 0.81 $\pm$ 0.1  \\
SN\,2019wka & $B$ & 0.45 $\pm$ 0.1 & 0.03 $\pm$ 0.1 	&  1.9 $\pm$ 6.4	& 0.45 $\pm$ 0.1 & 0.55  $\pm$ 0.07 & 0.06  $\pm$ 0.1 & 0.55  $\pm$ 0.07 & Yes & 0.0 $\pm$ 0.12 \\
SN\,2019wqz & $B$ & -0.08 $\pm$ 0.08 & -0.07 $\pm$ 0.08 &  110.6 $\pm$ 21.6	& 0.11 $\pm$ 0.08 & -0.12  $\pm$ 0.11 & 0.22  $\pm$ 0.12 & 0.25  $\pm$ 0.12 & No & 0.05 $\pm$ 0.08 \\
SN\,2020bpz & $B$ & -0.01 $\pm$ 0.08 & 0.03 $\pm$ 0.08 	&  54.2 $\pm$ 72.5	& 0.03 $\pm$ 0.08 & \dots & \dots & \dots & No & 0.0 $\pm$ 0.08 \\
SN\,2020ckp & $B$ & -0.17 $\pm$ 0.13 & -0.93 $\pm$ 0.13 &  129.8 $\pm$ 3.9	& 0.95 $\pm$ 0.13 & -0.22  $\pm$ 0.14 & -0.89  $\pm$ 0.18 & 0.92  $\pm$ 0.18 & Yes & 0.0 $\pm$ 0.2 \\
SN\,2020ckp & $V$ &  -0.23 $\pm$ 0.22 & -0.86 $\pm$ 0.23 & 127.6 $\pm$ 7.2  & 0.89 $\pm$ 0.23  & \dots & \dots & \dots & No & 0.83 $\pm$ 0.23 \\
SN\,2020dvr & $B$ & -0.04 $\pm$ 0.09 & -0.08 $\pm$ 0.09 &  121.7 $\pm$ 28.8	& 0.09 $\pm$ 0.09 & 0.08  $\pm$ 0.12 & 0.22  $\pm$ 0.23 & 0.23  $\pm$ 0.22 & No & 0.0 $\pm$ 0.09 \\
SN\,2020dyf & $B$ & -0.57 $\pm$ 0.06 & 1.12 $\pm$ 0.06 	&  58.5 $\pm$ 1.4	& 1.26 $\pm$ 0.06 & -0.87  $\pm$ 0.28 & 1.31  $\pm$ 0.19 & 1.57  $\pm$ 0.22 & Yes &0.16 $\pm$ 0.26 \\
SN\,2020dyf & $V$ &  -0.81 $\pm$ 0.08 & 1.12 $\pm$ 0.08  & 63.0 $\pm$ 1.6   & 1.38 $\pm$ 0.08  & \dots & \dots & \dots & No & 1.38 $\pm$ 0.08 \\
SN\,2020ejm & $B$ & -0.5 $\pm$ 0.07 & 0.59 $\pm$ 0.07 	&  65.1 $\pm$ 2.6	& 0.77 $\pm$ 0.07 & -0.43  $\pm$ 0.06 & 0.35  $\pm$ 0.06 & 0.55  $\pm$ 0.06 & Yes &0.22 $\pm$ 0.09  \\
SN\,2020ejm & $V$ &  -0.39 $\pm$ 0.13 & 0.54 $\pm$ 0.13  & 63.1 $\pm$ 5.7   & 0.66 $\pm$ 0.13  & \dots & \dots & \dots & No & 0.64 $\pm$ 0.13 \\
\hline
\multicolumn{11}{c}{SNe~Ia in elliptical host galaxies}\\
\hline
SN\,2018fhw & $B$ & -0.06 $\pm$ 0.11 & -0.14 $\pm$ 0.11 &  123.4 $\pm$ 20.7	& 0.15 $\pm$ 0.11 & -0.11  $\pm$ 0.16 & -0.09  $\pm$ 0.18 & 0.14  $\pm$ 0.17 & Yes & 0.0 $\pm$ 0.2 \\
SN\,2018htt & $B$ & 0.28 $\pm$ 0.08 & -0.05 $\pm$ 0.08 	&  174.9 $\pm$ 8.1	& 0.28 $\pm$ 0.08 & 0.36  $\pm$ 0.09 & -0.09  $\pm$ 0.19 & 0.37  $\pm$ 0.1 & Yes & 0.0 $\pm$ 0.14 \\
SN\,2018jgn & $B$ & -0.34 $\pm$ 0.41 & 0.1 $\pm$ 0.11 	&  81.8 $\pm$ 31.9	& 0.35 $\pm$ 0.39 & \dots  & \dots  & \dots & No & 0.0 $\pm$ 0.39 \\
SN\,2018jrn & $B$ & 0.24 $\pm$ 0.07 & 0.02 $\pm$ 0.1 	&  2.4 $\pm$ 8.4	& 0.24 $\pm$ 0.07 & \dots   & \dots   & \dots & No & 0.22 $\pm$ 0.07 \\
SN\,2019baq & $B$ & -0.05 $\pm$ 0.09 & 0.2 $\pm$ 0.09 	&  52.0 $\pm$ 12.5	& 0.21 $\pm$ 0.09 & 0.1  $\pm$ 0.06 & 0.08  $\pm$ 0.09 & 0.13  $\pm$ 0.07 & Yes &0.12 $\pm$ 0.12 \\
SN\,2019bjz & $B$ & -0.37 $\pm$ 0.13 & 0.13 $\pm$ 0.13 	&  80.3 $\pm$ 9.5	& 0.39 $\pm$ 0.13 & -0.02  $\pm$ 0.09 & 0.13  $\pm$ 0.17 & 0.13  $\pm$ 0.17 & Yes &0.28 $\pm$ 0.16 \\
SN\,2019ujy & $B$ & 0.71 $\pm$ 0.09 & -0.03 $\pm$ 0.09 	&  178.8 $\pm$ 3.6	& 0.71 $\pm$ 0.09 & 0.74  $\pm$ 0.28 & -0.09  $\pm$ 0.18 & 0.75  $\pm$ 0.28 & Yes & -0.0 $\pm$ 0.22 \\
SN\,2019urn & $B$ & 0.11 $\pm$ 0.06 & 0.08 $\pm$ 0.06 	&  18.0 $\pm$ 12.6	& 0.14 $\pm$ 0.06 & -0.06  $\pm$ 0.07 & 0.07  $\pm$ 0.07 & 0.09  $\pm$ 0.07 & No & 0.11 $\pm$ 0.06 \\
SN\,2020cpt & $B$ & -0.01 $\pm$ 0.13 & 0.17 $\pm$ 0.13 	&  46.7 $\pm$ 21.9	& 0.17 $\pm$ 0.13 & -0.28  $\pm$ 0.19 & 0.18  $\pm$ 0.19 & 0.33  $\pm$ 0.19 & No & 0.07 $\pm$ 0.13 \\
SN\,2020ctr & $B$ & -0.01 $\pm$ 0.06 & 0.21 $\pm$ 0.06 	&  46.4 $\pm$ 8.2	& 0.21 $\pm$ 0.06 & -0.09  $\pm$ 0.09 & 0.0  $\pm$ 0.1 & 0.09  $\pm$ 0.09 & No & 0.19 $\pm$ 0.06 \\
\hline
\multicolumn{11}{c}{SNe~Ia in S0 host galaxies}\\
\hline
SN\,2018evt & $B$ & -1.37 $\pm$ 0.09 & -0.32 $\pm$ 0.09 &  96.6 $\pm$ 1.8	& 1.41 $\pm$ 0.09 & -0.29  $\pm$ 0.21 & 0.0  $\pm$ 0.07 & 0.29  $\pm$ 0.21 & Yes &1.08 $\pm$ 0.22 \\
SN\,2018fhx & $B$ & 0.2 $\pm$ 0.35 & -0.06 $\pm$ 0.36 	&  171.7 $\pm$ 48.1	& 0.21 $\pm$ 0.35 & -0.01  $\pm$ 0.68 & -0.29  $\pm$ 0.29 & 0.29  $\pm$ 0.29 & No & 0.0 $\pm$ 0.35 \\
SN\,2018fnq & $B$ & 1.67 $\pm$ 0.06 & -0.32 $\pm$ 0.06 	&  174.6 $\pm$ 1.0	& 1.7 $\pm$ 0.06 & 0.37  $\pm$ 0.19 & -0.22  $\pm$ 0.13 & 0.43  $\pm$ 0.18 & Yes &1.27 $\pm$ 0.2 \\
SN\,2018fnq & $V$ &  0.07 $\pm$ 0.07  & -0.22 $\pm$ 0.07 & 144.3 $\pm$ 8.9  & 0.24 $\pm$ 0.07  & \dots & \dots & \dots & No & 0.21 $\pm$ 0.07 \\
SN\,2018gyr & $B$ & -0.16 $\pm$ 0.18 & 0.01 $\pm$ 0.18 	&  88.2 $\pm$ 32.2	& 0.16 $\pm$ 0.18 & \dots & \dots & \dots & No & 0.0 $\pm$ 0.18 \\
SN\,2019rzm & $B$ & -0.23 $\pm$ 0.08 & 0.07 $\pm$ 0.08 	&  81.5 $\pm$ 9.5	& 0.24 $\pm$ 0.08 & 0.03  $\pm$ 0.1 & -0.19  $\pm$ 0.1 & 0.19  $\pm$ 0.1 & No & 0.21 $\pm$ 0.08 \\
SN\,2019uoo & $B$ & 0.12 $\pm$ 0.31 & 0.0 $\pm$ 0.31 	&  0.0 $\pm$ 74.0	& 0.12 $\pm$ 0.31 & 0.31  $\pm$ 0.24 & -0.12  $\pm$ 0.11 & 0.33  $\pm$ 0.23 & No & 0.0 $\pm$ 0.31 \\
SN\,2019vv  & $V$ &  -0.11 $\pm$ 0.27 & 0.62 $\pm$ 0.29  & 50.2 $\pm$ 13.2  & 0.63 $\pm$ 0.29  & \dots & \dots & \dots & No & 0.49 $\pm$ 0.29 \\
SN\,2020clq & $B$ & -0.36 $\pm$ 0.07 & 0.16 $\pm$ 0.07 	&  78.0 $\pm$ 5.1	& 0.39 $\pm$ 0.07 & 0.25  $\pm$ 0.02 & -0.06  $\pm$ 0.05 & 0.26  $\pm$ 0.02 & No & 0.38 $\pm$ 0.07 \\
\hline
\multicolumn{11}{c}{SNe~Ia in irregular host galaxies}\\
\hline
SN\,2019rm  & $B$ & 0.5 $\pm$ 0.05 & 0.41 $\pm$ 0.05 	&  19.7 $\pm$ 2.2	& 0.65 $\pm$ 0.05 & 0.5  $\pm$ 0.12 & 0.4  $\pm$ 0.13 & 0.64  $\pm$ 0.12 & Yes & 0.0 $\pm$ 0.14 \\
\hline
\multicolumn{11}{c}{SNe~Ia in host galaxies of unknown morphological classification}\\
\hline
SN\,2018fqn & $B$ & 0.12 $\pm$ 0.05 & 0.0 $\pm$ 0.05 	&  0.0 $\pm$ 11.9	& 0.12 $\pm$ 0.05 & 0.18  $\pm$ 0.3 & 0.0  $\pm$ 0.3 & 0.18  $\pm$ 0.3 & Yes & 0.0 $\pm$ 0.3 \\
SN\,2018fsa & $B$ & 0.2 $\pm$ 0.08 & 0.36 $\pm$ 0.08 	&  30.5 $\pm$ 5.6	& 0.41 $\pm$ 0.08 & \dots & \dots & \dots & No & 0.4 $\pm$ 0.08 \\
SN\,2018hts & $B$ & -0.29 $\pm$ 0.05 & -0.1 $\pm$ 0.05 	&  99.5 $\pm$ 4.7	& 0.31 $\pm$ 0.05 & -0.15  $\pm$ 0.05 & -0.19  $\pm$ 0.01 & 0.24  $\pm$ 0.03 & Yes &0.14 $\pm$ 0.07 \\
SN\,2018koy & $B$ & -0.64 $\pm$ 0.08 & 0.91 $\pm$ 0.08 	&  62.6 $\pm$ 2.1	& 1.11 $\pm$ 0.08 & -0.35  $\pm$ 0.08 & 0.45  $\pm$ 0.14 & 0.57  $\pm$ 0.12 & Yes &0.5 $\pm$ 0.15 \\
SN\,2018koy & $V$ &  -0.66 $\pm$ 0.08 & 0.9 $\pm$ 0.08   & 63.0 $\pm$ 2.0   & 1.12 $\pm$ 0.08  & \dots & \dots & \dots & No & 1.11 $\pm$ 0.08 \\
SN\,2018kyi & $B$ & 0.0 $\pm$ 0.06 & -0.31 $\pm$ 0.06 	&  135.0 $\pm$ 5.5	& 0.31 $\pm$ 0.06 & 0.08  $\pm$ 0.15 & -0.36  $\pm$ 0.15 & 0.37  $\pm$ 0.15 & Yes & 0.0 $\pm$ 0.16 \\
SN\,2019bpb & $B$ & 0.04 $\pm$ 0.07 & -0.3 $\pm$ 0.07 	&  138.8 $\pm$ 6.6	& 0.3 $\pm$ 0.07 & 0.0  $\pm$ 0.01 & 0.15  $\pm$ 0.13 & 0.15  $\pm$ 0.13 & No & 0.29 $\pm$ 0.07 \\
SN\,2019kg  & $B$ & 0.76 $\pm$ 0.11 & -0.08 $\pm$ 0.11 	&  177.0 $\pm$ 4.1	& 0.76 $\pm$ 0.11 & 0.03  $\pm$ 0.25 & 0.11  $\pm$ 0.25 & 0.11  $\pm$ 0.25 & Yes &0.66 $\pm$ 0.27 \\
\hline
\end{longtable}
\footnotesize{Notes: SN\,2019vv has been observed in $V$ band only.} $p_{\rm corr}$ are the ISP subtracted (if so indicated in the ISP$_{\rm corr}$) and polarization bias corrected values.\\
}

\clearpage

\begin{table*}
	\centering
	\scriptsize
	\caption{\label{tab:archivalSNe}  Polarization of 35 archival SNe from \citet{2019MNRAS.490..578C} in the $B$ and $V$ bands.}
	\begin{tabular}{llccccccccc} 
	\hline
Name & MJD & Epoch & q$_B$ (\%) & u$_B$ (\%)  &  $\theta_B$ ($^\circ$) &  p$_B$ (\%) &  q$_V$ (\%) & u$_V$ (\%)  &  $\theta_V$ ($^\circ$) &  p$_V$ (\%)   \\
\hline
\multicolumn{11}{c}{SNe~Ia in spiral host galaxies}\\
\hline
SN\,2001dm & 52134.33168 & 6.3 & -0.23 $\pm$ 0.42 & -0.21 $\pm$ 0.46 & 111.5 $\pm$ 39.9 & 0.0 $\pm$ 0.44 & -0.04 $\pm$ 0.18 & -0.35 $\pm$ 0.22 & 131.8 $\pm$ 39.9 & 0.21 $\pm$ 0.22 \\
SN\,2001el & 52222.1011 & 39.6 & -0.03 $\pm$ 0.09 & 0.61 $\pm$ 0.09 & 46.2 $\pm$ 4.0 & 0.6 $\pm$ 0.09 & 0.02 $\pm$ 0.04 & 0.62 $\pm$ 0.03 & 43.9 $\pm$ 4.0 & 0.62 $\pm$ 0.03 \\
SN\,2001V & 51990.12065 & 17.5 & 0.14 $\pm$ 0.12 & 0.03 $\pm$ 0.24 & 5.1 $\pm$ 24.3 & 0.04 $\pm$ 0.12 & 0.14 $\pm$ 0.06 & 0.07 $\pm$ 0.05 & 12.3 $\pm$ 24.3 & 0.14 $\pm$ 0.06 \\
SN\,2002bo & 52369.05487 & 12.6 & -0.88 $\pm$ 0.1 & -0.36 $\pm$ 0.11 & 101.2 $\pm$ 3.1 & 0.94 $\pm$ 0.1 & -0.51 $\pm$ 0.08 & -0.09 $\pm$ 0.05 & 95.0 $\pm$ 3.1 & 0.51 $\pm$ 0.08 \\
SN\,2002fk & 52561.34352 & 13.4 & 0.28 $\pm$ 0.05 & -0.23 $\pm$ 0.04 & 160.3 $\pm$ 3.6 & 0.36 $\pm$ 0.05 & 0.3 $\pm$ 0.04 & -0.24 $\pm$ 0.04 & 160.5 $\pm$ 3.6 & 0.38 $\pm$ 0.04 \\
SN\,2003eh & 52782.01566 & 0.0 & 1.22 $\pm$ 0.49 & 1.19 $\pm$ 0.42 & 22.1 $\pm$ 7.6 & 1.58 $\pm$ 0.45 & 0.65 $\pm$ 0.28 & 0.98 $\pm$ 0.35 & 28.3 $\pm$ 7.6 & 1.08 $\pm$ 0.33 \\
SN\,2003W & 52673.26544 & -6.7 & -0.37 $\pm$ 0.14 & -0.35 $\pm$ 0.25 & 111.7 $\pm$ 11.2 & 0.43 $\pm$ 0.2 & -0.48 $\pm$ 0.17 & -0.31 $\pm$ 0.12 & 106.6 $\pm$ 11.2 & 0.53 $\pm$ 0.15 \\
SN\,2004dt & 53250.29035 & 10.3 & 0.17 $\pm$ 0.1 & -0.28 $\pm$ 0.12 & 150.1 $\pm$ 9.9 & 0.29 $\pm$ 0.11 & 0.03 $\pm$ 0.1 & -0.13 $\pm$ 0.08 & 142.3 $\pm$ 9.9 & 0.09 $\pm$ 0.08 \\
SN\,2004ef & 53259.147 & -5.3 & 0.04 $\pm$ 0.18 & -0.37 $\pm$ 0.28 & 137.9 $\pm$ 21.3 & 0.17 $\pm$ 0.28 & 0.01 $\pm$ 0.3 & -0.14 $\pm$ 0.27 & 136.1 $\pm$ 21.3 & 0.0 $\pm$ 0.27 \\
SN\,2004eo & 53268.10228 & -10.3 & -0.08 $\pm$ 0.14 & 0.15 $\pm$ 0.1 & 59.1 $\pm$ 18.7 & 0.1 $\pm$ 0.11 & -0.04 $\pm$ 0.17 & 0.09 $\pm$ 0.15 & 57.5 $\pm$ 18.7 & 0.0 $\pm$ 0.15 \\
SN\,2005de & 53594.00508 & -4.9 & 0.37 $\pm$ 0.09 & 0.02 $\pm$ 0.07 & 1.2 $\pm$ 6.8 & 0.35 $\pm$ 0.09 & 0.39 $\pm$ 0.1 & 0.03 $\pm$ 0.08 & 2.2 $\pm$ 6.8 & 0.37 $\pm$ 0.1 \\
SN\,2005df & 53641.30388 & 42.1 & -0.03 $\pm$ 0.05 & -0.03 $\pm$ 0.06 & 113.0 $\pm$ 39.0 & 0.0 $\pm$ 0.06 & 0.06 $\pm$ 0.04 & 0.01 $\pm$ 0.03 & 7.0 $\pm$ 39.0 & 0.04 $\pm$ 0.04 \\
SN\,2005hk & 53697.07734 & 11.7 & -0.02 $\pm$ 0.12 & -0.31 $\pm$ 0.16 & 133.6 $\pm$ 14.8 & 0.23 $\pm$ 0.16 & 0.08 $\pm$ 0.1 & -0.23 $\pm$ 0.07 & 144.2 $\pm$ 14.8 & 0.23 $\pm$ 0.07 \\
SN\,2005ke & 53691.08175 & -8.1 & 0.16 $\pm$ 0.12 & -0.13 $\pm$ 0.09 & 160.4 $\pm$ 15.4 & 0.15 $\pm$ 0.11 & 0.01 $\pm$ 0.11 & -0.18 $\pm$ 0.06 & 137.0 $\pm$ 15.4 & 0.16 $\pm$ 0.06 \\
SN\,2006X & 53825.15811 & 38.9 & 1.09 $\pm$ 0.28 & -7.17 $\pm$ 0.29 & 139.3 $\pm$ 1.1 & 7.24 $\pm$ 0.29 & 0.96 $\pm$ 0.06 & -6.03 $\pm$ 0.3 & 139.5 $\pm$ 1.1 & 6.09 $\pm$ 0.3 \\
SN\,2007fb & 54298.31065 & 9.9 & -0.66 $\pm$ 0.04 & -0.23 $\pm$ 0.05 & 99.8 $\pm$ 1.6 & 0.7 $\pm$ 0.04 & -0.6 $\pm$ 0.05 & -0.21 $\pm$ 0.04 & 99.6 $\pm$ 1.6 & 0.63 $\pm$ 0.05 \\
SN\,2007le & 54440.03806 & 40.7 & 0.92 $\pm$ 0.13 & -1.58 $\pm$ 0.1 & 150.1 $\pm$ 1.7 & 1.82 $\pm$ 0.11 & 0.85 $\pm$ 0.07 & -1.39 $\pm$ 0.05 & 150.7 $\pm$ 1.7 & 1.63 $\pm$ 0.06 \\
SN\,2007sr & 54511.2502 & 63.4 & 0.05 $\pm$ 0.05 & -0.03 $\pm$ 0.05 & 165.7 $\pm$ 24.9 & 0.01 $\pm$ 0.05 & 0.04 $\pm$ 0.05 & 0.0 $\pm$ 0.04 & 0.8 $\pm$ 24.9 & 0.0 $\pm$ 0.05 \\
SN\,2010ev & 55387.98468 & 2.9 & -1.8 $\pm$ 0.1 & 0.67 $\pm$ 0.08 & 79.8 $\pm$ 1.4 & 1.91 $\pm$ 0.09 & -1.52 $\pm$ 0.09 & 0.67 $\pm$ 0.06 & 78.1 $\pm$ 1.4 & 1.66 $\pm$ 0.09 \\
SN\,2010ko & 55539.19402 & -6.0 & -0.08 $\pm$ 0.13 & 0.12 $\pm$ 0.05 & 62.7 $\pm$ 17.1 & 0.09 $\pm$ 0.09 & 0.02 $\pm$ 0.88 & 0.1 $\pm$ 0.08 & 38.5 $\pm$ 17.1 & 0.0 $\pm$ 0.21 \\
SN\,2011ae & 55624.18558 & 4.0 & 0.25 $\pm$ 0.04 & 0.21 $\pm$ 0.03 & 19.6 $\pm$ 3.1 & 0.32 $\pm$ 0.04 & 0.19 $\pm$ 0.05 & 0.15 $\pm$ 0.04 & 19.3 $\pm$ 3.1 & 0.23 $\pm$ 0.05 \\
SN\,2012fr & 56267.31106 & 23.1 & 0.23 $\pm$ 0.06 & 0.06 $\pm$ 0.05 & 7.5 $\pm$ 7.5 & 0.23 $\pm$ 0.06 & 0.16 $\pm$ 0.01 & 0.0 $\pm$ 0.02 & 0.1 $\pm$ 7.5 & 0.16 $\pm$ 0.01 \\
SN\,2015ak & 57273.07036 & 4.9 & -0.36 $\pm$ 0.12 & -1.27 $\pm$ 0.08 & 127.1 $\pm$ 1.9 & 1.31 $\pm$ 0.09 & -0.34 $\pm$ 0.08 & -1.13 $\pm$ 0.05 & 126.7 $\pm$ 1.9 & 1.17 $\pm$ 0.06 \\
\hline
\multicolumn{11}{c}{SNe~Ia in elliptical host galaxies}\\
\hline
SN\,2004br & 53144.03228 & -3.9 & -0.07 $\pm$ 0.14 & 0.2 $\pm$ 0.1 & 53.9 $\pm$ 13.5 & 0.17 $\pm$ 0.1 & -0.03 $\pm$ 0.16 & 0.17 $\pm$ 0.13 & 49.7 $\pm$ 13.5 & 0.07 $\pm$ 0.14 \\
SN\,2008fl & 54736.11925 & 15.3 & 0.77 $\pm$ 0.24 & 0.53 $\pm$ 0.17 & 17.4 $\pm$ 6.6 & 0.89 $\pm$ 0.22 & 0.84 $\pm$ 0.08 & 0.65 $\pm$ 0.09 & 18.8 $\pm$ 6.6 & 1.05 $\pm$ 0.09 \\
SN\,2011iv & 55925.05084 & 19.5 & 0.09 $\pm$ 0.06 & -0.08 $\pm$ 0.05 & 159.2 $\pm$ 14.0 & 0.09 $\pm$ 0.06 & 0.08 $\pm$ 0.02 & -0.04 $\pm$ 0.03 & 166.1 $\pm$ 14.0 & 0.09 $\pm$ 0.03 \\
\hline
\multicolumn{11}{c}{SNe~Ia in S0 host galaxies}\\
\hline
SN\,2002el & 52501.18898 & -7.6 & 0.03 $\pm$ 0.13 & 0.23 $\pm$ 0.11 & 40.7 $\pm$ 14.1 & 0.17 $\pm$ 0.11 & 0.06 $\pm$ 0.15 & 0.16 $\pm$ 0.12 & 35.2 $\pm$ 14.1 & 0.09 $\pm$ 0.12 \\
SN\,2003hv & 52897.30336 & 6.1 & 0.03 $\pm$ 0.03 & 0.03 $\pm$ 0.05 & 25.0 $\pm$ 29.2 & 0.0 $\pm$ 0.04 & 0.02 $\pm$ 0.05 & 0.08 $\pm$ 0.03 & 38.5 $\pm$ 29.2 & 0.07 $\pm$ 0.03 \\
SN\,2003hx & 52911.34071 & 18.8 & -1.63 $\pm$ 0.39 & 1.0 $\pm$ 0.47 & 74.3 $\pm$ 6.2 & 1.83 $\pm$ 0.41 & -1.12 $\pm$ 0.14 & 0.8 $\pm$ 0.12 & 72.2 $\pm$ 6.2 & 1.36 $\pm$ 0.13 \\
SN\,2005cf & 53528.16586 & -5.8 & -0.61 $\pm$ 0.07 & 0.64 $\pm$ 0.04 & 66.8 $\pm$ 1.9 & 0.88 $\pm$ 0.06 & -0.7 $\pm$ 0.07 & 0.61 $\pm$ 0.04 & 69.4 $\pm$ 1.9 & 0.93 $\pm$ 0.06 \\
SN\,2005el & 53644.34433 & -2.7 & 0.35 $\pm$ 0.13 & -0.14 $\pm$ 0.11 & 169.2 $\pm$ 10.1 & 0.33 $\pm$ 0.13 & 0.41 $\pm$ 0.12 & -0.15 $\pm$ 0.15 & 169.7 $\pm$ 10.1 & 0.4 $\pm$ 0.13 \\
SN\,2007hj & 54361.10904 & 10.9 & -0.16 $\pm$ 0.16 & -0.26 $\pm$ 0.17 & 118.9 $\pm$ 15.2 & 0.22 $\pm$ 0.16 & -0.13 $\pm$ 0.09 & -0.14 $\pm$ 0.07 & 114.0 $\pm$ 15.2 & 0.16 $\pm$ 0.08 \\
SN\,2008ff & 54735.17383 & 31.0 & 0.59 $\pm$ 0.21 & -0.14 $\pm$ 0.14 & 173.4 $\pm$ 10.1 & 0.53 $\pm$ 0.21 & 0.46 $\pm$ 0.08 & -0.08 $\pm$ 0.09 & 175.1 $\pm$ 10.1 & 0.45 $\pm$ 0.08 \\
SN\,2008fp & 54736.33355 & 5.4 & 0.98 $\pm$ 0.06 & -1.99 $\pm$ 0.05 & 148.2 $\pm$ 0.7 & 2.22 $\pm$ 0.05 & 0.78 $\pm$ 0.04 & -1.62 $\pm$ 0.07 & 147.9 $\pm$ 0.7 & 1.8 $\pm$ 0.06 \\
\hline
\multicolumn{11}{c}{SNe~Ia in dwarf host galaxies}\\
\hline
SN\,2007if & 54363.17052 & 20.1 & 0.0 $\pm$ 0.7 & -0.8 $\pm$ 0.28 & 135.1 $\pm$ 9.9 & 0.7 $\pm$ 0.28 & -0.12 $\pm$ 0.14 & -0.57 $\pm$ 0.11 & 129.0 $\pm$ 9.9 & 0.57 $\pm$ 0.11 \\
		\hline
	\end{tabular}
\end{table*}

\clearpage
\onecolumn

{\scriptsize
\begin{longtable}{lllcclc}
\caption{\label{tab:hostinfo} Host-galaxy information for the SN sample and the normalised SN positions in terms of flux percentile.} \\
\hline
SNe Name & Host & Morph. & Morph. Source & Flux Percentile & Image Source & Fitted? \\
\hline
\endfirsthead
\caption{continued.}\\
\hline
SNe Name & Host & Morph. & Morph. Source & Flux Percentile & Image Source & Fitted? \\
\hline
\endhead
\hline
\endfoot
SN 2001dm & NGC 749 & S & NED & 0.286 & Legacy Survey & Yes \\
SN 2001el & NGC 1448 & Scd & NED & 0.829 & Legacy Survey & No \\
SN 2001V & NGC 3987 & Sb & NED & 0.931 & Legacy Survey & No \\
SN 2002bo & NGC 3190 & Sa & NED & 0.464 & Legacy Survey & Yes \\
SN 2002el & NGC 6986 & S$0^-$ & NED & 0.955 & Pan-STARRS & Yes \\
SN 2002fk & NGC 1309 & Sbc & NED & 0.264 & Legacy Survey & Yes \\
SN 2003eh & MCG +01-29-3 & Sb & NED & 0.476 & Legacy Survey & Yes \\
SN 2003hv & NGC 1201 & S$0^\degree$ & NED & 0.895 & Legacy Survey & Yes \\
SN 2003hx & NGC 2076 & S$0^+$ & NED & 0.070 & DSS & Yes \\
SN 2003W & UGC 5234 & Sc & NED & 0.070 & Legacy Survey & No \\
SN 2004br & NGC 4493 & E & NED & 0.269 & Legacy Survey & Yes \\
SN 2004dt & NGC 799 & Sa & NED & 0.298 & Legacy Survey & Yes \\
SN 2004ef & UGC 12158 & Sb & NED & 0.284 & Legacy Survey & Yes \\
SN 2004eo$^a$ & NGC 6928 & Sab & NED & 0.955 & Legacy Survey & Yes \\
SN 2005cf & MCG -01-39-3 & S0 & NED & 0.999 & Pan-STARRS & Yes \\
SN 2005de & UGC 11097 & S & NED & 0.910 & Legacy Survey & No \\
SN 2005df & NGC 1559 & Scd & NED & 0.904 & Legacy Survey & Yes \\
SN 2005el & NGC 1819 & S0 & NED & 0.995 & DSS & No \\
SN 2005hk & UGC 272 & Sd & NED & 0.853 & Legacy Survey & No \\
SN 2005ke & NGC 1371 & Sa & NED & 0.664 & Legacy Survey & Yes \\
SN 2006X & NGC 4321 & Sbc & NED & 0.380 & Legacy Survey & Yes \\
SN 2007fb & UGC 12859 & Sbc & NED & 0.343 & Legacy Survey & No \\
SN 2007hj & NGC 7461 & S0 & NED & 0.659 & Legacy Survey & Yes \\
SN 2007if$^b$ & ... & Dwarf & \citet{2010ApJ...713.1073S} & 0.500 &  & Yes \\
SN 2007le & NGC 7721 & Sc & NED & 0.127 & Legacy Survey & Yes \\
SN 2007sr & NGC 4038 & Sm & NED & 1.000 & DSS & No \\
SN 2008ff & ESO 284- G 032 & S0/a & NED & 0.991 & DSS & Yes \\
SN 2008fl & NGC 6805 & E & NED & 0.652 & DSS & Yes \\
SN 2008fp$^c$ & ESO 428-G14 & S$0^\degree$ & NED & 0.671 & Pan-STARRS & Yes \\
SN 2010ev & NGC 3244 & Scd & NED & 0.105 & DSS & No \\
SN 2010ko & NGC 1954 & Sbc & NED & 1.000 & DSS & Yes \\
SN 2011ae & MCG -03-30-19 & Scd & NED & 0.199 & Pan-STARRS & Yes \\
SN 2011iv & NGC 1404 & E & NED & 0.343 & Legacy Survey & Yes \\
SN 2012fr & NGC 1365 & Sb & NED & 0.345 & Legacy Survey & Yes \\
SN 2015ak & ESO 108-21 & S & NED & 0.873 & Legacy Survey & No \\
SN 2018cif & IC 1422 & S & This work & 0.952 & Pan-STARRS & Yes \\
SN 2018evt & MCG -01-35-011 & S$0^-$ & NED & 0.433 & Pan-STARRS & Yes \\
SN 2018fhw & WISEA J041805.98-633651.7 & E & This work & 0.564 & Legacy Survey & Yes \\
SN 2018fhx & WISEA J062438.18-234354.5 & S$0^\degree$ & NED & 0.409 & Pan-STARRS & Yes \\
SN 2018fnq & ESO 284- G 026 & S$0^-$ & LEDA & 0.664 & Legacy Survey & No \\
SN 2018fqd & 2MASX J20515213-3650127 & S & LEDA & 0.697 & DSS & Yes \\
SN 2018fqn$^d$ & GALEXASC J223938.25-150510.0 & ... & ... & 0.279 & Pan-STARRS & Yes \\
SN 2018fsa & WISEA J222536.90-130431.5 & S or E & This work & 0.865 & Pan-STARRS & Yes \\
SN 2018fuk & ESO 016- G 011 & S & NED & 0.911 & DSS & Yes \\
SN 2018fvi & AM 0156-672 & S & NED & 0.715 & DSS & No \\
SN 2018fvy & WISEA J013216.87-330602.3 & S & This work & 0.899 & Legacy Survey & Yes \\
SN 2018fzm$^e$ & WISEA J210319.95-513250.7 & S & NED & 0.332 & Pan-STARRS & Yes \\
SN 2018gyr & WISEA J004947.23-613914.0 & S0 & LEDA & 0.526 & Legacy Survey & Yes \\
SN 2018hsa & NGC 7038 & Sc & NED & 0.996 & Pan-STARRS & Yes \\
SN 2018hsy & CGCG 091-100 & Sa & LEDA & 0.106 & Pan-STARRS & Yes \\
SN 2018hts$^f$ & SDSS J232432.96+170537.9 & ... & ... & 0.617 & Legacy Survey & No \\
SN 2018htt & NGC 1209 & E & NED & 0.093 & Pan-STARRS & Yes \\
SN 2018jbm & LCRS B014013.3-453958 & S & This work & 0.997 & DSS & No \\
SN 2018jdq & WISEA J112525.69-082759.8 & S & This work & 0.390 & Pan-STARRS & Yes \\
SN 2018jeo & ESO 564- G 014 & S & NED & 1.000 & Pan-STARRS & Yes \\
SN 2018jff & LCRS B234528.8-451528 & S & This work & 0.112 & Legacy Survey & Yes \\
SN 2018jgn & WISEA J000255.82-265450.6 & E & NED & 0.140 & Pan-STARRS & Yes \\
SN 2018jky & NGC 1329 & Sa & NED & 1.000 & Pan-STARRS & Yes \\
SN 2018jny & SDSS J092438.14+252352.7 & S & LEDA & 0.979 & Pan-STARRS & Yes \\
SN 2018jrn & ESO 575-IG 016 NED01 & E & This work & 0.355 & Pan-STARRS & Yes \\
SN 2018jtj & SDSS J124103.32+080417.4 & Sc & LEDA & 0.624 & Pan-STARRS & Yes \\
SN 2018kav & WISEA J045054.61-175913.1 & S & This work & 0.189 & Pan-STARRS & Yes \\
SN 2018koy$^g$ & ... & ... & ... & 0.971 & DSS & Yes \\
SN 2018kyi & WISEA J070709.55+261930.2 & ... & ... & 0.202 & Pan-STARRS & Yes \\
SN 2019axg & MCG -01-27-027 & Sab & LEDA & 0.756 & Pan-STARRS & Yes \\
SN 2019bak & WISEA J101331.71+293304.9 & S & LEDA & 0.041 & Pan-STARRS & Yes \\
SN 2019baq & WISEA J122201.57+202030.1 & E & LEDA & 0.657 & Pan-STARRS & Yes \\
SN 2019bdz & CGCG 048-018 & Sbc & LEDA & 0.901 & Pan-STARRS & Yes \\
SN 2019bff & WISEA J141432.08+171358.6 & S & This work & 0.181 & Pan-STARRS & No \\
SN 2019bjw & WISEA J105547.49+274016.3 & Sb & LEDA & 0.036 & Pan-STARRS & Yes \\
SN 2019bjz$^h$ & III Zw 075 NOTES01 & E & This work (SDSS Spectrum) & 0.007 & Pan-STARRS & Yes \\
SN 2019bkh & MCG +05-31-141 & Sb & LEDA & 0.125 & Pan-STARRS & Yes \\
SN 2019bpb & WISEA J051209.21-435515.8 & S or E & This work & 0.996 & DSS & Yes \\
SN 2019kg & WISEA J114245.72+214253.3 & S or E & This work & 0.016 & Pan-STARRS & Yes \\
SN 2019np & NGC 3254 & Sbc & NED & 0.853 & Pan-STARRS & Yes \\
SN 2019rm & WISEA J055313.73-730655.3 & Ir & This work (FORS2 Image) & 0.780 & DSS & No \\
SN 2019rx & WISEA J052200.10-071132.9 & S & This work & 0.643 & Pan-STARRS & Yes \\
SN 2019rzm & MCG -03-04-038 & S0a & LEDA & 0.777 & Pan-STARRS & Yes \\
SN 2019shq & ESO 432-IG 006 NED02 & Sa & LEDA & 0.241 & DSS & Yes \\
SN 2019ubs & UGC 01895 & Sc & NED & 0.721 & Pan-STARRS & Yes \\
SN 2019ubt & WISEA J232957.47+294943.5 & S & This work & 0.556 & Pan-STARRS & Yes \\
SN 2019udk & CGCG 383-067 & Sbc & LEDA & 0.378 & Pan-STARRS & Yes \\
SN 2019uhz & WISEA J014714.00-082745.2 & Sab & LEDA & 0.685 & Legacy Survey & Yes \\
SN 2019ujy & 2MASX J19415466-2010042 & E & This work & 0.135 & Pan-STARRS & Yes \\
SN 2019ulp & UGC 04431 & Sbc & NED & 0.176 & Legacy Survey & Yes \\
SN 2019umr & ESO 533- G 006 & S & NED & 0.204 & Legacy Survey & Yes \\
SN 2019uoo & MCG -02-22-024 & S0a & LEDA & 0.523 & Pan-STARRS & Yes \\
SN 2019upw & NGC 0109 & Sa & NED & 0.587 & Legacy Survey & Yes \\
SN 2019urn & WISEA J005749.28-173826.7 & E & This work & 0.091 & Legacy Survey & Yes \\
SN 2019vju & NGC 3514 & Sc & NED & 0.469 & Pan-STARRS & Yes \\
SN 2019vnj & WISEA J110901.05-143809.5 & S & This work & 0.224 & Pan-STARRS & Yes \\
SN 2019vrq & GALEXASC J030421.67-160124.7 & S & This work & 0.012 & Legacy Survey & No \\
SN 2019vsa & IC 2995 & Sc & NED & 0.238 & Pan-STARRS & No \\
SN 2019vv & CGCG 022-015 & S0a & LEDA & 0.581 & Legacy Survey & No \\
SN 2019wka & SDSS J031440.16-005213.8 & S & This work & 0.375 & Legacy Survey & No \\
SN 2019wqz & WISEA J114717.39-291104.3 & S & LEDA & 0.987 & Pan-STARRS & Yes \\
SN 2020bpz & 2MFGC 09157 & S & This work & 0.363 & Pan-STARRS & No \\
SN 2020ckp & WISEA J083621.44-253200.8 & Sbc & LEDA & 0.283 & Pan-STARRS & Yes \\
SN 2020clq & NGC 1613 & S$0^+$ & NED & 0.307 & Legacy Survey & Yes \\
SN 2020cpt & CGCG 076-111 & E & LEDA & 0.762 & Legacy Survey & Yes \\
SN 2020ctr & WISEA J085837.55+201131.6 & E & LEDA & 0.016 & Legacy Survey & Yes \\
SN 2020dvr & ESO 566- G 027 & Sbc & NED & 0.659 & Pan-STARRS & Yes \\
SN 2020dyf & WISEA J153030.68-123612.4 & Sb & NED & 0.400 & Pan-STARRS & Yes \\
SN 2020ejm & IC 2560 & SB(r)b & NED & 0.029 & DSS & Yes  \\
\end{longtable}
\noindent
\footnotesize{Notes: The morphological classifications are taken from NED \citep{1991ASSL..171...89H}, the HyperLEDA database \citep{Makarov2014_HyperLEDA}, or have been determined in this work. The ``Image Source'' indicates the source of the images used for the SN location determination, and the ``Fitted?'' column indicates whether the isophote fitting converged.}\\
\footnotesize{$^a$ Used red filter.}\\
\footnotesize{$^b$ Host not listed in NED, the flux percentile was visually estimated.}\\
\footnotesize{$^c$ Had to use $z$ filter, $gri$ filters had issues with artifacts.}\\
\footnotesize{$^d$ Host data from the  \href{http://stella.sai.msu.ru/~pavlyuk/sncat/snentireinfo.php?snname=2018fqn}{Sternberg Astronomical Institute Supernova Group}.}\\
\footnotesize{$^e$ Bright objects in proximity may affect the result.}\\
\footnotesize{$^f$ Host data from \href{http://vizier.u-strasbg.fr/viz-bin/VizieR?-source=&-out.add=_r&-out.add=_RAJ\%2C_DEJ&-sort=_r&-to=&-out.max=20&-meta.ucd=2&-meta.foot=1&-c=351.13731848+17.09368216&-c.rs=10}{Vizier}.}\\
\footnotesize{$^g$ Unknown host galaxy.}\\
\footnotesize{$^h$ Too many objects in area, flux profile did not flatten.}\\
}

\bsp	
\label{lastpage}
\end{document}